\documentclass[11pt]{article}
\pdfoutput=1
\usepackage{jheppub}

\usepackage{amssymb,amsmath}
\usepackage{amsfonts}
\usepackage{mathtools}
\allowdisplaybreaks

\usepackage{physics}
\usepackage{cancel}
\usepackage{enumitem}
\usepackage{bm}
\usepackage{mathrsfs}

\let\originalleft\left
\let\originalright\right
\renewcommand{\left}{\mathopen{}\mathclose\bgroup\originalleft}
\renewcommand{\right}{\aftergroup\egroup\originalright}

\makeatletter
\newenvironment{equations}[1][]{\subequations\ifx\relax#1\relax\else\label{#1}\fi\align\ignorespaces}{\endalign\ignorespacesafterend\endsubequations}
\def\@spliteq#1{\begin{equation}\begin{split}#1\end{split}\end{equation}}
\def\splitequation{\collect@body\@spliteq}

\makeatother

\usepackage{xspace}
\renewcommand{\d}{\ifmmode\operatorname{d}\!\else\textrm{d}\xspace\fi}

\newcommand\de{\partial}
\newcommand\str{\text{str}}

\newcommand{\rvline}{\hspace*{-\arraycolsep}\vline\hspace*{-\arraycolsep}}

\newcommand{\nn}{\nonumber}
\newcommand{\diff}{\mathrm{d}}

\makeatletter
\gdef\@fpheader{}
\makeatother

\title{Superconformal anomalies from superconformal Chern-Simons polynomials}

\author[a,b]{Camillo Imbimbo,}
\author[c,d]{Davide Rovere,}
\author[e]{and Alison Warman}
\affiliation[a]{Dipartimento di Fisica, Universit\`a di Genova, Via Dodecaneso 33, 16146 Genoa, Italy}
\affiliation[b]{INFN, Sezione di Genova, Genoa, Italy}
\affiliation[c]{Dipartimento di Fisica e Astronomia ``Galileo Galilei'', Universit\`a di Padova, Via F. Marzolo 8, 35131 Padua, Italy}
\affiliation[d]{INFN, Sezione di Padova, Padua, Italy}
\affiliation[e]{Mathematical Institute, University of Oxford, Woodstock Road, Oxford, OX2 6GG, United Kingdom}

\emailAdd{camillo.imbimbo@ge.infn.it}
\emailAdd{davide.rovere@pd.infn.it}
\emailAdd{warman@maths.ox.ac.uk}

\abstract{
We consider the 4-dimensional  $\mathcal{N}=1$ Lie superconformal algebra and search for completely ``symmetric'' (in the graded sense)  3-index invariant tensors. The solution we find is unique and we show that the corresponding invariant polynomial cubic in the generalized curvatures of superconformal gravity vanishes.  Consequently, the associated Chern-Simons polynomial is a non-trivial anomaly cocycle. We explicitly compute this cocycle  to all orders in the independent fields of superconformal gravity and establish that it is BRST equivalent to the so-called superconformal $a$-anomaly.  We briefly discuss the possibility that the superconformal $c$-anomaly also admits a similar Chern-Simons formulation and the  potential holographic, 5-dimensional, interpretation of our results.

}


\begin{document}
\maketitle

\section{Introduction and summary}

Historically, anomalies were first discovered by means of perturbative computations \cite{Adler:1969gk}, \cite{Bell:1969ts}.  The BRST formulation of gauge theories uncovered a cohomological, non-perturbative,  interpretation of anomalies as  cocycles of ghost number 1 of  the BRST operator acting on the infinite-dimensional space of  fields \cite{Becchi:1975nq}. Later on it was understood that the anomaly cocycles  of both Yang-Mills and gravitational theories are simply related to  (a natural extension of the)   secondary Chern-Simons characteristic classes \cite{Stora:1976LM}, \cite{Langouche:1984gn}. This beautiful topological understanding of  Yang-Mills and gravitational anomalies simplified enormously  their computation in arbitrary dimensions  and for general gauge groups. It was also fruitful in many  applications to string theory \cite{Green:1984sg} and holography \cite{Maldacena:1997re}, \cite{Witten:1998qj}.

While the BRST cohomological interpretation of anomalies is universal, the link between BRST anomaly cocycles and Chern-Simons classes is not. To date, neither Weyl anomalies nor supersymmetry anomalies have been associated with Chern-Simons invariants. This is at least one reason why their computation and classification are both less comprehensive and more intricate compared to Yang-Mills and gravitational theories \cite{Piguet:1981rr,Bonora:1983ff,Bonora:1984pz,Bonora:1984pn,Bonora:1985cq,Bonora:1985ug,Deser:1993yx,Brandt:1996au,Brandt:2002pa,Boulanger:2007ab,Bonora:2013rta}. In this article we extend  to conformal and supersymmetric theories the connection between anomalies and secondary Chern-Simons classes. Specifically, we will show that the generalized Chern-Simons invariant associated to the $d=4$, $\mathcal{N}=1$ Lie superconformal algebra computes one of the two independent anomalies  of 4-dimensional superconformal gravity.

To put this result in the appropriate context, let us review the connection between Yang-Mills anomalies and Chern-Simons polynomials as uncovered by Stora and Zumino  \cite{Stora:1976LM,Stora:1984,Zumino:1983ew,Manes:1985df}.\footnote{ In appendix \ref{appendix:SZ} we review the details of the relation between the Stora-Zumino formulation of anomalies by means of generalized forms and the so-called ``two-step descent'' procedure in which one extends ordinary forms to higher-dimensions. }   Their  idea is to introduce on $d$-dimensional  space-time $M_d$  a \emph{generalized connection}  $ \bm{A}$ with values in the Lie algebra  $\mathfrak{g}$ of the gauge group:  $ \bm{A}$  is defined to be the sum of the gauge field $A$ and its corresponding ghost $c$,  
\begin{equation} 
\label{eq:intro_A_c}
 \bm{A}=c+A.
\end{equation}
They also  introduced  a  \emph{generalized BRST operator $\delta$}
 \begin{equation} \label{eq:intro_delta}
   \delta=s+\dd,
 \end{equation}
where $s$ is the BRST operator and $\dd$ the de Rham exterior differential acting on forms.  $\bm{A}$  is a generalized form of \emph{total degree } --- defined to be the sum of ghost number and form degree ---  equal to +1.\footnote{Generalized forms $\bm{\Omega}_n =\sum_{p+q=n} \Omega^{(p)}_q$ of  total degree $n$ are the sum of ordinary forms $\Omega^{(p)}_q$ of different form degrees $p$ and ghost numbers $q$, such that $q+p=n$.}  $\delta$ increases the total degree by 1. It is essential to keep in mind that, unlike ordinary forms,
generalized forms of total degree $n$ greater than the space-time dimension $d$ do not in general vanish.  

$\dd$ and $s$ are taken to anticommute with each other: hence the nilpotency of $\delta$ is equivalent to the nilpotency of the BRST operator $s$.  The cohomology of $\delta$ on the space of generalized local  forms is isomorphic to the  cohomology of $s$ modulo $\dd$ on the space of local ordinary forms. Therefore  anomalies are obtained by integrating $\delta$-cocycles with total degree $d+ 1$ on the space-time manifold $M_d$.  These are local functionals of ghost number 1.

Given the generalized connection  $\bm{A}$ and  the generalized BRST operator $\delta$ one can define the corresponding curvature 
\begin{equation}\label{eq:intro_curvature}
\bm{F} = \delta \,\bm{A} + \bm{A}^2,
\end{equation}
which is a generalized form of total degree +2 with values in the adjoint representation of the gauge Lie algebra $\mathfrak{g}$. The generalized curvature satisfies the Bianchi identity
 \begin{equation}\label{intro:bianchi}
\delta\, \bm{F} + [\bm{A}, \bm{F}] =0,
\end{equation}
by virtue of the nilpotency of $\delta$.  Therefore, when $d$ is even, $\mathfrak{g}$-invariant polynomials $P_{\frac{d}{2}+1}( \bm{F})$ of  $\bm{F}$ of degree $\tfrac{d}{2}+1$ are $\delta$-cocycles of total degree $d+2$
\begin{equation}
\delta \, P_{\frac{d}{2}+1}(\bm{F})=0.
\end{equation}
Because of the curvature definition (\ref{eq:intro_curvature}),  $P_{\frac{d}{2}+1}( \bm{F})$ is also $\delta$-exact
\begin{equation}
P_{\frac{d}{2}+1}(\bm{F})= \delta \, Q_{d+1}(\bm{A} , \bm{F}),
\end{equation}
where  $Q_{d+1}(\bm{A} , \bm{F})$ are the celebrated (generalized) Chern-Simons polynomials. They  are (non-gauge invariant) generalized forms of total  degree $d+1$. The dependence of  $Q_{d+1}(\bm{A} , \bm{F})$ on the generalized connection $\bm{A}$ and curvature $\bm{F}$ is  just the same as the dependence of ordinary Chern-Simons polynomials in  $d+1$ dimensions on the ordinary connection $A$ and curvature $F$. However, as stressed above, generalized  Chern-Simons  polynomials $Q_{d+1}(\bm{A} , \bm{F})$ {\it do not} in general vanish in $d$ dimensions. 

The relevance of Chern-Simons polynomials in ordinary form cohomology is the following: ordinary  Chern-Simons forms $Q_{d+1}(A , F)$ are  not in general closed and, as such,  they do {\it not} define de Rham cohomology classes. However there are situations in which some ordinary curvature characteristic class  $P_{\frac{d}{2}+1}(F)$,  ``accidentally''  {\it vanishes} on manifolds $M_n$ of dimension $n\ge d+2$: in that case  the corresponding $Q_{d+1}(A , F)$ is closed and it defines a characteristic class on $M_n$ of form degree $d+1$,  which is called \emph{secondary} for this reason. 

Going back to the BRST cohomology, the central observation of Stora and Zumino was that, for Yang-Mills  (and gravitational \cite{Bardeen:1984pm}) gauge theories, the generalized curvature $ \bm{F}$  is  actually ``horizontal'',  which means that its higher ghost number components vanish
\begin{equation} \label{eq:intro_genF}
    \bm{F}=F.
\end{equation}
It follows that
\begin{equation} 
 P_{\frac{d}{2}+1}(\bm{F}) =  P_{\frac{d}{2}+1}(F)=0, 
\end{equation} 
as ordinary forms of degree $d+2$ do vanish  in dimension  $d$.  Hence, in the Yang-Mills BRST context one finds oneself in the precise analogue of the situation in which ordinary secondary characteristic classes arise in ordinary form cohomology: the (``primary'') characteristic class $P_{\frac{d}{2}+1}(\bm{F})$ vanishes and hence the generalized  secondary $Q_{d+1}(\bm{A} , \bm{F})$ is $\delta$-closed
\begin{equation}
\delta \,Q_{d+1}(\bm{A} , \bm{F}) =0.
\end{equation}
By integrating $Q_{d+1}(\bm{A} , \bm{F})$   on $M_d$  one obtains an anomaly  cocycle. For Yang-Mills and gravitational theories all anomaly cocycles can be obtained in this way \cite{Brandt:1989rd}.

 The basic novelty one encounters when considering  either  supersymmetry or conformal symmetries is that the generalized curvature $ \bm{F}$ defined by the corresponding BRST transformations ceases to be horizontal. Characteristic classes $ P_{\frac{d}{2}+1}(\bm{F})$ are then not guaranteed to vanish and this potentially negates the relevance of the generalized Chern-Simons polynomial $Q_{d+1}(\bm{A} , \bm{F})$ to anomalies. 
 
Although the non-horizontality of the generalized  $ \bm{F}$ is a generic feature of both conformal and supersymmetry transformations,
 let us illustrate how it comes about in $d=4\,,\,\mathcal{N}=1$ conformal supergravity, the field theory we are going to explore in this paper \cite{Kaku:1977rk,Kaku:1978nz,vanNieuwenhuizen:2004rh,VanNieuwenhuizen:1981ae,Fradkin:1985am}.\footnote{For a recent review of  4-dimensional $\mathcal{N}=1$ superconformal gravity see also \cite{Freedman:2012zz}. The BRST formulation of the same theory was first worked out in \cite{Baulieu:1986ab}. }  Conformal supergravity is a ``pure gauge'' theory: it has neither auxiliary nor ``matter'' fields.
 Its gauge fields are 1-forms with values in the appropriate bundles
 \begin{equation}
      A^i=\;\{\;e^a,\;\omega^{ab},\;b,\;a,\; f^a,\;\psi^\alpha,\;\Tilde{\psi}^\alpha \},
\end{equation}
each one in correspondence with  the generators of  the $\mathfrak{su}(2,2|1)$ Lie superconformal algebra:\footnote{We will use the index $i$ as the index running along all the 24  generators  $T_i$ of $\mathfrak{su}(2,2|1)$. $P^a$ and $J^{ab}$ are the generators of translations and Lorentz transformations, the Weyl  (dilatation) generator is denoted by $W$, the R-symmetry charge by $R$, $K^a$ denotes the special conformal generators, $Q^\alpha$ and $S^\alpha$ are, respectively,  the supersymmetry charges and conformal supersymmetry charges. Our conventions for the $d=4$, $\mathcal{N}=1$ superconformal algebra are reviewed in appendix \ref{appendixB} and properties of the supertrace in appendix \ref{appendix:super-Lie-algebras}.}
 \begin{equation}\label{intro:Ti}
 T_i=\;\{P^a,\;J^{ab},\;W,\; R,\; K^a,\;Q^\alpha,\; S^\alpha \}.
\end{equation}
When one attempts to define  the analogue of the generalized connection (\ref{eq:intro_A_c}) one faces a complication which is common to
all theories which include gravity: the ghosts associated to translations $P^a$ are not valued in the frame bundle but are instead valued in the space-time tangent bundle. We denote the BRST ghosts of conformal supergravity as:
 \begin{equation}\label{intro:superformalghosts}
 c^i=\{\xi^\mu,\; \Omega^{ab},\;\sigma,\;\alpha,\;\theta^a,\;\zeta^\alpha,\; \eta^\alpha\}.
 \end{equation}
Diffeomorphisms must therefore be treated differently to the rest of the Lie superalgebra transformations.
As we will explain in Section \ref{sec:2}, this has a twofold effect \cite{Baulieu:1985md,Baulieu:1986ab,Imbimbo:2018duh,Frob:2021sao}. First, one has to introduce a novel BRST operator $\hat{s}$, 
 ``equivariant'' with respect to diffeomorphisms
\begin{equation}\label{intro:hats}
\hat{s} = s + \mathcal{L}_\xi, 
\end{equation}
where $\mathcal{L}_\xi$ denotes the Lie derivative along the vector field $\xi^\mu$. The sign in (\ref{intro:hats}) is chosen so that  the (diffeomorphism) equivariant $\hat{s} $ includes all transformations {\it other than} diffeomorphisms.  Correspondingly, the generalized connection (\ref{eq:intro_A_c}) is defined to be\footnote{We will denote with bold letters the generalized forms: $\bm{A}^i$  for the generalized connections and $\bm{F}^i$ for the generalized curvatures.} 
\begin{equation}
    \bm{A}^i=\;\{ e^a,\;  \Omega^{ab}+ \omega^{ab},\;\sigma+b,\;\alpha+a,\; \theta^a+f^a,\;\zeta^\alpha+\psi^\alpha,\; \eta^\alpha+\Tilde{\psi}^\alpha \}.
\end{equation}
In other words, the ghost number 1 component of the generalized connection along $P^a$  is taken to vanish.  Let us observe that the (1-form) components of the generalized connection along the bosonic (fermionic)  generators  of the Lie superalgebra are respectively anti-commuting (commuting). Hence it is convenient to introduce
\begin{equation}\label{intro:boldh}
    \bm{A} \equiv    \bm{A}^i\, T_i
    \end{equation}
    and take (fermionic) bosonic generators $T_i$ to be (anti)commuting: in this way    $\bm{A}$ is {\it anti-commuting}.
    
Supersymmetric theories require one more step: the definition of the generalized BRST operator (\ref{eq:intro_delta}) must be modified to include one third term \cite{Imbimbo:2018duh}
 \begin{equation}\label{eq:intro_delta_new}
  \delta = \hat{s}+\dd-i_\gamma.
 \end{equation}
 $i_\gamma$ is the nilpotent operator which contracts an ordinary form along the {\it commuting} vector field $\gamma^\mu$ bilinear
 in the supersymmetry ghost $\zeta_\alpha$:\footnote{In this article Dirac gamma matrices are denoted by $\Gamma^\mu$ to avoid confusing them with the ghost bilinear $\gamma^\mu$.}
 \begin{equation}\label{intro:gammamu}
 \gamma^\mu \equiv \bar\zeta\, \Gamma^\mu\, \zeta.
 \end{equation}
 Nilpotency of the generalized $\delta$ is equivalent to 
that of the BRST operator $s$:\footnote{ This is a consequence of the BRST transformation of the diffeomorphism ghost $\xi^\mu$
\begin{equation}
s\,\xi^\mu=-\tfrac{1}{2}\mathcal{L}_\xi \xi^\mu+\gamma^\mu
\end{equation} 
  and the relations, valid on forms, for $\hat{s}$, $\dd$, and $i_\gamma$
 \begin{equation}
\hat{s}^2 = \mathcal{L}_\gamma, \qquad \mathcal{L}_\gamma = \{ \dd, i_\gamma\},\qquad i_\gamma^2=0.
 \end{equation}}
\begin{equation}\label{eq:intro_deltanilpotence}
\delta^2=0\qquad  \Leftrightarrow \qquad  s^2=0.
 \end{equation}

 Given these ingredients, one  proceeds to define the generalized curvatures associated to the $\mathfrak{su}(2,2|1)$ Lie superconformal algebra exactly as in (\ref{eq:intro_curvature}) 
  \begin{equation}\label{intro:supercurvature}
 \bm{F} = \delta\, \bm{A} +  \bm{A}^2,
 \end{equation}
 where 
 \begin{equation}
  \bm{F}\equiv\bm{F}^i\,T_i
  \end{equation} 
  is a {\it commuting}  generalized form of total fermionic number +2 which satisfies the generalized Bianchi identity (\ref{intro:bianchi}).  However, unlike the Yang-Mills and gravitational case,  $\bm{F}$ does not turn out to be ``horizontal'': rather one finds that
 \begin{equation}\label{intro:lambda0}
    \bm{F}=F+\lambda_0.
\end{equation}
$F$ is an ordinary 2-form of ghost number 0 and $\lambda_0$ is a (non-vanishing)  1-form with ghost number 1, with values in the Lie superconformal algebra $\mathfrak{su}(2,2|1)$.  We will denote the components $\lambda_0^i$  of $\lambda_0$  as 
\begin{equation}
  \lambda_0^i = \{\lambda_0^P, \; \lambda_0^J, \; \lambda_0^W, \;\lambda_0^R,\;\lambda_0^K,\; \lambda_0^Q, \;\lambda_0^S\},
  \end{equation}
 following the same order of the generators as in eq. (\ref{intro:Ti}).

The emergence of a  non-vanishing non-horizontal curvature component $\lambda_0$ is intimately tied with  presence of the extra term $i_\gamma$ in the definition of the generalized BRST operator (\ref{eq:intro_delta_new}): this term  encodes the effect of  coupling supergravity to YM gauge fields. The BRST transformations of the ghost fields 
are ---  for both bosonic YM and conformal supergravity --- ``geometric'': they are  fixed by the structure constants of the underlying Lie (super)algebra and  nilpotency is ensured by the (super)Jacobi identities of the corresponding Lie (super)algebras. 
In the bosonic YM  and gravitational case,  the BRST transformation rules for the ghosts {\it also} uniquely  fix  the familiar, ``geometric'' BRST transformation rules for the connections: those transformation rules  are not deformable.  In short, the BRST transformations of both ghosts and connections are, for  both YM and gravity, completely dictated by the geometry of the underlying Lie algebra. This ceases to be true in the supersymmetric context: as we will explain in Section \ref{sec:2},  nilpotency of the BRST operator on the ghosts of conformal supergravity determines  the transformations of the connections {\it only up to 1-forms of ghost number 1 which are $i_\gamma$-closed} --- precisely because of the presence of $i_\gamma$ in the definition of the generalized $\delta$, eq. (\ref{eq:intro_delta_new}). These 1-forms of ghost number 1 are the $\lambda_0$'s  which appear in eq. (\ref{intro:lambda0}), which  indeed  do satisfy
\begin{equation}\label{intro:igammalambda0}
i_\gamma(\lambda_0)=0.
\end{equation}
Eq. (\ref{intro:igammalambda0}) restricts the general form of the  $\lambda_0$'s to be
\begin{equation}
\lambda_0 = e^a \,\bar\zeta\, \Gamma_a\, X,
\end{equation}
where $X$ has ghost number 0 and is valued in $\mathfrak{su}(2,2|1)$. $X$  is  fixed by  the requirement of nilpotency of BRST transformations on connections themselves, as we will explain in Sections \ref{sec:2} and \ref{sec:sol_for_constraints}:  it  turns out to be non-vanishing.  

Nilpotency of BRST transformations on gauge fields has one more implication: the ghost number 0 components $F^i$ of the  generalized curvatures must satisfy certain constraints, which we will also review in Section   \ref{sec:2}. These are  supersymmetric extensions of the familiar zero-torsion constraint of general relativity. Superconformal gravity constraints are algebraic equations for the gauge fields $\{\omega^{ab}, \;f^a,\;\Tilde{\psi}^\alpha \}$, which can be solved to express them locally  in terms of the physical fields $\{e^a, b,\;a,\;\psi^\alpha \}$. It is an interesting fact that the non-horizontal components $\lambda_0^i$ take values only in the ``unphysical'' directions $\{J^{ab},\;K^a,\; S^\alpha \}$ of the Lie superconformal algebra $\mathfrak{su}(2,2|1)$.

   A priori,  the lack of horizontality of the generalized curvature jeopardizes the  Stora-Zumino mechanism to produce BRST anomaly cocycles.  However,  horizontality of the generalized curvature is a sufficient but not  necessary condition for the existence of secondary Chern-Simons classes. It is the vanishing of the characteristic classes $P_{3}(\bm{F})$ that is strictly necessary for the secondary classes to emerge. We therefore searched for  $\mathfrak{su}(2,2|1)$ invariant cubic polynomials and found that there exists only one of them, up to a multiplicative constant. We computed the corresponding Chern class $P_{3}(\bm{F})$
\begin{equation}
P_{3}(\bm{F})=\Tilde{d}_{ijk}\,\bm{F}^i\,\bm{F}^j\,\bm{F}^k
\end{equation}
and found, remarkably, that it indeed vanishes --- despite the non-horizontality of $\bm{F}$!  The corresponding Chern Simons generalized form $Q_5( \bm{A},\bm{F})$ does therefore define, upon integration over space-time $M_4$, a superconformal anomaly  which we compute explicitly, in components and exactly to all orders in the number of fields, and present in Section \ref{sec:superanomaly}, eqs. (\ref{superconfinvariantanomaly}-\ref{sec4: CS3}). 

The Chern-Simons anomaly cocycle $Q_5(\bm{A}, \bm{F})$ is, by construction,  invariant under {\it rigid} $\mathfrak{su}(2,2|1)$  transformations. It depends on all the ghosts (\ref{intro:superformalghosts}) of the $\mathfrak{su}(2,2|1)$ Lie superalgebra, with the exception of the diffeomorphism ghosts $\xi^\mu$.\footnote{There are no diffeomorphism anomalies in 4-dimensions, so this is  expected from the start. In our scheme, the functional space does not contain $\xi^\mu$ at all.   In other dimensions, diffeomorphism anomalies would translate into Lorentz anomalies.} In particular it also depends on the ghosts $\Omega^{ab}$  and $\theta^a$ associated, respectively,  with local Lorentz and special conformal transformations. In Section \ref{sec:equivariantcocycle}  we will  show that one can add BRST-trivial cocycles to the Chern-Simons cocycle $Q_5(\bm{A},\bm{F})$ to obtain   {\it equivalent} anomaly cocycles  $Q^{equiv}_5(\bm{A},\bm{F})$  (eq.(\ref{sec4:Qequiv5final})) which are  independent of the Lorentz ghosts  $\Omega^{ab}$. We call the $\Omega^{ab}$-independent representatives of the anomaly ``Lorentz-equivariant'' cocycles. They lead to anomalous Ward identities which involve a  symmetric, conserved but not traceless stress-energy tensor $\mathcal{T}_{\mu\nu}$. 

In Section \ref{sec:equivariantcocycle}  we will show that one can also choose representatives of the anomaly, which beyond being independent of  $\Omega^{ab}$, are  also independent   of  $\theta^a$.   Such anomaly cocycles do  not depend on the Weyl gauge connection $b$ either:  this is so since $b$ and (a suitable completion of) $\theta^a\, e_a$ form a BRST trivial pair.    

It should be emphasized that the $\Omega^{ab}$ and $\theta^a$  independent cocycles  are no longer invariant under the full rigid $\mathfrak{su}(2,2|1)$  Lie superconformal algebra. They lead to anomalous Ward identities which involve the symmetric, conserved but not traceless stress-energy tensor $\mathcal{T}_{\mu\nu}$, the R-symmetry current $\mathcal{J}_\mu$ and the supersymmetry current $\mathcal{S}_\mu$ associated with the supersymmetry $Q^\alpha$. These are the (anomalous) Ward identities which are usually discussed in the literature.

We will show that the superconformal  Chern-Simons anomaly cocycle  is equivalent to the  so-called $a$-anomaly of superconformal gravity.  
We select among all the equivalent  Lorentz equivariant and $\theta^a$ independent cocycles a particular one that simplifies the explicit expressions for the supersymmetry anomalies.  We write it down  in components, to all order in the number of both fermionic and bosonic fields, in appendix \ref{appendix:lorentzcocycle}.  


The rest of this paper is organized as follows: 

In Section \ref{sec:2} we review the BRST formulation of $d=4\,,\;\mathcal{N}=1$ superconformal gravity, which was first worked out in \cite{Baulieu:1986ab},
by following a slightly different logic and formalism. This will allow us to describe the ingredients relevant to the computation of the anomaly. Our formalism will keep manifest the underlying covariance under the full Lie superalgebra $\mathfrak{su}(2,2|1)$ of the equations that determine the $\lambda_0^i$'s. In this Section we also take the opportunity to  elucidate why  and how translations $P^a$  must be dealt with differently than other symmetries in the BRST context and why this entails, in the supersymmetric case,  introducing the $i_\gamma$ term in the definition of the generalized BRST operator. 

In Section \ref{sec:sol_for_constraints}, which also reproduces  results already presented in \cite{Baulieu:1986ab},  we describe how to solve the BRST nilpotency equations that both determine $\lambda_0^i$'s  and generate the constraints on the ordinary curvatures of superconformal gravity. Our presentation possibly clarifies why the solution to the BRST nilpotency conditions found in \cite{Baulieu:1986ab}  is the only possible solution. We solve the constraints to express the fields $\{\omega^{ab}, \;f^a,\;\Tilde{\psi}^\alpha \}$ explicitly in terms of the physical fields $\{e^a, b,\;a,\;\psi^\alpha \}$.  The main purpose of this section  is pointing out that while  the superconformal algebra uniquely fixes the BRST rules of the ghosts, it determines the BRST rules of the gauge fields only up to the $\lambda_0$ terms, which in turn are fixed by BRST nilpotency.    

In Section \ref{sec:superanomaly} we describe the unique completely symmetric (in the graded sense) $\mathfrak{su}(2,2|1)$ invariant tensor and show that the corresponding generalized characteristic class $P_{3}(\bm{F})$ vanishes. To perform this latter computation we made use of  \textsc{FieldsX} \cite{Frob:2020gdh}. We then present the ensuing secondary generalized  Chern-Simons class which captures a superconformal anomaly.  This is our main result.

In Section \ref{sec:equivariantcocycle} we describe an anomaly cocycle equivalent to the Chern-Simons $\mathfrak{su}(2,2|1)$  cocycle, 
which is independent of $\Omega^{ab}$, $\theta^a$ and $b$.  We show, by working out its explicit form, that it is equivalent to the superconformal $a$-anomaly. 

In Section \ref{sec:conclusions} we draw our conclusions and describe open problems.

\section{BRST  formulation of conformal supergravity}\label{sec:2}

As mentioned in the Introduction, $d=4$, $\mathcal{N}=1$ conformal supergravity is a ``pure'' gauge theory:  all of its fields  are 1-form connections taking values in the appropriate bundles
 \begin{equation}
      A^i=\;\{ e^a,\;\omega^{ab},\;\;b,\;a,\; f^a,\;\psi^\alpha,\;\Tilde{\psi}^\alpha \},
\end{equation}
in correspondence to the generators of the $\mathfrak{su}(2,2|1)$ Lie superconformal algebra:
 \begin{equation}
 T_i=\;\{P^a,\;J^{ab},\;W,\; R,\; K^a,\;Q^\alpha,\; S^\alpha \}.
\end{equation}
The generators $T_i$ are graded:\footnote{We  denote by $(-1)^{|i|}$ the grading of the generator $T_i$, i.e.  $(-1)^{|i|}=+1$ for the bosonic  generators $\{P^a,\;J^{ab},\;W,\; R,\; K^a\}$ and $(-1)^{|i|}=-1$ for  the fermionic ones $\{Q^\alpha,\; S^\alpha \}$. We denote with the bracket the (anti)-commutator:   $[T_i,T_j]\equiv T_iT_j-(-)^{|i||j|}T_jT_i$. } they satisfy (anti)commutation relations
\begin{equation}
[T_i, T_j ] = f_i{}^k{}_j \, T_k,
\end{equation}
where $ f_i{}^k{}_j$ are the structure constants of the $d=4\,,\mathcal{N}=1$ Lie superconformal algebra.\footnote{We list them in appendix \ref{appendixB}.} 

The BRST formulation of conformal supergravity differs from that of pure (super) Yang-Mills theories in one crucial aspect. Let us delve a bit deeper into this distinction.

In (super)YM theories, one introduces in correspondence to each generator $T_i$ a ghost field $c^i$ with opposite statistics  $-(-1)^{|i|}$. The  resulting Lie superalgebra valued combination
\begin{equation}
c = c^i\, T_i
\end{equation}
is anti-commuting, and its BRST transformations are completely fixed by the structure constants of the Lie superalgebra:
\begin{equation}\label{superYMbrstg}
s\,c = -\tfrac{1}{2}\,[c,c],
\end{equation}
or, equivalently, 
\begin{equation}\label{superYMbrstgbis}
s\,c^i =-\tfrac{1}{2}\Tilde{f}_j{}^i{}_k\, c^j\, c^k,
\end{equation}
where, as reviewed in appendix \ref{appendixB},
\begin{equation} \label{eq:f_tilde_def}
    \Tilde{f}^{\;i}_{j\;k}\equiv (-)^{|j|(|k|+1)}f^{\;i}_{j\;k}.
\end{equation}
The (super)Jacobi identity
\begin{equation}
[c,[c,c]]=0
\end{equation}
ensures that the BRST rules (\ref{superYMbrstgbis}) are nilpotent.  Furthermore, the BRST transformations for the (anti-commuting) Lie superalgebra valued  connection 
\begin{equation}
A = A^i\, T_i
\end{equation}
are also completely specified by the structure constants of the Lie superalgebra 
\begin{equation}
s\, A = - \,\dd\, c - [ A, c].
 \end{equation}
 
For  conformal supergravity --- and for any theory which includes gravity ---  one has to proceed differently.  In correspondence to diffeomorphisms one introduces an  anti-commuting ghost $ \xi^\mu$ which is a vector field:  there is no ghost valued in the $P^a$ sub-algebra. 
The BRST operator $s$ acts  on generic tensor  fields $\phi$ via the Lie derivative $\mathcal{L}_\xi$\footnote{The minus sign in front of the Lie derivative is traditional in a certain stream of literature.}
\begin{equation}
s\, \phi = -\mathcal{L}_\xi\, \phi + \mathrm{other\; gauge\; transformations},
\end{equation} 
and  on the ghost $\xi^\mu$ as follows
\begin{equation}\label{secone:sxi}
s\, \xi^\mu = - \tfrac 1 2 \,  \mathcal{L}_\xi\,\xi^\mu + \gamma^\mu.
\end{equation}
$\gamma^\mu$ is a quadratic function of the other ghosts whose precise form depends on the details of the gravitational theory one considers. We are going to exhibit its expression for superconformal gravity momentarily.   Nilpotency of $s$ requires that 
\begin{equation}\label{secone:sgamma}
s\, \gamma^\mu = -\mathcal{L}_\xi\, \gamma^\mu.
\end{equation}
The way to deal with this situation is to disentangle translations from the other local symmetries. One introduces  an ``equivariant'' (with respect to diffeomorphisms) BRST operator  $\hat{s}$, whose action is defined on the smaller  functional space of  ghosts and connections  which does not  include $\xi^\mu$:
 \begin{equation}\label{secone:hats}
\hat{s} = s + \mathcal{L}_\xi.
\end{equation}
$\hat{s}$ involves only the ghosts $c^I$ corresponding to the  gauge transformations other than translations. In the superconformal case these ghosts are:
\begin{equation}
 c^I=\{ \Omega^{ab},\;\sigma,\;\alpha,\;\theta^a,\;\zeta^\alpha,\; \eta^\alpha\}. 
\end{equation}
Nilpotency of $s$ is equivalent to the following relation for the equivariant BRST operator:
\begin{equation}
\hat{s}^2 = \mathcal{L}_\gamma,
\end{equation}
valid on the reduced field space which does not involve $\xi^\mu$.

The action of $\hat{s}$ on the ghosts $c^I$ cannot be simply defined by truncating the BRST transformation rule for the ghosts  (\ref{superYMbrstgbis}) to the $c^I$: since the $\{T_{I}\}$'s  do not span a subalgebra,  the  truncated BRST transformations
\begin{equation} \label{eq:s_0gi}
\hat{s}_0\,c^I = -\tfrac{1}{2}\,\Tilde{f}_J{}^I{}_K\,c^J\,c^K
\end{equation}
would  not be  in general nilpotent.  Indeed,  let $\{i\} =\{ a, I\}$  be  the index running along the full Lie superalgebra and $a$ the index running along the translations subalgebra: the Jacobi identity relevant for the nilpotency of (\ref{eq:s_0gi})  writes
\begin{equation}
\Tilde{f}_J{}^I{}_K\,\Tilde{f}_L{}^J{}_M = -\Tilde{f}_a{}^I{}_K\,\Tilde{f}_L{}^a{}_M.
\end{equation}
Hence
\begin{equation}
\hat{s}_0^2\,c^I =- \Tilde{f}_J{}^I{}_K\,\hat{s}_0 \,c^J\,c^K 
= -\Tilde{f}_a{}^I{}_K\,\gamma^a\,c^K,
\end{equation}
where we introduced the ghost bilinear with values in the translations subalgebra
\begin{equation}
\gamma^a \equiv \tfrac{1}{2}\,\Tilde{f}_L{}^a{}_M\,c^L\,c^M.
\end{equation}
We  need therefore to introduce a suitable deformation of $\hat{s}_0$. One can start from the ans\"atz, dictated by ghost number conservation, which includes a term proportional to the gauge connection $A^I$: 
\begin{equation}\label{secone:sgI}
\hat{s}\,c^I = \hat{s}_0\,c^I + i_\gamma (A^I) =  -\tfrac{1}{2}\,\Tilde{f}_J{}^I{}_K\,c^J\,c^K + i_\gamma(A^I),
\end{equation}
where $\gamma = \gamma^\mu\,\de_\mu$ is the ghost number 2 vector field which appears in the BRST transformations of the diffeomorphism ghost (\ref{secone:sxi}) and $i_\gamma$ is the contraction of a form with the commuting vector field $\gamma^\mu$. Note that
\begin{equation}
i_\gamma^2 =0,
\end{equation} \label{secone:shatgamma}
 since $\gamma^\mu$ is commuting. Moreover we must impose
\begin{equation}
\hat{s} \, \gamma^\mu= 0,
\end{equation}
as consequence of (\ref{secone:sgamma}).  Therefore
\begin{align}
\hat{s}^2\,c^I  &= - \Tilde{f}_J{}^I{}_K\,\hat{s}\,c^J\,c^K + i_{\hat{s}\gamma}\,(A^I) -i_\gamma\,(\hat{s}A^I)= \nonumber\\
& =  -\Tilde{f}_a{}^I{}_K\,(\gamma^a-i_\gamma\,e^a)\,c^K -i_\gamma(\hat{s}\,A^I+\Tilde{f}_j{}^I{}_K\, A^j\,c^K). 
\end{align}
We see therefore that we must take
\begin{equation}\label{secone:gammadef}
\gamma^a= \tfrac{1}{2}\,\Tilde{f}_L{}^a{}_M\,c^L\,c^M=  i_\gamma\,e^a
\end{equation}
and 
\begin{equation}\label{secone:shI}
\hat{s}\, A^I = - \dd\, c^I  - \Tilde{f}_j{}^I{}_K\, A^j\,c^K + \lambda_0^I,
\end{equation}
where $\lambda_0^I$ are $i_\gamma$-closed 1-forms which take value in the Lie superalgebra
\begin{equation}\label{secone:igammalambda0}
i_\gamma(\lambda^I_0)=0.
\end{equation}
Eq. (\ref{secone:gammadef}) fixes the vector field  $\gamma^\mu$  which appears in the BRST transformation (\ref{secone:sxi}) of the ghost $\xi^\mu$  in terms of the structure constants of the Lie superalgebra: for $\mathfrak{su}(2,2|1)$ we obtain\footnote{The $\gamma$ deformation is a signal of topological gravity \cite{Imbimbo:2018duh} or supersymmetry \cite{Baulieu:1985md,Frob:2021sao}. Note that in the bosonic conformal case, $\Tilde{f}_L{}^a{}_M = 0$, because no commutator of generators $T_I$ gives $P^a$ (unlike the  supersymmetric case, where $\{Q,Q\}\sim P$). Therefore, even if the truncation does not define an algebra, the truncated BRST operator is nilpotent and the $\gamma$ deformation does not arise.}
\begin{equation}
\gamma^\mu = \bar\zeta\, \Gamma^a\, \zeta\, e^\mu{}_a.
\end{equation}
Condition (\ref{secone:sgamma}) fixes the BRST rule for the connection $e^a$, which is therefore ``universal'' for supergravity theories:
\begin{equation}\label{secone:sea}
\hat{s} \, e^a = -\Omega^a{}_b\,e^b-\sigma\, e^a-2\,\overline{\zeta}\,\Gamma^a\psi.
\end{equation}
 In conclusion, the requirement of nilpotency of the BRST transformations on  both the $c^I$'s ghosts and the diffeomorphisms ghost $\xi^\mu$ completely determines  the BRST transformations of  the ghosts, eqs. (\ref{secone:sxi}) and (\ref{secone:sgI}), which can be read off from the structure constants of the gauge superalgebra.    On the other hand,  nilpotency of the BRST transformations on ghosts determines BRST rules for the connections $A^I$, eqs. (\ref{secone:shI}), {\it only up to}  $i_\gamma$-closed 1-forms $\lambda_0^I$: we will see shortly that 
 the $\lambda_0^I$ are determined by the requirement of nilpotency of $s$ on the connections $A^I$: the $\lambda_0^I$'s do not have  an immediate  interpretation in terms of the geometry of the gauge superalgebra.

 We can now introduce the  generalized-connection $\bm{A}^i$:
 \begin{equation}
 \bm{A}^i=A^i+c^i,
 \end{equation}
 where $i$ runs along all the generators $T_i$ of the Lie superalgebra, with the understanding that the generalized connection along translations has no ghost number 1 component
 \begin{equation}
 A^a = e^a.
 \end{equation}
Moreover eqs. (\ref{secone:sgI}) and (\ref{secone:shI}),  dictate the form of the generalized BRST operator
\begin{equation} \label{delta_operator}
\delta=\hat{s}+\dd-i_\gamma,
\end{equation}
which differs from the Stora-Zumino analogue  (\ref{eq:intro_delta})  for  the $i_\gamma$ term, which encodes, in the BRST formalism,  the ``coupling''  to supergravity.
Generalized curvatures are defined in terms of the generalized differential $\delta$ and generalized connections in the usual way
 \begin{equation} \label{eq:super_h_super_H}
   \bm{F}^i=\delta\, \bm{A}^i+\tfrac{1}{2}[\bm{A},\bm{A}]^i=\delta\, \bm{A}^i+\tfrac{1}{2}\Tilde{f}_j{}^i{}_k\,\bm{A}^j\,\bm{A}^k.
\end{equation}
We can compute $\bm{F}$ by making use of eqs. (\ref{secone:sgI}) and (\ref{secone:shI}) to obtain
\begin{equation} \label{eq:sH_hor}
    \bm{F}^i=F^i+\lambda_0^i.
\end{equation}
In other words the generalized-curvatures fail to be ``horizontal''  because of the  $\lambda_0^I$   which were left undetermined by the condition of nilpotency of the BRST operator on the ghosts.\footnote{Note that the BRST transformation rules for the vierbein, eq. (\ref{secone:sea}), which are universal, imply however that $\lambda_0^P=0$.} One must therefore investigate the restrictions on the $\lambda_0^I$'s coming from nilpotency of BRST transformations on the generalized connections:
\begin{equation}\label{secone:delta2h}
\delta^2\,\bm{A}^i=\delta^2\, A^i=-i_\gamma(F^i)+\hat{s}\,\lambda_0^i-\Tilde{f}_j{}^i{}_k\,\lambda_0^j\,c^k  =0,
\end{equation}
where $F^i$ are ordinary  2-form curvatures
\begin{equation}\label{secone:ordinaryH}
F^i = \dd\, A^i+\tfrac{1}{2}\,\Tilde{f}_j{}^i{}_k\,A^j\,A^k.
\end{equation}
Equation (\ref{secone:delta2h}) shows that not all the $\lambda_0^i$'s can be taken to vanish, unless we impose $F^i=0$ for all curvatures, which would eliminate all propagating degrees of freedom from the theory. 

As we made clear, eq.  (\ref{secone:delta2h}) is quite general: it is  valid for any gravitational theory based on a Lie superalgebra with generators $\{T_i\}$.  A solution of this equation for the $d=4,\,\mathcal{N}=1$ superconformal algebra $\mathfrak{su}(2,2|1)$  was found in  \cite{Baulieu:1986ab}.  This solution for the $\lambda_0^i$'s also requires a set of constraints on the ordinary (both bosonic and fermionic) curvatures $F^i$.  We conducted with the help of \textsc{FieldsX} \cite{Frob:2020gdh} a somewhat more systematic analysis of  (\ref{secone:delta2h}), which we summarize in Section \ref{sec:sol_for_constraints} with the intent to ascertain if  more general solutions exist. We recovered the same solution of  \cite{Baulieu:1986ab} and nothing more. 

Let us conclude this section by presenting the details of this solution. The BRST rules for  the ghosts of $\mathfrak{su}(2,2|1)$ can be read off from (\ref{secone:sxi}) and (\ref{secone:sgI}):
 \begin{subequations} \label{eq:s_hat_ghosts_explicit}
    \begin{align}
        s\,\xi&=-\tfrac{1}{2}\mathcal{L}_\xi\,\xi +\overline{\zeta}\,\Gamma^a\,\zeta\, e^\mu{}_a,\\
        \hat{s}\,\Omega^{ab}&=i_\gamma(\omega^{ab})-(\Omega^2)^{ab}+2\,i\,\overline{\zeta}\,\Gamma^{ab}\,\eta, \\
        \hat{s}\,\sigma&=i_\gamma(b)+2\, i\,\overline{\zeta}\,\eta, \\
        \hat{s}\,\alpha&=i_\gamma(a)+2\,\overline{\zeta}\,\Gamma_5\,\eta, \\
        \hat{s}\,\theta^a&=i_\gamma(f^a) -\Omega^{a}{}_b\,\theta^b+\sigma\,\theta^a+\overline{\eta}\,\Gamma^a\,\eta,  \\
        \hat{s}\,\zeta&=i_\gamma(\psi)-(\tfrac{1}{4}\,\Omega^{ab}\,\Gamma_{ab}+\tfrac{1}{2}\,\sigma-\tfrac{3}{2}\,i\,\alpha\,\Gamma_5)\,\zeta, \\
        \hat{s}\,\eta&=i_\gamma(\Tilde{\psi})-(\tfrac{1}{4}\,\Omega^{ab}\,\Gamma_{ab}-\tfrac{1}{2}\,\sigma+\tfrac{3}{2}\,i\,\alpha\,\Gamma_5)\,\eta+i\,\theta^a\,\Gamma_a\,\zeta.
    \end{align}
\end{subequations}
The BRST rules for the connections follow from (\ref{secone:sea}) and (\ref{secone:shI}):\footnote{We describe in the next Section how to compute the $\lambda_0^i$'s  by imposing BRST nilpotency on the gauge fields. The resulting expression for the non-vanishing $\lambda_0^i$'s  are  listed in the next page.}
\begin{subequations} \label{eq:s_hat_fields}
    \begin{align}
        \hat{s}\,e^a&=-\Omega^a{}_b\,e^b-\sigma e^a-2\,\overline{\zeta}\,\Gamma^a\,\psi, \\
        \hat{s}\,\omega^{ab}&=-(\dd\,\Omega^{ab}+\omega^{a}{}_c\,\Omega^{cb}-\omega^{b}{}_c\,\Omega^{ca})-2\,e^{[a}\theta^{b]}+2\,i\,(\overline{\psi}\,\Gamma^{ab}\eta+\overline{\zeta}\,\Gamma^{ab}\,\Tilde{\psi})+(\lambda_0^J)^{ab}, \\
        \hat{s}\,b&=-\dd\, \sigma-2\,e^a\theta_a+2\,i\,(\overline{\psi}\,\eta+\overline{\zeta}\,\Tilde{\psi})+\lambda_0^W, \\
        \hat{s}\,a&=-\dd\, \alpha+2\,(\overline{\psi}\,\Gamma_5\,\eta+\overline{\zeta}\,\Gamma_5\,\Tilde{\psi}) + \lambda_0^R, \\
        \hat{s}\,f^a&=-(\dd\,\theta^a+\omega^a{}_c\,\theta^c-b\,\theta^a)-\Omega^a{}_b\,f^b+\sigma f^a+2\,\overline{\eta}\,\Gamma^a\,\Tilde{\psi}+(\lambda_0^K)^a, \\
        \hat{s}\,\psi&=-(\dd+\tfrac{1}{4}\,\omega^{ab}\Gamma_{ab}+\tfrac{1}{2}\,b-\tfrac{3}{2}\,i\,a\,\Gamma_5)\,\zeta-(\tfrac{1}{4}\,\Omega^{ab}\,\Gamma_{ab}+\tfrac{1}{2}\,\sigma-\tfrac{3}{2}\,i\,\alpha\,\Gamma_5)\,\psi\,+\nonumber\\
        &\qquad-i\,e^a\,\Gamma_a\,\eta+\lambda_0^Q, \\
        \hat{s}\,\Tilde{\psi}&=-(\dd+\tfrac{1}{4}\,\omega^{ab}\,\Gamma_{ab}-\tfrac{1}{2}\,b+\tfrac{3}{2}\,i\,a\,\Gamma_5)\,\eta-\left(\tfrac{1}{4}\Omega^{ab}\,\Gamma_{ab}-\tfrac{1}{2}\,\sigma+\tfrac{3}{2}i\,\alpha\,\Gamma_5\right)\,\Tilde{\psi}\,+\nonumber\\
        &\qquad + if^a\,\Gamma_a\,\zeta+i\theta^a\,\Gamma_a\,\psi+\lambda_0^S,
    \end{align}
\end{subequations}
where the square brackets denote anti-symmetrization (with no numerical factors). 

The explicit expressions for the two-form curvatures
\footnote{The $\Tilde{\;\;}$ on the superconformal curvatures $\Tilde{R}^{ab}$, $ \Tilde{F}^W$ and $ \Tilde{F}^R$ is meant to distinguish them from the standard curvatures, $R^{ab}$, $\dd\, b $ and $\dd\,a$.  The $\Tilde{\;\;}$ on  $ \Tilde{\rho}$ and $\Tilde{T}^a$ is a reminder that these are the conformal partners of the usual  torsion $T^a$ and gravitino curvature $\rho$.   $\Tilde{\;\;}$   should not be confused with the Hodge dual which we denote by $\star$.}
\begin{equation}
F^i=\{T^a,\Tilde{R}^{ab},\;\Tilde{F}^W,\;\Tilde{F}^R,\;\Tilde{T}^a,\;\rho,\;\Tilde{\rho}\}\;,
\end{equation}
which include contributions from the full superconformal algebra, are: 
\begin{equations} \label{eq:curvatures}
  T^a &=(\dd\, e^a+\omega^a{}_b\,e^b+b\,e^a)+\overline{\psi}\,\Gamma^a\,\psi=D\,e^a+\overline{\psi}\,\Gamma^a\,\psi, \\
\Tilde{R}^{ab}&={R}^{ab}(\omega)+2\,e^{[a}f^{b]}-2\,i\,\overline{\psi}\,\Gamma^{ab}\,\Tilde{\psi},\\
 \Tilde{F}^W &=\dd\, b+2\,e^a\,f_a-2\,i\,\overline{\psi}\,\Tilde{\psi}, \\
  \Tilde{F}^R &=\dd \,a-2\,\overline{\psi}\,\Gamma_5\,\Tilde{\psi}, \\
 \rho&=(\dd+\tfrac{1}{4}\,\omega^{ab}\,\Gamma_{ab}+\tfrac{1}{2}\,b-\tfrac{3}{2}\,i\,a\Gamma_5)\,\psi+i\,e^a\Gamma_a\Tilde{\psi}=D\,\psi+i\,e^a\,\Gamma_a\,\Tilde{\psi}, \\
\Tilde{T}^a &=(\dd f^a+\omega^a{}_b\,f^b-b\,f^a)-\overline{\Tilde{\psi}}\,\Gamma^a\,\Tilde{\psi}=D\,f^a-\overline{\Tilde{\psi}}\,\Gamma^a\,\Tilde{\psi}, \\
 \Tilde{\rho}&=(\dd+\tfrac{1}{4}\,\omega^{ab}\,\Gamma_{ab}-\tfrac{1}{2}\,b+\tfrac{3}{2}\,i\,a\Gamma_5)\,\Tilde{\psi}-i\,f^a\,\Gamma_a\,{\psi}=D\,\Tilde{\psi}-i\,f^a\,\Gamma_a\,{\psi},  
   \end{equations}
where $D$ is the covariant derivative with respect to Lorentz, Weyl and $U(1)_R$ symmetries.
The non-vanishing $\lambda_0^I$'s turn out to be:
\begin{equations} \label{eq:lambda_0_a}
 (\lambda_0^J)^{ab}&=2\, e^c\,\overline{\zeta}\,\Gamma_c\,\rho^{ab},\\
 \lambda_0^S&=\tfrac{1}{4}\,\Gamma_5\,\Gamma^{mn}\,\Gamma_c\,\zeta\,\Tilde{F}^R_{mn}e^c,\\
(\lambda_0^K)^a &=-i\,e^c\,\overline{\zeta}\,\Gamma_c\Gamma_b\,\Tilde{\rho}^{\prime ab},
        \label{eq:lambda_0_c}
\end{equations}
where we defined 
\begin{equation}
\rho \equiv \frac 1 2 \,\rho_{ab}\, e^a\, e^b,  \qquad \Tilde{\rho} \equiv \frac 1 2\, \Tilde{\rho}_{ab}\, e^a\, e^b, \qquad\Tilde{F}^R \equiv \frac{1}{2}\, \Tilde{F}^R_{ab}\, e^a\, e^b, 
\end{equation}
and  introduced the  ``modified" 2-form curvatures 
\begin{equations} \label{eq:mod_curv}
\Tilde{R'}^{ab}&=  \Tilde{R}^{ab}-2\,e^e\,\overline{\psi}\,\Gamma_c\, \rho^{ab},  \\
        \Tilde{\rho}'&=\Tilde{\rho}-\tfrac{1}{4}\,\Gamma_5\,\Gamma^{mn}\,\Gamma_c\,\psi\Tilde{F}^R_{mn}\,e^c, \\
      \Tilde{T'}^a&=\Tilde{T}^a+i\,e^c\,\overline{\psi}\,\Gamma_c\,\Gamma_b\,\Tilde{\rho}^{\prime ab},
\end{equations}
which have the property of transforming without derivatives of the supersymmetry ghost $\zeta$ under BRST transformations.  

Eqs. (\ref{secone:delta2h}) which ensure the nilpotency of the generalized BRST operator $\delta$ on all fields, are satisfied by the $\lambda_0^i$'s in (\ref{eq:lambda_0_a}-\ref{eq:lambda_0_c}) only on the subspace of fields defined by the set of constraints
\begin{equations}  
T^a&=0,  \label{eq:curv_constraints_a}\\
\Tilde{F}^W&=-{\star\Tilde{F}^R}, \\
\Gamma^a\, \rho_{ab}&=0,\\
 \Tilde{\mathcal{R}}^\prime_{\mu\nu}&=- \Tilde{F}^W_{\mu\nu},      \label{eq:curv_constraints_d}
   \end{equations}
where  $\Tilde{\mathcal{R}}^\prime_{\mu\nu}$ is the Ricci tensor constructed with the modified curvature $\Tilde{R'}^{ab}$:
\begin{equation}
\Tilde{\mathcal{R}}^\prime_{\mu\nu} \equiv  \Tilde{R'}_{\mu\rho}{}^{ab}\, e_b{}^\rho\, e_{\nu a}.
\end{equation}
As we will review in the next section,  these constraints can be solved algebraically to express the fields $\{\omega^{ab}, \Tilde{\psi}, f^a\}$ as local functions of the independent fields $\{e^a, \psi, a, b\}$. 

\section{Non-horizontal components of the curvatures and constraints} \label{sec:sol_for_constraints}
This section, which can be skipped at a first reading and whose results reproduce those found in  \citep{Baulieu:1986ab}, is devoted to solving eqs. (\ref{secone:delta2h}). In the generalized form approach, the failure of BRST  nilpotency in the ``big'' field space of unconstrained generalized connections is the failure of the \emph{generalized Bianchi identity}:
\begin{equation}
\delta\,\bm{F} + [\bm{A},\bm{F}] = \delta^2\,\bm{A},
\end{equation}
where $\delta$ is defined in (\ref{delta_operator}). Since the BRST rules of the ghosts are  nilpotent in the ``big'' field space, the previous equation simplifies to
\begin{equation}
\delta\,\bm{F} + [\bm{A},\bm{F}] =\delta^2\,A,
\end{equation}
or equivalently, in components,
\begin{equation}
\delta \, \bm{F}^i - f^i_{kj} \, \bm{A}^j \, \bm{F}^k=\delta^2\,A^i.
 \end{equation}
Filtering in the ghost number, one gets: 

a)  the Bianchi identities for the ordinary curvatures (ghost number zero); 

b)  the BRST transformation rules for the ordinary curvatures (ghost number one);  

c) $s^2$ on the gauge fields or equivalently the BRST transformation rules for the $\lambda^i_0$'s (ghost \\
\indent number two), and 

d) the $i_\gamma$-closeness of the  $\lambda^i_0$'s  (ghost number three): 
\begin{equations}
& \dd F^i -f_k{}^i{}_j\,A^j\,F^k = 0,\\
& \hat{s}\,F^i-  f_k{}^i{}_j\,c^j\,F^k = - \dd \lambda_0^i +f_k{}^i{}_j\,A^j\,\lambda_0^k,\\
& s^2\,A^i = - i_\gamma(F^i)  + \hat{s}\,\lambda_0^i - f_k{}^i{}_j\,c^j\, \lambda_0^k =0, \label{ghost2condition}\\
& i_\gamma\,(\lambda_0^i) = 0. \label{igammacondition}
\end{equations}
The equations at ghost number two are the same as  eqs.  (\ref{secone:delta2h}).  The trilinear Fierz identity for the commuting spinor $\zeta$:
\begin{equation}
 \Gamma_\mu \, \zeta\,\bar\zeta\, \Gamma^\mu\, \zeta  =0, 
\end{equation}
together with  eq. (\ref{igammacondition}),   fixes the general structure of the $\lambda_0^i$'s:
\begin{equations}
& \lambda_0^i=e^a\,\overline{\zeta}\,\Gamma_a\,X^i,\;\;\text{for the bosonic fields}, \label{sec3:Xstructurea}\\
& \lambda_0^i = X^i\,\Gamma_a\,\zeta\,e^a,\;\;\text{for the fermionic fields.}\label{sec3:Xstructureb}
\end{equations}
$X^i$ is a zero-form of ghost number 0. When  $\lambda_0^i$ is associated to a bosonic generator, $X^i$ is a Majorana  spinor;  when  $\lambda_0^i$  is associated to a fermionic generator, $X^i$ is a matrix acting on the spinorial indices of $\zeta$. 

Eq. (\ref{ghost2condition})  can be  projected onto one component quadratic in the supersymmetry ghosts $\zeta$ and one component linear in $\zeta$.  Let us therefore correspondingly separate the part $S$ in the BRST operator $\hat{s}$ which is proportional to $\zeta$ and  associated to local supersymmetry \cite{Frob:2021sao}:
\begin{equation}
\hat{s} = S + \hat{s}^\prime.
\end{equation}
The projection of eq. (\ref{ghost2condition}) onto the component linear in $\zeta$ becomes
\begin{equation}\label{sec3:lambda0superconformalcov}
\hat{s}^\prime \,\lambda_0^i - \sum_{j\not=\zeta}\,f^{\;i}_{k\;j}\,c^j\, \lambda_0^k  =0.
\end{equation}
This equation states simply that the $\lambda_0^i$'s transform covariantly under all the transformations of the superconformal algebra other than the supersymmetry transformations. 
The projection of eq. (\ref{ghost2condition}) onto the component quadratic in $\zeta$, after taking into account   eqs. (\ref{sec3:Xstructurea}-\ref{sec3:Xstructureb}), writes 
\begin{equation}
 - i_\gamma\,(F^{\prime\;i})  - e^a\,\overline{\zeta}\,\Gamma_a\, S\,X^i  -f^{\;i}_{k \;\zeta}\,\overline{\zeta}\, \lambda_0^k=0,
\end{equation}
where we introduced  the ``modified''  curvatures
\begin{equation}
F^{\prime\;i} \equiv F^i - e^c\,\overline{\psi}\,\Gamma_c X^i.
\end{equation}
The dependence on the derivative of the ghost $\zeta$ in the BRST variation of $F^{\prime\;i}$ cancels between the first term and the BRST  variation of $\psi$. Hence the modified curvatures $F^{\prime\;i}$ are supercovariant --- i.e. their variations under (local) supersymmetry do not depend on derivatives of the supersymmetry ghosts --- if we take  $X^i$ proportional to the modified curvatures themselves.  The possible modified curvatures involved in each $X^i$ are fixed by superconformal covariance (\ref{sec3:lambda0superconformalcov}).\footnote{Weyl weights and R-charges of ghosts and connections are summarized in appendix \ref{appendixB}.}  In particular the mass dimension of $X^i$ must be the same as that of $A^i$ increased by one half. 

We already determined the BRST rule of the vierbein in eq. (\ref{secone:sea}), which implies that  the corresponding $\lambda_0$ vanishes:
\begin{equation}
\lambda_0^P = 0.
\end{equation} 
Therefore $T'^a = T^a$,  and the nilpotency equation for the vierbein 
reads:
\begin{equation}\label{sec3:s2ea}
s^2\,e^a = -i_\gamma\,(T^a) - 2\,\bar{\zeta}\,\Gamma^a\,\lambda_0^Q.
\end{equation}
$\lambda_0^Q$ is necessarily proportional to the torsion, because the mass dimension of $X^\psi$ is 1 and the torsion is the unique curvature with the required mass dimension. The most general ans\"atz for  $\lambda_0^Q$ consistent with superconformal covariance  is
\begin{equation}\label{sec3:lambda0psiansatz}
\lambda_0^Q = \tfrac{1}{2}\,(t_1\,T_{mn}{}^m\,\Gamma^n + t_2\,T_{mn}{}^p\,\varepsilon^{mn}{}_{pq}\,\Gamma_5\,\Gamma^q)\,\Gamma^c\,\zeta\,e_c,
\end{equation}
where
\begin{equation}
T^a = \tfrac{1}{2}\,T_{mn}{}^a\,e^m\,e^n.
\end{equation}
By plugging (\ref{sec3:lambda0psiansatz}) into the nilpotency equation (\ref{sec3:s2ea}), one obtains
\begin{align}
s^2\,e^a &= 
t_2 \, e^b \,T_{bc}{}^{c} \,\gamma^{a } + t_2\, T^{ac }{}_{c }\, \gamma^{\beta } + 2\,i \,t_1\, e_b\, T^{a b }{}_{c} \gamma^{c }  - 2\,i\, t_1 \, e^b\,T^{ac}{}_{b } \,\gamma_{c } \,+ \nonumber\\
& \quad + (1 + 2\, i \, t_1) \, e^b\, T_{bc}{}^{a} \,\gamma^{c} -  t_2\, e^a\, T_{cb}{}{}^{b }\, \gamma^{c}.
\end{align}
Thus, BRST nilpotency on $e^a$ requires both the vanishing of the torsion
\begin{equation}\label{sec3:torsionconstraint}
T^a = 0
\end{equation}
and of $\lambda_0^Q$
\begin{equation}
\lambda_0^Q = 0.
\end{equation}
It follows  that 
\begin{equation}
\rho'=\rho.
\end{equation} 
The torsion constraint  (\ref{sec3:torsionconstraint}) can be algebraically solved  for the spin connection, expressing it in terms of $e^a$, $b$ and $\psi$:
\begin{equation}\label{sec3:compositeomega}
\omega_\mu{}^{ab} = \tfrac{1}{2}\,e^{\nu[a}\,\de_{[\mu}\,e_{\nu]}{}^{b]} - \tfrac{1}{2}\,e^{\nu a}\,e^{\rho b}\,e_\mu{}^c\,\de_{[\nu}\,e_{\rho]c} + e_\mu{}^{[a}\,b^{b]} + \psi_\mu\,\Gamma^{[a}\,\psi^{b]} + \bar{\psi}^a\,\Gamma_\mu\,\psi^b.
\end{equation}
The torsion constraint  is necessary to ensure BRST nilpotency in any supergravity theory, independently of any equations of motion.  Note  that bosonic connections of conformal supergravity have 48 off-shell degrees of freedom, while the fermionic connections have 24.  Since the spin connection has precisely 24 components, the torsion constraint (\ref{sec3:torsionconstraint})  ensures the matching between  bosonic and fermionic degrees of freedom. 

Let us turn to the $\lambda_0$'s associated to the Lorentz, Weyl and R-symmetry generators. Superconformal covariance dictates their form to be
\begin{equations}
& (\lambda_0^J)^{ab} = e_c\,\bar{\zeta}\,\Gamma^c\,(x_1\,\rho^{ab}+x_2\,\Gamma_5\,\varepsilon^{abmn}\,\rho_{mn}), \label{sec3:lambda0Lansatz}\\
&\lambda_0^W = e_c\,\bar{\zeta}\,\Gamma_c\,(y_1\,\Gamma^{ab}\,\rho_{ab} + y_2\,\Gamma_5\,\Gamma^{ab}\,\varepsilon_{abmn}\,\rho^{mn}), \label{sec3:lambda0Wansatz}\\
&\lambda_0^R = e_c\,\bar{\zeta}\,\Gamma^c\,(z_1\,\Gamma_5\,\Gamma^{ab}\,\rho_{ab} + z_2\,\Gamma_{ab}\,\varepsilon^{abmn}\,\rho_{mn}).\label{sec3:lambda0Ransatz}
\end{equations}
Nilpotency of the BRST operator implies that the BRST variation of a constraint is a linear combination of  constraints.
Hence
\begin{equation}
0 = \hat{s}\,T^a + \Omega^a{}_b\,T^b + \sigma\,T^a = - (\lambda_0^J)^a{}_b\,e^b + \lambda_0^W\,e^a + 2\,\bar{\zeta}\,\Gamma^a\,\rho.
\end{equation}
By plugging the ans\"atz for $\lambda_0^J$ and $\lambda_0^W$ into this equation, one obtains:
\begin{align}
0 = e_c\,\bar{\zeta}\,&\,\big{(}\Gamma^a\,\rho^{cb} + \tfrac{x_1}{2}\,\Gamma^c\,\rho^{ba} + \tfrac{x_1}{2}\,\Gamma^b\,\rho^{ac} \,+\nonumber\\
& - x_2\,\Gamma^c\,\Gamma_5\,\varepsilon^{abmn}\,\rho_{mn} + 2\,i\,y_2\,\Gamma^{mn}\,\eta^{ab}\,\rho_{mn}\,+\nonumber\\
& + i\,y_1\,\Gamma_5\,\Gamma_p\,\varepsilon^{cmnp}\,\eta^{ab}\,\rho_{mn} - y_1\,\eta^{c[m}\,\Gamma^{n]}\,\eta^{ab}\,\rho_{mn}\big{)}\,e_b.
\end{align}
This is equivalent to
\begin{equation}\label{sec3:rhoconstraint1}
0 = \Gamma^a\,\rho^{cb} + \tfrac{x_1}{2}\,\Gamma^c\,\rho^{ba} + \tfrac{x_1}{2}\,\Gamma^b\,\rho^{ac}
\end{equation}
and
\begin{equation} \label{eq:x2_y1_y2}
 x_2 = 0, \quad y_1 = y_2 = 0 \Rightarrow \lambda_0^W = 0.
\end{equation}
Eq. (\ref{sec3:rhoconstraint1}) is consistent  with $\rho$ not identically vanishing only if 
\begin{equation} \label{eq:x1_2}
x_1=2,
\end{equation}
which in turns implies
\begin{equation}\label{sec3:rhoconstraint3}
0 = \Gamma^{[a}\,\rho^{bc]}.
\end{equation}
This equation,  which we will call \emph{fermionic constraint}, is equivalently written as 
\begin{equation}\label{sec3:rhoconstraint2}
\Gamma^a\,\rho_{ab} = 0.
\end{equation}
From eqs. (\ref{sec3:rhoconstraint3})-(\ref{sec3:rhoconstraint2})  and (\ref{sec3:lambda0Wansatz})-(\ref{sec3:lambda0Ransatz}) we deduce  that
\begin{equation}
\lambda_0^R =0.
\end{equation}
A Majorana spinorial two-form $\rho$ carries the following representation of the Lorentz group:
\begin{equation}
\rho \sim \bm{4} \oplus \bm{12} \oplus \bm{8},
\end{equation}
where $\bm{4}=(\tfrac{1}{2},0)\oplus (0,\tfrac{1}{2})$ is the Dirac representation, $ \bm{8}= (\tfrac{3}{2},0)\oplus(0,\tfrac{3}{2})$ and $\bm{12} =(\tfrac{1}{2},1)\oplus (1,\tfrac{1}{2})$. 
The fermionic constraint  imposes 16 equations which  put the $\bm{4}\oplus\bm{12}$ to zero.\footnote{The $\bm{4}$  corresponds to the spinor $ \Gamma^{ab}\,\rho_{ab}$.  The self-dual combination $\star\rho+ i\,\Gamma_5\,\rho$ is the $\bm{12}$ and the anti-self-dual part $\star\rho- i\,\Gamma_5\,\rho $ is the  $\bm{4}\oplus\bm{8}$, where ${\star\rho_{ab}} \equiv \tfrac{1}{2}\,\varepsilon_{abcd}\,\rho^{cd}$.}

These 16 equations can be solved to express the conformal gravitino $\tilde{\psi}$ algebraically in terms of the other fields:
\begin{equation}\label{sec3:psitildesol}
\tilde{\psi}_a = \tfrac{i}{2}\,\Gamma^b\,D_{[b}\psi_{a]}  -\tfrac{i}{12}\,\Gamma_a\,\Gamma^{bc}D_{[b}\psi_{c]}.
\end{equation}
Therefore matching fermionic and bosonic degrees of freedom requires that 16 bosonic off-shell degrees of freedom also be eliminated: we will see momentarily that these composite degrees of freedom are  the  $f^a$ fields.

Since eqs. (\ref{eq:x2_y1_y2}) and (\ref{eq:x1_2}) have determined the $\lambda_0$ associated to Lorentz transformations (\ref{sec3:lambda0Lansatz}) to be
\begin{equation}\label{sec3:lambda0L}
(\lambda_0^J)^{ab} = 2\, e_c\,\bar{\zeta}\,\Gamma^c\,\rho^{ab},
\end{equation}
BRST nilpotency on the gravitino $\psi$  is equivalent to
\begin{equation}
\label{sec3:s2psi}
s^2\,\psi = - i_\gamma\,(\rho) - \tfrac{1}{4}\,(\lambda_0^J)^{ab}\,\Gamma_{ab}\,\zeta  = \rho_{bc}\,\gamma^b\,e^c - \tfrac{1}{2}\,e_c\,\bar{\zeta}\,\Gamma^c\,\rho^{ab}\,\Gamma_{ab}\,\zeta  = 0,
\end{equation}
where $\rho \equiv  \tfrac{1}{2}\,\rho_{bc}\,e^b\,e^c$.  One can verify that this equation is indeed satisfied by using both  the Fierz identity for $\zeta$ and the fermionic constraint (\ref{sec3:rhoconstraint3}) for $\rho$.

By taking the covariant derivative of the torsion we obtain 
\begin{equation}
0=\text{D}\,T^a = R^a{}_b\,e^b + F^W\,e^a + 2\,\bar{\psi}\,\Gamma^a\,\text{D}\,\psi = \tilde{R}^a{}_b\,e^b + \tilde{F}^W\,e^a + 2\,\bar{\psi}\,\Gamma^a\,\rho,
\end{equation}
 or in components \footnote{We take the last two indices of  $\tilde{R}_{mn,ab}$ as valued in the Lorentz bundle.}
\begin{equation}
\tilde{R}_{[mn}{}^a{}_{b]} + \delta^a{}_{[b}\,\tilde{F}^W_{mn]} - 2\,\bar{\psi}_{[b}\,\Gamma^a\rho_{mn]} = 0.
\end{equation}
 We call this equation \emph{Bianchi constraint}, because it is a modified algebraic Bianchi identity for $\tilde{R}^{ab}$. This equation, together with the fermionic constraint,   allows one to write the antisymmetric part of the Ricci tensor of $\tilde{R}_{mn}{}^{ab}$ in terms of the Weyl and fermionic curvatures as 
\begin{equation}
\tfrac{1}{2}\,\Tilde{\mathcal{R}}_{[ab]} - \tilde{F}^W_{ab} + \bar{\psi}^c\,\Gamma_c\,\rho_{ab} = 0,
\end{equation}
or equivalently, in terms of the modified superconformal Ricci tensor,
\begin{equation} \label{RicciAntiConstraint}
\tilde{\mathcal{R}}'_{ab}- \tilde{\mathcal{R}}'_{ba} = -2\,\tilde{F}^W_{ab}.
\end{equation}
Superconformal covariance dictates the following  form for $\lambda_0^S$: 
\begin{equation}\label{sec3:lambda0tildeansatz}
\lambda_0^S = \tfrac{i}{4}\,\big{(}x\,{\star{\Tilde{F}^R_{ab}}} + y\,\Tilde{F}^W_{ab}\big{)}\,\Gamma^{ab}\,\Gamma_c\,\zeta\,e^c + \tfrac{i}{4}\,z\,\Tilde{\mathcal{R}}'\,\Gamma_c\,\zeta\,e^c,
\end{equation}
where $x,y,z$ are constants and 
\begin{equation}
{\star{\Tilde{F}^R_{ab}}} = \tfrac{1}{2}\,\varepsilon_{ab}{}^{mn}\,\Tilde{F}^R_{ab}.
\end{equation}  
We did not include a term proportional to $\Tilde{\mathcal{R}}'_{[ab]}\,\Gamma^{ab}$ in ans\"atz (\ref{sec3:lambda0tildeansatz})  since this term is equivalent to the one proportional to $\Tilde{F}^W_{ab}$ thanks to eq. (\ref{RicciAntiConstraint}).  Inserting (\ref{sec3:lambda0tildeansatz})  into the  BRST nilpotency equations for  $a$ and $b$,   one arrives at
\begin{equations}
s^2\,a &= -i_\gamma\,\big{(}(1-x)\,\Tilde{F}^R +y\, {\star{\Tilde{F}^W}}\big{)} -\tfrac{1}{2}\,t\,\Tilde{\mathcal{R}}'\,\gamma_c\,e^c,\\
s^2\,b &= -i_\gamma\,\big{(}x\,{\star{\Tilde{F}^R}} +(1+y)\,\Tilde{F}^W\big{)}-\tfrac{1}{2}\,t\,\Tilde{\mathcal{R}}'\,\gamma_c\,e^c.
\end{equations}
These equations impose the following constraint on the Weyl and R-symmetry superconformal curvatures
\begin{equation} \label{sec3:Wchiralconstraint}
\Tilde{F}^W = -\,{\star{\Tilde{F}^R}},
\end{equation}
together with
\begin{equation}\label{sec3;xyt}
y = x-1,\qquad  t = 0.
\end{equation}
We will call (\ref{sec3:Wchiralconstraint}) the  \emph{Weyl-chiral constraint}. By plugging  both  (\ref{sec3;xyt}) and the Weyl-chiral constraint  into eq. (\ref{sec3:lambda0tildeansatz}) we determine  $\lambda_0^S$ to be
\begin{equation}\label{sec3:lambda0psitilde}
\lambda_0^S = \tfrac{1}{4}\,\Gamma_5\,\Gamma^{mn}\,\Gamma_c\,\zeta\,\Tilde{F}^R_{mn}\,e^c.
\end{equation}

The BRST variation of the fermionic constraint leads to the equation for the Ricci tensor of $\tilde{R}^{\prime}_{ab}$
\begin{equation}\label{sec3:ricciconstraint}
\tilde{\mathcal{R}}'_{ab} = \star{\tilde{F}}^R_{ab},
\end{equation}
which  we will call the  \emph{Ricci constraint}. This equation, together with the Weyl-chiral constraint,  again implies eq. (\ref{RicciAntiConstraint})  for the antisymmetric part of the  Ricci tensor of $\tilde{R}^\prime_{ab}$,  but it also sets  its  symmetric part  to zero.


 The tensor $\Tilde{R}^\prime_{mn,ab}$ has $6\times 6 = 36$ components. It transforms in  the following representation of the Lorentz group:
\begin{equation}
\Tilde{R}^\prime_{mn,ab} \sim \bm{10}_s\oplus\bm{9}_s\oplus\bm{1}_s\oplus\bm{1}_s'\oplus\bm{9}_a\oplus\bm{6}_a,
\end{equation}
where the suffix $s$ ($a$) denotes that the representation is  symmetric (anti-symmetric) with respect to the exchange of the two pairs of indices of   $\Tilde{R}^\prime_{mn,ab}$. $\bm{1}_s$ is  the Ricci scalar $\Tilde{\mathcal{R}}^\prime$,  $\bm{1}'_s$ its dual $\varepsilon^{mnab}\,\Tilde{R}_{mn,ab}^\prime$,  $\bm{9}_s\oplus \bm{1}_s\oplus\bm{6}_a$ is  the Ricci tensor  $\Tilde{\mathcal{R}}^\prime_{ab}$ and  $\bm{10}_s$ is the Weyl tensor representation. 

The Ricci constraint  puts the  $\bm{9}_s\oplus\bm{1}_s$ to zero and the $\bm{6}_a$ equal to $\tilde{F}^W_{ab}$.
The Bianchi constraint sets the $\bm{1}'_s\oplus\bm{9}_a$ to zero, beyond also putting the  $\bm{6}_a$ equal to the $\tilde{F}^W_{ab}$.  The independent components of $\Tilde{R}^\prime_{mn,ab}$ are hence captured by the Weyl tensor $\bm{10}_s$.

The 16 independent equations (\ref{sec3:ricciconstraint}) associated to the Ricci constraint can be solved algebraically for the 16 independent  $f_\mu{}^a$:
\begin{align}\label{fabsolved}
f_{ab} \equiv  e_b{}^\mu\, f_\mu{}^a = &\, -\tfrac{1}{4}\,\mathcal{R}_{ab} + \tfrac{1}{24}\,\mathcal{R}\,\eta_{ab} + \tfrac{1}{4}\,(\star\tilde{F}^R)_{ab} \,+\nonumber \\
& -\tfrac{1}{2}\,\bar{\psi}^c\,\Gamma_b\,\rho_{ac} 
-\tfrac{i}{2}\,\bar{\psi}^c\,\Gamma_{ca}\,\tilde{\psi}_b + \tfrac{i}{2}\,\bar{\psi}_b\,\Gamma_{ca}\,\tilde{\psi}^c + \tfrac{i}{6}\,\eta_{ab}\,\bar{\psi}^c\,\Gamma_{cd}\,\tilde{\psi}^d.
\end{align}
In conclusion, the superconformal Lorentz curvature $\Tilde{R}^\prime_{mn}{}^{ab}$, upon constraints, describes the $\bm{10}_s$ Weyl tensor degrees of freedom of the physical (non-superconformal) curvature $R_{mn}{}^{ab}$. The  Ricci degrees of freedom of the physical (non-superconformal) Riemann tensor $R_{mn}{}^{ab}$, which sit in the  $\bm{9}_s\oplus\bm{1}_s$ representation, are instead described by the symmetric part of $f_{ab}$. The remaining independent (off-shell) bosonic curvatures  are the physical  (non-superconformal)  curvature tensors $F^W_{mn}$ and $F^R_{mn}$.\footnote{ Indeed,  $R_{mn}{}^{ab}$, $F^W_{mn}$, $F^R_{mn}$, $f_{ab}$  have, before constraints,  respectively, $36$, $6$, $6$ and $16$ components,  for a total of  $64$ bosonic components. The Bianchi, Ricci and Weyl-chiral constraints impose $(1+9+6)+(1+9)+6 = 32$ conditions. Of the $64-32=32$ free components,   $20$ are the  components of the physical Riemann tensor, $6$ are the components of $F^R_{mn}$ and $6$ those of $F^W_{mn}$.}

Nilpotency of the BRST transformation for $\omega^{ab}$ 
\begin{equation}
s^2\omega^{ab} = -i_\gamma (\tilde{R}^{ab}) + \hat{s}\,(\lambda_0^J)^{ab} +\Omega^a{}_c\,(\lambda_0^J)^{cb} + 2\,i\,\bar{\zeta}\,\Gamma^{ab}\,\lambda_0^S
\end{equation}
is now ensured  thanks to  the  Bianchi  and  Weyl-chiral constraints, along with expressions (\ref{sec3:lambda0L}) and (\ref{sec3:lambda0psitilde}) for  $\lambda_0^J$ and $\lambda_0^S$. 
The nilpotency equation for $\tilde{\psi}$ 
\begin{equation}\label{sec3:s2psitilde}
s^2\tilde{\psi} =\, -i_\gamma\,(\tilde{\rho}') + \hat{s}\,\lambda_0^S +\tfrac{1}{4}\,\Omega^{ab}\,\Gamma_{ab}\,\lambda_0^S 
+ \tfrac{1}{2}\,\sigma\,\lambda_0^S - \tfrac{3}{2}\,i\,\alpha\,\Gamma_5\,\lambda_0^S - \tfrac{1}{4}(\lambda_0^J)^{ab} \Gamma_{ab} \eta + i(\lambda_0^K)_a\Gamma^a\zeta
\end{equation}
involves the yet to be determined $\lambda_0^K$, for which superconformal invariance dictates the following ans\"atz:
\begin{equation}
(\lambda_0^K)^a = -i\,x\,e_c\,\bar{\zeta}\,\Gamma^c\,\Gamma_b\,\tilde{\rho}'^{ab},
\end{equation}
with $x$ constant. By plugging this expression into the nilpotency equation (\ref{sec3:s2psitilde}) one obtains
\begin{align}
s^2\,\tilde{\psi} =&\, -i_\gamma\,(\tilde{\rho}') + 
\tfrac{1}{2}\,(\bar{\zeta}\,\Gamma_5\,\tilde{\rho}'_{ab})\,\Gamma^{ab}\,\Gamma_5\,\Gamma^c\,\zeta\,e_c - x\,\bar{\zeta}\,\Gamma^c\,\Gamma_b\,\tilde{\rho}'^{ba}\,\Gamma_a\,\zeta\,e_c \,+\nonumber\\
& - \tfrac{1}{2}\,e_c\,\bar{\zeta}\,\Gamma^c\,\rho^{ab}\,\Gamma_{ab}\,\eta + \tfrac{1}{2}\,(\bar{\eta}\,\Gamma_5\,\rho_{ab})\,\Gamma^{ab}\,\Gamma_5\,\Gamma^c\,\zeta\,e_c.
\end{align}
The $\zeta\eta$ terms cancel out thanks to the fermionic constraint. The remaining terms $\bar\zeta\zeta $ terms all vanish thanks to the identity
\begin{equation}
\Gamma^{ab}\,\tilde{\rho}'_{ab} = 0,
\end{equation}
which descends from the solution (\ref{sec3:psitildesol}) of the fermionic constraint, if one also takes $x=1$, that is 
\begin{equation}
(\lambda_0^K)^a = -i\,e_c\,\bar{\zeta}\,\Gamma^c\,\Gamma_b\,\tilde{\rho}'^{ab}.
\end{equation}
Finally, the nilpotency equation for $f^a$ 
\begin{equation}
s^2 f^a = - i_\gamma(\tilde{T}^a) + \hat{s}\,(\lambda_0^K)^a + \Omega^a{}_b\,(\lambda_0^K)^b - \sigma\,(\lambda_0^K)^a +2\,\bar{\eta}\,\Gamma^a\,\lambda_0^S
\end{equation}
holds thanks to  the Ricci constraint, which also ensures that the trace of the $f^a$ curvature $\tilde{T}$ vanishes
\begin{equation}\label{Ttildeconstraint}
\tilde{T}'_{ab}{}^a = 0.
\end{equation}

\section{The Chern-Simons superconformal anomaly}\label{sec:superanomaly}

When the constraints are satisfied, the superconformal generalized curvatures 
\begin{equation} \label{sec4:sH_hor}
  \bm{F}=   \bm{F}^i \, T_i
\end{equation}
satisfy the generalized Bianchi identities
\begin{equation}\label{sec4:genBianchi}
\delta \, \bm{F} + [\bm{A},   \bm{F}] =0.
\end{equation}
We can therefore construct generalized Chern classes of total degree 6  by considering cubic  polynomials of the curvatures 
\begin{equation}\label{sec4:P3dtilde}
    \bm{P}_3(\bm{F}^i)=\Tilde{d}_{ijk}\bm{F}^i\bm{F}^j\bm{F}^k,
\end{equation}
which are {\it superconformal invariant}.  From its definition, $\Tilde{d}_{ijk}$ is a ``completely symmetric''  (in the graded sense) 
tensor
\begin{equation} \label{sec4eq:dijk_simm1}
     \Tilde{d}_{ijk} =(-)^{|i||j|}\, \Tilde{d}_{jik},\qquad \Tilde{d}_{ijk}  =(-)^{|j||k|}\, \Tilde{d}_{ikj}. 
\end{equation}
$ \bm{P}_3(\bm{F}^i)$ is superconformal invariant if $\Tilde{d}_{ijk}$ is a superconformal invariant tensor, that is if it satisfies:
\begin{equation}\label{sec4:dinvarianceeq}
f_m{}^l{}_i\,\tilde{d}_{ljk}+(-)^{|j||m|}f_m{}^l{}_j\,\tilde{d}_{ilk}+(-)^{|m||k|+|m||j|}\,f_m{}^l{}_k\,\tilde{d}_{ijl}=0.
\end{equation}
Superconformal invariance of $\Tilde{d}_{ijk}$, together with the generalized Bianchi identity (\ref{sec4:genBianchi}), ensures that 
$\bm{P}_3(\bm{F}^i)$ is $\delta$-closed:
\begin{equation}
\delta\,  \bm{P}_3(\bm{F}^i)=0.
\end{equation}
We searched for solutions of eq. (\ref{sec4:dinvarianceeq})  with symmetry properties (\ref{sec4eq:dijk_simm1}) and found
a single solution, up to a multiplicative constant:
\begin{align} \label{eq:inv_poly}
 \bm{P}_3(\bm{F}^i) =& + 15 \,(\bm{\tilde{F}^{R}{}})^3+ 3 \,\bm{\tilde{F}^{R}{}} \,(\bm{\tilde{F}^{W}{}})^2- \tfrac{3}{4} \varepsilon_{abcd} \,\bm{\tilde{F}^{W}{}} \,\bm{\tilde{R}}^{ab} \,\bm{\tilde{R}}^{cd}- \tfrac{3}{2} \,\bm{\tilde{F}^{R}{}} \,\bm{\tilde{R}}_{ab} \,\bm{\tilde{R}}^{ab}\,+\nonumber\\
&-6 \,\Bar{\bm{\rho}} \,\Gamma^{ab}{}{} \Gamma_5{}{}  \,\bm{\tilde{\rho}} \,\bm{\tilde{R}}_{ab}+60\, i \,\Bar{\bm{\rho}} \,\bm{\tilde{\rho}} \,\bm{\tilde{F}^{R}{}}+12 \,\Bar{\bm{\rho}}\,\Gamma_5{}{}\,\bm{\tilde{\rho}}  \,\bm{\tilde{F}^{W}{}}-12\,i \,\Bar{\bm{\rho}}\,\Gamma^{a}{}{} \Gamma_5{}{} \,\bm{\rho}  \,\bm{\tilde{T}}_{a}\,+\nonumber\\
&+12 \,\bm{\tilde{F}^{R}{}} \,\bm{T}^{a} \,\bm{\tilde{T}}_{a}-6\, \varepsilon_{abcd} \,\bm{\tilde{R}}^{cd} \,\bm{T}^{a} \,\bm{\tilde{T}}^{b}-12\,i \,\Bar{\bm{\tilde{\rho}}} \,\Gamma^{a}{}{} \Gamma_5{}{} \,\bm{\tilde{\rho}} \,\bm{T}_{a}.
\end{align}
Since the super-covariant generalized curvatures are not horizontal 
\begin{equation} \label{eq:gen_curv_H}
    \bm{F}^i=F^i+\lambda_0^i,
\end{equation}
it is not ``a priori'' guaranteed that  the BRST-invariant generalized polynomial $P_3(\bm{F}^i)$ gives rise  to  a secondary generalized Chern-Simons class of degree 5, i.e. to an anomaly cocycle.  Since the non-horizontal components of the generalized curvatures are 1-forms, $\bm{P}_3(\bm{F}^i)$ has, in principle, components of form degrees  4 and 3:
\begin{equation}
 \bm{P}_3 = P^{(4)}_2 + P^{(3)}_3.
\end{equation}
However, it is easy to see that   $P^{(3)}_3=0$, due to the specific form of the superconformal invariant (\ref{eq:inv_poly}) that we found, and the fact that only $\{\lambda_0^K, \lambda_0^J, \lambda_0^S\}$ are non-vanishing.  Hence
\begin{equation}\label{sec4:P42}
\bm{P}_3(\bm{F}^i)=-\tfrac{3}{4}\,\varepsilon_{abcd}\,(\lambda_0^J)^{ab}\,(\lambda_0^J)^{cd}\,\Tilde{F}^W-\tfrac{3}{2}\,(\lambda_0^J)^{ab}\,(\lambda_0^J)_{ab}\,\Tilde{F}^R-6\,\Bar{\rho}\, \Gamma_{ab}\, \Gamma_5\lambda_0^S(\lambda_0^J)^{ab}.
\end{equation}
It is quite remarkable that,  by taking into account both the expressions for $\lambda_0^i$'s (\ref{eq:lambda_0_a}-\ref{eq:lambda_0_c}) and the constraints on curvatures (\ref{eq:curv_constraints_a})-(\ref{eq:curv_constraints_d}), this 4-form turns out to vanish
\begin{equation}
   \bm{P}_3(\bm{F}^i)=0.
\end{equation}
The vanishing of $   \bm{P}_3(\bm{F}^i)$ triggers the Chern-Simons secondary class mechanism:  the generalized Chern-Simons polynomial of total degree 5 is an anomaly cocycle
\begin{equation}
\bm{P}_3(\bm{F}) = \delta\,Q^{(5)}(\bm{A},\bm{F}) =0.
\end{equation}


The Chern-Simons polynomial is completely determined by the super-invariant tensor $\Tilde{d}_{ijk}$ and the structure constants $\Tilde{f}^i_{jk}$ according to the universal formula
   \begin{equation}\label{superconfinvariantanomaly}
   Q^{(5)}(\bm{A}, \bm{F}) =\Tilde{d}_{ijk}\,\bm{A}^i\,\bm{F}^j\,\bm{F}^k- \tfrac{1}{4}\, \Tilde{f}^i_{mn}\, \Tilde{d}_{ijk}\,\bm{A}^m\,\bm{A}^n \, \bm{A}^j\, \bm{F}^k + \tfrac{1}{40}\, \Tilde{f}^i_{pq}\,\Tilde{f}^l_{nj}\,\Tilde{d}_{ilk}\,\bm{A}^p\,\bm{A}^q \, \bm{A}^n\, \bm{A}^j\, \bm{A}^k.
    \end{equation}
If we denote by  $Q_{5,N}$ the part of the Chern-Simons polynomial (\ref{superconfinvariantanomaly}) of degree $N$ in the number of  generalized forms $\bm{A}$ and  $\bm{F}$, we obtain the following explicit expressions, to all orders in all the fermionic fields \footnote{We simplified these expressions slightly by inserting the torsion constraint $T^a=0$.}
\begin{align} \label{sec4: CS1}
Q_{5; 3}=&+\,15\, \bm{a}\, (\bm{\tilde{F}^R}{})^2\,+\, 2\, \bm{\tilde{F}^R}{}\bm{b} \,\bm{\tilde{F}^W}{} \,+\, \bm{a} (\bm{\tilde{F}^W}{})^2 \,+\nonumber\\
&-\, \tfrac{1}{2} \varepsilon_{abcd} \,\bm{\omega}^{ab}\,  \bm{\tilde{R}}^{cd}\, \bm{\tilde{F}^W}{} -  \tfrac{1}{4} \varepsilon_{abcd} \,\bm{b}\, \bm{\tilde{R}}^{ab} \,\bm{\tilde{R}}^{cd}\,+\nonumber\\
&-\, \bm{\omega}^{ab} \,  \bm{\tilde{R}}_{ab}\,\bm{\tilde{F}^R}{} -  \tfrac{1}{2}\, \bm{a}\, \bm{\tilde{R}}_{ab}\, \bm{\tilde{R}}^{ab} \,+\nonumber\\
&-\, 2\, \overline{\bm{\tilde{\rho}}}\, \Gamma^{ab} \Gamma_5 \,\bm{\psi}\, \bm{\tilde{R}}_{ab} \,+\, 2 \,\bar{\bm{\rho}}\, \Gamma^{ab} \Gamma_5{} \,\bm{\tilde{\psi}}\, \bm{\tilde{R}}_{ab} - 2 \,\bar{\bm{\rho}}\,\Gamma^{ab}{} \Gamma_5{}\,\bm{\tilde{\rho}}\, \bm{\omega}_{ab} \,+\nonumber\\
&+\, 20i\, \overline{\bm{\psi}}\,\bm{\tilde{\rho}}\,\bm{\tilde{F}^R}{}\,-\,20i\, \bar{\bm{\rho}}\,\bm{\tilde{\psi}}\, \bm{\tilde{F}^R}{}\,+\,20i \,\bar{\bm{\rho}}\,\bm{\tilde{\rho}}\,\bm{a} \,+\,\nonumber\\
&+\, 4\, \overline{\bm{\psi}}\, \Gamma_5 \,\bm{\tilde{\rho}}\,\bm{\tilde{F}^W}{}\,-\, 4 \,\bar{\bm{\rho}}\, \Gamma_5{} \,\bm{\tilde{\psi}}\, \bm{\tilde{F}^W}{}  \,+\, 4 \,\bar{\bm{\rho}}\,\Gamma_5{}\, \bm{\tilde{\rho}} \,\bm{b}\,+\nonumber\\
&+\, 8i\, \bar{\bm{\rho}}\,\Gamma_{a}{} \Gamma_5{} \,\bm{\psi} \,\bm{\tilde{T}}^{a} \,-\, 4i\,\bm{f}^a\, \bar{\bm{\rho}}\,\Gamma_{a}{} \Gamma_5{} \,\bm{\rho}  \,+\nonumber\\
&+\, 4 \,e^{a}\,\bm{\tilde{F}^R}{} \,\bm{\tilde{T}}_{a}\,- \,2\, \varepsilon_{abcd} \,e^{a}\,\bm{\tilde{T}}^{b}\, \bm{\tilde{R}}^{cd}\,-\, 4i \, e^a\,\overline{\bm{\tilde{\rho}}}\,\Gamma_{a}{} \Gamma_5{}\,\bm{\tilde{\rho}},
\end{align}
\begin{align} \label{sec4: CS2}
\!\! Q_{5; 4} = &-4i \,\overline{\bm{\psi}}\,\bm{\tilde{\rho}} \,\bm{a} \,\bm{b}  - 8 \,\overline{\bm{\psi}}\,\Gamma^{a}{}{}\,\bm{\rho} \,\bm{f}_{a} \,\bm{a}   - 8 \,\overline{\bm{\tilde{\psi}}}\,\Gamma^{a}{}{}\,\bm{\tilde{\rho}} \,\bm{a} e_{a}   - \tfrac{1}{4} \,\overline{\bm{\psi}}\,\Gamma^{a}{}{}\,\bm{\rho} \,\bm{f}_{b}  \,\bm{\omega}_{cd} \varepsilon_{a}{}^{bcd} \,+\nonumber\\ 
&+  \tfrac{1}{4} \,\overline{\bm{\tilde{\psi}}}\,\Gamma^{a}{}{} \,\bm{\tilde{\rho}}\,e_{b}  \,\bm{\omega}_{cd} \varepsilon_{a}{}^{bcd}  + 2i \,\overline{\bm{\psi}}\,\Gamma^{ab}{}{} \,\bm{\tilde{\rho}} \,\bm{a}  \,\bm{\omega}_{ab}  + \tfrac{3}{2}i \,\overline{\bm{\psi}}\,\Gamma^{ab}{}{}\,\bm{\tilde{\rho}} \,\bm{f}_{c} e_{d}  \varepsilon_{ab}{}^{cd}  \,+\nonumber\\
&-  \tfrac{1}{2}i \,\overline{\bm{\psi}} \,\Gamma^{ab}{}{} \,\bm{\tilde{\rho}} \,\bm{\omega}_{c}{}^{e} \,\bm{\omega}_{de} \varepsilon_{ab}{}^{cd}  - \tfrac{1}{4} \,\overline{\bm{\psi}} \,\Gamma^{a}{}{} \,\Gamma^{bc}{}{}\,\bm{\rho} \,\bm{f}_{a}  \,\bm{\omega}_{de} \varepsilon_{bc}{}^{de}  + 3 \,\overline{\bm{\psi}}\,\Gamma_5{}{}\,\bm{\tilde{\rho}}  \,\bm{f}^{a} e_{a} \,+\nonumber\\
& - 4i \,\overline{\bm{\psi}} \,\Gamma_5{}{}\Gamma^{a}{}{}\,\bm{\rho}\,\bm{f}_{a} \,\bm{b}    +  \tfrac{5}{2}i \,\overline{\bm{\psi}}\,\Gamma_5{}{}\Gamma^{a}{}{}\,\bm{\rho} \,\bm{f}^{b}  \,\bm{\omega}_{ab}   + 4i \,\overline{\bm{\tilde{\psi}}}\,\Gamma_5{}{}\Gamma^{a}{}{} \,\bm{\tilde{\rho}} e_{a} \,\bm{b}   \,+\nonumber\\
&+  \tfrac{7}{2}i \,\overline{\bm{\tilde{\psi}}}\,\Gamma_5{}{}\Gamma^{a}{}{} \,\bm{\tilde{\rho}} e^{b}  \,\bm{\omega}_{ab}  + \,\overline{\bm{\psi}}\,\Gamma_5{}{}\Gamma^{a}{}{} \,\Gamma^{b}{}{}\,\bm{\tilde{\rho}}  \,\bm{f}_{a} e_{b}   +  \tfrac{1}{4}i \,\overline{\bm{\psi}}\,\Gamma_5{}{}\Gamma^{bc}{}{}\Gamma^{a}{}{}\,\bm{\rho} \,\bm{f}_{a}  \,\bm{\omega}_{bc}    \,+\nonumber\\
&+  \tfrac{1}{4}i \,\overline{\bm{\tilde{\psi}}} \,\Gamma_5{}{}\Gamma^{bc}{}{}\Gamma^{a}{}{}\,\bm{\tilde{\rho}}\,e_{a}  \,\bm{\omega}_{bc}    - 4i \,\overline{\bm{\rho}}\,\bm{\tilde{\psi}} \,\bm{a} \,\bm{b}  + 2i  \,\bm{a} \,\bm{\omega}_{ab} \,\overline{\bm{\rho}}\,\Gamma^{ab}{}{} \,\bm{\tilde{\psi}} \,+\nonumber\\
&-  \tfrac{3}{2}i  \,\bm{f}_{c} e_{d} \varepsilon_{ab}{}^{cd} \,\overline{\bm{\rho}}\,\Gamma^{ab}{}{} \,\bm{\tilde{\psi}} + \tfrac{1}{4}i  \,\bm{\omega}_{c}{}^{e} \,\bm{\omega}_{de} \varepsilon_{ab}{}^{cd} \,\overline{\bm{\rho}}\,\Gamma^{ab}{}{} \,\bm{\tilde{\psi}} + \tfrac{1}{16}i  \,\bm{\omega}_{ab} \,\bm{\omega}_{ef} \varepsilon_{cd}{}^{ef} \,\overline{\bm{\rho}} \,\Gamma^{cd}{}{}\,\Gamma^{ab}{}{} \,\bm{\tilde{\psi}} \,+\nonumber\\
&- 3  \,\bm{f}^{a} e_{a} \,\overline{\bm{\rho}}\,\Gamma_5{}{} \,\bm{\tilde{\psi}} - 2i \,\overline{\bm{\psi}}\,\bm{\tilde{\psi}} \,\overline{\bm{\rho}} \Gamma_5{}{}  \,\bm{\tilde{\psi}} - 10i \,\overline{\bm{\psi}}\,\Gamma_5{}{} \,\bm{\tilde{\psi}} \,\overline{\bm{\rho}}  \,\bm{\tilde{\psi}} \,+\nonumber\\
&-   \,\bm{f}_{a} e_{b}\,\overline{\bm{\rho}} \,\Gamma^{a}{}{} \,\Gamma^{b}{}{} \Gamma_5{}{} \,\bm{\tilde{\psi}} -  \tfrac{1}{2}  \varepsilon_{abcd}\,\overline{\bm{\psi}}\,\Gamma^{ab}{}{}  \,\bm{\tilde{\psi}}\,\overline{\bm{\rho}}  \,\Gamma^{cd}{}{} \,\bm{\tilde{\psi}} + 3 \,\bm{f}^{a} e_{a} \,\bm{b} \,\bm{\tilde{F}^{R}{}} \,+\nonumber\\
&+ 3 \,\bm{f}^{a} e^{b} \,\bm{\omega}_{ab} \,\bm{\tilde{F}^{R}{}} -  \tfrac{1}{4} \,\bm{\omega}^{ab} \,\bm{\omega}_{a}{}^{c} \,\bm{\omega}_{bc} \,\bm{\tilde{F}^{R}{}} - 3i \,\overline{\bm{\psi}} \,\Gamma^{ab}{}{}\,\bm{\tilde{\psi}} \,\bm{\omega}_{ab}  \,\bm{\tilde{F}^{R}{}} - 6 \,\overline{\bm{\psi}} \,\Gamma^{a}{}{}  \,\bm{\psi}\,\bm{f}_{a} \,\bm{\tilde{F}^{R}{}} \,+\nonumber\\
&+ 30 \,\overline{\bm{\psi}} \Gamma_5{}{} \,\bm{\tilde{\psi}} \,\bm{a} \,\bm{\tilde{F}^{R}{}} + 6 \,\overline{\bm{\tilde{\psi}}} \,\Gamma^{a}{}{} \,\bm{\tilde{\psi}} \,\bm{\tilde{F}^{R}{}} e_{a} + 6i \,\overline{\bm{\psi}} \,\bm{\tilde{\psi}} \,\bm{\tilde{F}^{R}{}} \,\bm{b} -  \,e_{a}\,\bm{f}^{a} \,\bm{a}  \,\bm{\tilde{F}^{W}{}} \,\nonumber\\
&+ \tfrac{1}{2} \,\bm{f}_{a} e_{b} \,\bm{\omega}_{cd} \varepsilon^{abcd} \,\bm{\tilde{F}^{W}{}} -  \tfrac{1}{8} \,\bm{\omega}_{ab} \,\bm{\omega}_{c}{}^{e} \,\bm{\omega}_{de} \varepsilon^{abcd} \,\bm{\tilde{F}^{W}{}} -  \tfrac{1}{2}i \,\overline{\bm{\psi}} \varepsilon_{abcd} \,\Gamma^{ab}{}{} \,\bm{\tilde{\psi}}\,\bm{\omega}^{cd}  \,\bm{\tilde{F}^{W}{}} \,+\nonumber\\
&- i \,\overline{\bm{\psi}} \, \Gamma_5{}{} \Gamma^{a}{}{}\,\bm{\psi} \,\bm{f}_{a} \,\bm{\tilde{F}^{W}{}} - 2i \,\overline{\bm{\psi}} \,\bm{\tilde{\psi}} \,\bm{a} \,\bm{\tilde{F}^{W}{}} + i \,\overline{\bm{\tilde{\psi}}} \,\Gamma^{a}{}{} \Gamma_5{}{} \,\bm{\tilde{\psi}} e_{a} \,\bm{\tilde{F}^{W}{}} \,+\nonumber\\
&+ 2 \,\overline{\bm{\psi}} \Gamma_5{}{} \,\bm{\tilde{\psi}} \,\bm{b} \,\bm{\tilde{F}^{W}{}} -  \,\overline{\bm{\psi}} \Gamma_5{}{}\,\bm{\tilde{\psi}} \,\bm{\omega}^{ab}  \,\bm{\tilde{R}}_{ab} + i \,\overline{\bm{\psi}} \, \Gamma_5{}{} \Gamma^{a}{}{}\,\bm{\psi}\,\bm{f}^{b}  \,\bm{\tilde{R}}_{ab} \,+\nonumber\\
&+ i \,\overline{\bm{\psi}} \,\Gamma^{ab}{}{} \,\bm{\tilde{\psi}} \,\bm{a} \,\bm{\tilde{R}}_{ab} + i \,\overline{\bm{\tilde{\psi}}} \,\Gamma^{a}{}{} \Gamma_5{}{} \,\bm{\tilde{\psi}} e^{b} \,\bm{\tilde{R}}_{ab} -  \,e_{b}\,\bm{f}_{a} \,\bm{a}  \,\bm{\tilde{R}}^{ab} -  \tfrac{1}{4} \,\bm{a} \,\bm{\omega}_{a}{}^{c} \,\bm{\omega}_{bc} \,\bm{\tilde{R}}^{ab} \,+\nonumber\\
& + \tfrac{3}{2} \,\bm{f}_{c} e_{d} \,\bm{b} \varepsilon_{ab}{}^{cd} \,\bm{\tilde{R}}^{ab} + \tfrac{1}{2} \,\bm{f}_{c} e^{e} \,\bm{\omega}_{de} \varepsilon_{ab}{}^{cd} \,\bm{\tilde{R}}^{ab} + \tfrac{1}{2} \,\bm{f}^{e} e_{c} \,\bm{\omega}_{de} \varepsilon_{ab}{}^{cd} \,\bm{\tilde{R}}^{ab} -  \tfrac{1}{4} \,\bm{f}^{e} e_{e} \,\bm{\omega}_{cd} \varepsilon_{ab}{}^{cd} \,\bm{\tilde{R}}^{ab} \,+\nonumber\\
&-  \tfrac{1}{8} \,\bm{b} \,\bm{\omega}_{c}{}^{e} \,\bm{\omega}_{de} \varepsilon_{ab}{}^{cd} \,\bm{\tilde{R}}^{ab} -  \tfrac{1}{2}i \,\overline{\bm{\psi}}\,\bm{\tilde{\psi}} \varepsilon_{abcd} \,\bm{\omega}^{ab}  \,\bm{\tilde{R}}^{cd} +  \varepsilon_{abcd} \,\overline{\bm{\psi}}\,\Gamma^{a}{}{}\,\bm{\psi} \,\bm{f}^{b}  \,\bm{\tilde{R}}^{cd} \,+\nonumber\\
&+ \varepsilon_{abcd}\,\overline{\bm{\tilde{\psi}}}  \,\Gamma^{a}{}{} \,\bm{\tilde{\psi}} e^{b} \,\bm{\tilde{R}}^{cd} -  \tfrac{1}{2}i \varepsilon_{abcd}\,\overline{\bm{\psi}}  \,\Gamma^{ab}{}{} \,\bm{\tilde{\psi}} \,\bm{b} \,\bm{\tilde{R}}^{cd} + 2i \,\overline{\bm{\tilde{\psi}}}\,\Gamma^{a}{}{}\,\bm{\tilde{\psi}} \,\overline{\bm{\psi}} \,\Gamma_5{}{}\Gamma_{a}{} \,   \,\bm{\rho} \,+\nonumber\\
&+ 10i \,\overline{\bm{\psi}}\,\Gamma_5{}{} \,\bm{\tilde{\psi}} \,\overline{\bm{\psi}}  \,\bm{\tilde{\rho}} + 2i \,\overline{\bm{\psi}}\Gamma_5{}{} \,\bm{\tilde{\rho}}\,\overline{\bm{\psi}}  \,\bm{\tilde{\psi}} + \tfrac{1}{2}\varepsilon_{abcd} \,\overline{\bm{\psi}}   \,\Gamma^{ab}{}{} \,\bm{\tilde{\psi}}\,\overline{\bm{\psi}}\,\Gamma^{cd}{}{}  \,\bm{\tilde{\rho}} \,+\nonumber\\
&- 2i \,\overline{\bm{\tilde{\psi}}}\,\Gamma_5{}{}\Gamma_{a}{}\,\bm{\tilde{\rho}} \,\overline{\bm{\psi}}  \,\Gamma^{a}{}{}  \,\bm{\psi}  + 2 \,\overline{\bm{\psi}} \,\Gamma^{a}{}{} \,\bm{\psi} \,\bm{a} \,\bm{\tilde{T}}_{a} + 4 \,\overline{\bm{\psi}} \Gamma_5{}{} \,\bm{\tilde{\psi}} e^{a} \,\bm{\tilde{T}}_{a} \,+\nonumber\\
&+ i \,\overline{\bm{\psi}} \,\Gamma_5{}{}\Gamma^{a}{}{}  \,\bm{\psi} \,\bm{b} \,\bm{\tilde{T}}_{a} + \,\bm{a} e_{a} \,\bm{b} \,\bm{\tilde{T}}^{a} + \,\bm{a} e^{b} \,\bm{\omega}_{ab} \,\bm{\tilde{T}}^{a} - 2 \,\bm{f}_{b} e_{c} e_{d} \varepsilon_{a}{}^{bcd} \,\bm{\tilde{T}}^{a} + \tfrac{1}{2} e_{b} \,\bm{\omega}_{c}{}^{e} \,\bm{\omega}_{de} \varepsilon_{a}{}^{bcd} \,\bm{\tilde{T}}^{a} \,+\nonumber\\
&+ \tfrac{1}{2} e_{b} \,\bm{b} \,\bm{\omega}_{cd} \varepsilon_{a}{}^{bcd} \,\bm{\tilde{T}}^{a} -  \tfrac{1}{2} e^{e} \,\bm{\omega}_{bc} \,\bm{\omega}_{de} \varepsilon_{a}{}^{bcd} \,\bm{\tilde{T}}^{a} +  \varepsilon_{abcd}\,\overline{\bm{\psi}} \,\Gamma^{a}{}{}\,\bm{\psi} \,\bm{\omega}^{cd}  \,\bm{\tilde{T}}^{b} \,+\nonumber\\
&- i \,\overline{\bm{\psi}} \,\Gamma_5{}{}\Gamma^{a}{}{} \,\bm{\psi} \,\bm{\omega}_{ab}  \,\bm{\tilde{T}}^{b} - 2i\varepsilon_{abcd} \,\overline{\bm{\psi}}  \,\Gamma^{cd}{}{} \,\bm{\tilde{\psi}}\, e^{a} \,\bm{\tilde{T}}^{b}, 
\end{align}
\begin{align} \label{sec4: CS3}
Q_{5; 5} =&-2 \,\bm{f}_{a} \,\bm{f}_{b} e_{c} e_{d} \,\bm{b} \varepsilon^{abcd} -  \tfrac{4}{5} \,\bm{f}_{a} \,\bm{f}_{b} e_{c} e^{e} \,\bm{\omega}_{de} \varepsilon^{abcd}+\nn\\
& + \tfrac{4}{5} \,\bm{f}_{a} \,\bm{f}^{e} e_{b} e_{c} \,\bm{\omega}_{de} \varepsilon^{abcd} -  \tfrac{2}{5} \,\bm{f}_{a} \,\bm{f}^{e} e_{b} e_{e} \,\bm{\omega}_{cd} \varepsilon^{abcd} +\nonumber\\
&+ \tfrac{3}{5} \,\bm{f}_{a} e_{b} \,\bm{b} \,\bm{\omega}_{c}{}^{e} \,\bm{\omega}_{de} \varepsilon^{abcd} + \tfrac{1}{5} \,\bm{f}_{a} e^{e} \,\bm{\omega}_{be} \,\bm{\omega}_{c}{}^{l} \,\bm{\omega}_{dl} \varepsilon^{abcd} -  \tfrac{1}{5} \,\bm{f}_{a} e^{e} \,\bm{b} \,\bm{\omega}_{bc} \,\bm{\omega}_{de} \varepsilon^{abcd} \,+\nonumber\\
&+ \tfrac{1}{5} \,\bm{f}^{e} e_{a} \,\bm{\omega}_{be} \,\bm{\omega}_{c}{}^{l} \,\bm{\omega}_{dl} \varepsilon^{abcd} + \tfrac{1}{5} \,\bm{f}^{e} e_{a} \,\bm{b} \,\bm{\omega}_{bc} \,\bm{\omega}_{de} \varepsilon^{abcd} -  \tfrac{1}{10} \,\bm{f}^{e} e_{e} \,\bm{\omega}_{ab} \,\bm{\omega}_{c}{}^{l} \,\bm{\omega}_{dl} \varepsilon^{abcd} \,+\nonumber\\
&-  \tfrac{1}{5} \,\bm{f}^{e} e^{l} \,\bm{\omega}_{ab} \,\bm{\omega}_{ce} \,\bm{\omega}_{dl} \varepsilon^{abcd} -  \tfrac{1}{40} \,\bm{b} \,\bm{\omega}_{a}{}^{e} \,\bm{\omega}_{be} \,\bm{\omega}_{c}{}^{l} \,\bm{\omega}_{dl} \varepsilon^{abcd} + 6i \,\overline{\bm{\psi}}\,\bm{\tilde{\psi}} \,\bm{f}^{a} \,\bm{a} e_{a}  \,+\nonumber\\
&+ i \,\overline{\bm{\psi}} \,\bm{\tilde{\psi}}\,\bm{f}_{a} e_{b} \,\bm{\omega}_{cd} \varepsilon^{abcd}  -  \tfrac{1}{4}i \,\overline{\bm{\psi}} \,\bm{\tilde{\psi}}\,\bm{\omega}_{ab} \,\bm{\omega}_{c}{}^{e} \,\bm{\omega}_{de} \varepsilon^{abcd}  - 3 \,\overline{\bm{\tilde{\psi}}}\,\Gamma^{a}{}{} \,\bm{\tilde{\psi}} \,\bm{a} e_{a} \,\bm{b}  \,+\nonumber\\
&- 3 \,\overline{\bm{\tilde{\psi}}}\,\Gamma^{a}{}{} \,\bm{\tilde{\psi}} \,\bm{a} e^{b} \,\bm{\omega}_{ab}  - 2 \,\overline{\bm{\tilde{\psi}}}\,\Gamma^{a}{}{} \,\bm{\tilde{\psi}} \,\bm{f}_{b} e_{c} e_{d} \varepsilon_{a}{}^{bcd}  + \tfrac{7}{20} \,\overline{\bm{\tilde{\psi}}}\,\Gamma^{a}{}{} \,\bm{\tilde{\psi}} e_{b} \,\bm{\omega}_{c}{}^{e} \,\bm{\omega}_{de} \varepsilon_{a}{}^{bcd}  \,+\nonumber\\
&+ \tfrac{1}{2} \,\overline{\bm{\tilde{\psi}}} \,\Gamma^{a}{}{} \,\bm{\tilde{\psi}} \,e_{b} \,\bm{b} \,\bm{\omega}_{cd} \varepsilon_{a}{}^{bcd}  -  \tfrac{23}{40} \,\overline{\bm{\tilde{\psi}}}\,\Gamma^{a}{}{} \,\bm{\tilde{\psi}}\, e^{e} \,\bm{\omega}_{bc} \,\bm{\omega}_{de} \varepsilon_{a}{}^{bcd}  \,+\nonumber\\
&+ \tfrac{3}{40} \,\overline{\bm{\tilde{\psi}}}\,\Gamma^{a}{}{} \,\bm{\tilde{\psi}}\, e_{b} \,\bm{\omega}_{ac} \,\bm{\omega}_{de} \varepsilon^{bcde}  + 6i \,\overline{\bm{\psi}} \,\Gamma^{ab}{}{} \,\bm{\tilde{\psi}}\,\bm{f}_{a} \,\bm{a} e_{b}  + \tfrac{3}{2}i \,\overline{\bm{\psi}} \,\Gamma^{ab}{}{} \,\bm{\tilde{\psi}}\,\bm{a} \,\bm{\omega}_{a}{}^{c} \,\bm{\omega}_{bc}  \,+\nonumber\\
&+ 3i \,\overline{\bm{\psi}} \,\Gamma^{ab}{}{} \,\bm{\tilde{\psi}}\,\bm{f}_{c} e_{d} \,\bm{b} \varepsilon_{ab}{}^{cd}  + \tfrac{19}{20}i \,\overline{\bm{\psi}} \,\Gamma^{ab}{}{} \,\bm{\tilde{\psi}}\,\bm{f}_{c} e^{e} \,\bm{\omega}_{de} \varepsilon_{ab}{}^{cd}  + \tfrac{17}{20}i \,\overline{\bm{\psi}} \,\Gamma^{ab}{}{} \,\bm{\tilde{\psi}}\,\bm{f}^{e} e_{c} \,\bm{\omega}_{de} \varepsilon_{ab}{}^{cd}  \,+\nonumber\\
&-  \tfrac{3}{5}i \,\overline{\bm{\psi}}\,\Gamma^{ab}{}{} \,\bm{\tilde{\psi}} \,\bm{f}^{e} e_{e} \,\bm{\omega}_{cd} \varepsilon_{ab}{}^{cd}  -  \tfrac{9}{40}i \,\overline{\bm{\psi}}\,\Gamma^{ab}{}{} \,\bm{\tilde{\psi}} \,\bm{b} \,\bm{\omega}_{c}{}^{e} \,\bm{\omega}_{de} \varepsilon_{ab}{}^{cd}  \,+\nonumber\\
&+ \tfrac{3}{20}i \,\overline{\bm{\psi}}\,\Gamma^{ab}{}{} \,\bm{\tilde{\psi}} \,\bm{f}_{b} e_{c} \,\bm{\omega}_{de} \varepsilon_{a}{}^{cde}  + \tfrac{1}{20}i \,\overline{\bm{\psi}}\,\Gamma^{ab}{}{} \,\bm{\tilde{\psi}} \,\bm{f}_{c} e_{b} \,\bm{\omega}_{de} \varepsilon_{a}{}^{cde}  \,+\nonumber\\
&-  \tfrac{1}{20}i \,\overline{\bm{\psi}}\,\Gamma^{ab}{}{} \,\bm{\tilde{\psi}} \,\bm{\omega}_{bc} \,\bm{\omega}_{d}{}^{f} \,\bm{\omega}_{ef} \varepsilon_{a}{}^{cde}  + \tfrac{1}{40}i \,\overline{\bm{\psi}}\,\Gamma^{ab}{}{} \,\bm{\tilde{\psi}} \,\bm{b} \,\bm{\omega}_{bc} \,\bm{\omega}_{de} \varepsilon_{a}{}^{cde} \, +\nonumber\\
&+ 6 \,\overline{\bm{\psi}}\Gamma_5{}{} \,\bm{\tilde{\psi}} \,\bm{f}^{a} e_{a} \,\bm{b}  + 6 \,\overline{\bm{\psi}}\Gamma_5{}{} \,\bm{\tilde{\psi}} \,\bm{f}^{a} e^{b} \,\bm{\omega}_{ab}  -  \tfrac{1}{2} \,\overline{\bm{\psi}}\Gamma_5{}{} \,\bm{\tilde{\psi}} \,\bm{\omega}^{ab} \,\bm{\omega}_{a}{}^{c} \,\bm{\omega}_{bc}  \,+\nonumber\\
&+ \tfrac{3}{5} \,\overline{\bm{\psi}}\bm{\tilde{\psi}} \,\overline{\bm{\psi}}\,\Gamma^{ab}{}{}\,\bm{\tilde{\psi}}  \varepsilon_{abcd}  \,\bm{\omega}^{cd} \, -  \tfrac{14}{5}i \,\overline{\bm{\psi}}\,\Gamma_5{}{}\,\bm{\tilde{\psi}} \,\overline{\bm{\psi}} \,\Gamma^{ab}{}{} \,\bm{\tilde{\psi}} \,\bm{\omega}_{ab}   \,+\nonumber\\
&+ 3 \,\overline{\bm{\psi}} \,\Gamma^{a}{}{} \,\bm{\psi}\,\bm{f}_{a} \,\bm{a} \,\bm{b}  - 3 \,\overline{\bm{\psi}}\,\Gamma^{a}{}{} \,\bm{\psi} \,\bm{f}^{b} \,\bm{a} \,\bm{\omega}_{ab}  - 2 \,\overline{\bm{\psi}}\,\Gamma^{a}{}{} \,\bm{\psi}  \,\bm{f}_{b} \,\bm{f}_{c} e_{d} \varepsilon_{a}{}^{bcd} \,+\nonumber\\
&+ \tfrac{9}{20} \,\overline{\bm{\psi}}\,\Gamma^{a}{}{} \,\bm{\psi} \,\bm{f}_{b} \,\bm{\omega}_{c}{}^{e} \,\bm{\omega}_{de} \varepsilon_{a}{}^{bcd}  -  \tfrac{1}{2} \,\overline{\bm{\psi}}\,\Gamma^{a}{}{} \,\bm{\psi} \,\bm{f}_{b} \,\bm{b} \,\bm{\omega}_{cd} \varepsilon_{a}{}^{bcd}  -  \tfrac{21}{40} \,\overline{\bm{\psi}}\,\Gamma^{a}{}{} \,\bm{\psi} \,\bm{f}^{e} \,\bm{\omega}_{bc} \,\bm{\omega}_{de} \varepsilon_{a}{}^{bcd}  \,+\nonumber\\
&+ \tfrac{1}{40} \,\overline{\bm{\psi}} \,\Gamma^{a}{}{} \,\bm{\psi}\,\bm{f}_{b} \,\bm{\omega}_{ac} \,\bm{\omega}_{de} \varepsilon^{bcde}  -  \tfrac{28}{5} \,\overline{\bm{\psi}}\,\Gamma^{a}{}{} \,\bm{\psi}\,\overline{\bm{\psi}}  \Gamma_5{}{} \,\bm{\tilde{\psi}}\,\bm{f}_{a}   \,+\nonumber\\
&+ \tfrac{6}{5}i \,\overline{\bm{\psi}} \,\Gamma^{a}{}{}\,\bm{\psi}\,\overline{\bm{\psi}}\,\Gamma^{cd}{}{} \,\bm{\tilde{\psi}}\varepsilon_{abcd}   \,\bm{f}^{b}   -  \tfrac{3}{5} \,\overline{\bm{\tilde{\psi}}} \,\Gamma^{a}{}{}\,\bm{\tilde{\psi}}\,\overline{\bm{\psi}}\,\Gamma^{b}{}{}\,\bm{\psi}  \varepsilon_{abcd}   \,\bm{\omega}^{cd}  \,+\nonumber\\
&+ \tfrac{2}{5}i \,\overline{\bm{\tilde{\psi}}}\,\Gamma^{a}{}{}\Gamma_5{}{}\,\bm{\tilde{\psi}} \,\overline{\bm{\psi}}  \,\Gamma^{b}{}{}\,\bm{\psi}  \,\bm{\omega}_{ab}   + 2 \,\overline{\bm{\psi}} \,\bm{\tilde{\psi}}\,\overline{\bm{\psi}}  \,\bm{\tilde{\psi}} \,\bm{a} \,+\nonumber\\
&-  \,\overline{\bm{\psi}}\,\Gamma^{ab}{}{} \,\bm{\tilde{\psi}} \,\overline{\bm{\psi}} \,\Gamma_{ab}{}  \,\bm{\tilde{\psi}} \,\bm{a} + 18 \,\overline{\bm{\psi}}\,\Gamma_5{}{} \,\bm{\tilde{\psi}} \,\overline{\bm{\psi}} \Gamma_5{}{}  \,\bm{\tilde{\psi}} \,\bm{a} \,+\nonumber\\
&+ 2 \,\overline{\bm{\tilde{\psi}}}\,\Gamma_{a}{}\,\bm{\tilde{\psi}} \,\overline{\bm{\psi}}  \,\Gamma^{a}{}{}  \,\bm{\psi} \,\bm{a} -  \tfrac{4}{5} \,\overline{\bm{\tilde{\psi}}}\,\Gamma^{a}{}{} \Gamma_5{}{} \,\bm{\tilde{\psi}} \,\overline{\bm{\psi}}  \,\bm{\tilde{\psi}} e_{a} \,+\nonumber\\
&+ \tfrac{28}{5} \,\overline{\bm{\tilde{\psi}}}\,\Gamma^{a}{}{} \,\bm{\tilde{\psi}}\,\overline{\bm{\psi}}  \Gamma_5{}{} \,\bm{\tilde{\psi}}  e_{a} + \tfrac{6}{5}i \,\overline{\bm{\tilde{\psi}}}\,\Gamma^{a}{}{}\,\bm{\tilde{\psi}}  \,\overline{\bm{\psi}} \Gamma^{cd}{}{} \,\bm{\tilde{\psi}}\, e^{b}\,\varepsilon_{abcd}   \,+\nonumber\\
&-  \tfrac{4}{5} \,\overline{\bm{\tilde{\psi}}}\,\Gamma^{a}{}{}\Gamma_5{}{} \,\bm{\tilde{\psi}} \,\overline{\bm{\psi}}  \,\Gamma_{ab}{}  \,\bm{\tilde{\psi}} e^{b} + \tfrac{28}{5}i \,\overline{\bm{\psi}}\,\Gamma_5{}{} \,\bm{\tilde{\psi}} \,\overline{\bm{\psi}}  \,\bm{\tilde{\psi}} \,\bm{b} \,+\nonumber\\
&+ \tfrac{3}{10}   \,\overline{\bm{\psi}}\,\Gamma^{ab}{}{}\,\bm{\tilde{\psi}} \,\overline{\bm{\psi}} \,\Gamma^{cd}{}{}  \,\bm{\tilde{\psi}} \,\varepsilon_{abcd}\,\bm{b} + \tfrac{2}{5}i \,\overline{\bm{\tilde{\psi}}}\,\Gamma_{a}{}\Gamma_5{}{} \,\bm{\tilde{\psi}} \,\overline{\bm{\psi}}  \,\Gamma^{a}{}{} \,\bm{\psi}  \,\bm{b}.
\end{align}

\section{An equivalent anomaly cocycle}\label{sec:equivariantcocycle}


Anomalies are BRST equivalence classes.  In this section we want to describe the class of all the anomaly  representatives equivalent to the Chern-Simons  cocycle (\ref{superconfinvariantanomaly}) which can be obtained by adding  to it $\delta$-exact polynomials of generalized connections $\bm{A}^i$ and generalized curvatures $\bm{F}^i$.  We will also restrict ourselves to polynomials of generalized connections $\bm{A}^i$ and generalized curvatures $\bm{F}^i$ which are invariant under rigid Lorentz transformations. It turns out that the space of  $\delta$-trivial Lorentz invariant generalized polynomials of total degree 5 has dimension 29. There are therefore 29 gauge parameters that describe this class of anomaly cocycles equivalent to (\ref{superconfinvariantanomaly}).\footnote{Let us make clear that these are not all the possible equivalent representatives of the anomaly cocycle (\ref{superconfinvariantanomaly}). ``A priori''  one could also consider trivial cocycles  which are the $\delta$ variation of polynomials of ordinary connections and curvatures which cannot be written as the $\delta$ variation of polynomials of generalized connections and curvatures.}

It is easily seen that  the superconformal invariant cocycle (\ref{superconfinvariantanomaly}) is the {\it unique} anomaly  representative in this class which enjoys full {\it rigid} $\mathcal{N}=1$ superconformal invariance. Indeed any other superconformal invariant equivalent cocycle must be  the $\delta$-variation of a superinvariant cocycle of (generalized) degree 4. This cocycle of degree 4 would necessarily involve an even number of  generalized connections $\bm{A}^i$: but  there are no superconformal tensors (super)-antisymmetric with an even number of indices $i$. Hence there are no superconformal invariant representatives other than (\ref{superconfinvariantanomaly}). 

The superconformal invariant anomaly cocycle (\ref{superconfinvariantanomaly}) depends on  the ghosts $\{c^I\}$ of the superconformal algebra. Therefore  the corresponding anomalous Ward identities involve   all the  currents associated to  superconformal algebra generators $\{T_I\}$.   One can ask  if one can pick representatives which put to zero  anomalies relative to specific subalgebras of the superconformal gauge symmetry.   In the BRST formalism this is equivalent to choosing  anomaly representatives  which are independent of a subset of the ghosts $\{c^I\}$.

The superconformal invariant  anomalous cocycle  (\ref{superconfinvariantanomaly})  does not depend on  the diffeomorphism ghost $\xi^\mu$: as mentioned in the introduction this  reflects the fact that there are no diffeomorphism anomalies in 4-dimensions.  General arguments suggest that, for the same reason, it should be possible to choose an equivalent cocycle which does not contain the Lorentz ghosts $\Omega^{ab}$ \citep{Bardeen:1984pm}.  To our knowledge this has been formally proven only for bosonic theories. In the next subsection we therefore present  a general proof that one can  choose representatives in the same $\delta$-cohomology class as (\ref{superconfinvariantanomaly}) which are  both invariant under rigid Lorentz transformations and independent of the  Lorentz generalized connection $\bm{\omega}^{ab}$.  In the following, we will refer to such representatives  as   {\it Lorentz equivariant} cocycles.  The anomalous Ward identities associated to  Lorentz equivariant  representatives describe stress-energy tensors which are both conserved and symmetric.  

Requiring that the anomaly representative be  Lorentz equivariant  does not uniquely fix it. All  Lorentz-equivariant  representatives differ by  the $\delta$-variation of a degree 4 Lorentz-invariant  polynomial  of the $\bm{A}^i$'s and  the $\bm{F}^i$'s not involving $\bm{\omega}^{ab}$. It can be checked that this is a vector space of dimension 19. Hence there are 19 gauge parameters, out of the original 29, that one can choose  still preserving both local reparametrizations and local Lorentz symmetry.   

One can further fix these 19 gauge parameters by imposing  renormalization conditions on perturbative diagrams involving the (non)-conserved currents.  To efficiently describe these renormalization conditions it is useful to introduce the concept of {\it perturbative  degree} of a given monomial obtained by  expanding the generalized connections and curvatures of the anomaly polynomials into ordinary forms. The perturbative degree is  defined by assigning degree 1 to all ordinary field forms, with the exception of the vierbein form $e^a$ which is given  degree 0. Therefore generalized connections $\bm{A}^i= c^i + A^i$ other than the vierbein have perturbative degree 1.  Generalized ``horizontal''  non-vanishing curvatures  $\{ \bm{\tilde{F}^R}, \bm{\tilde{F}^W}, \,\bm{\rho}\}$  associated to physical connections  also have perturbative degree 1.
Non-horizontal generalized curvatures $\{\bm{\tilde{R}}^{ab}, \bm{\tilde{T}}^{a},\bm{\tilde\rho}\}$ have a component of perturbative degree 1,  i.e.  the ordinary curvatures  $\{\tilde{R}^{ab}, \tilde{T}^{a},\tilde\rho\}$,  and a component of perturbative degree 2,
i.e.  $(\lambda_0^J)^{ab}$, $ \lambda_0^S$, $(\lambda_0^K)^a$.

The usefulness of the concept of perturbative degree is the following. By expanding a generalized anomaly polynomial into ordinary forms one obtains monomials of perturbative degree 3, 4 and 5. Monomials of perturbative degree $n$ describe anomalous Feynman diagrams involving $n$ currents. 

For example we verified that the coefficients of the monomials of perturbative degree 3 describing ``triangular'' anomalies of  $U(1)_R$ and Weyl symmetries  involving two additional bosonic currents are independent of the 19 gauge parameters describing  Lorentz equivariant anomalies. However ``triangular'' $U(1)_R$ and Weyl  anomalies involving two  fermionic currents do depend on (some of) the 19 gauge parameters. Their specific values are renormalization prescription choices, compatible with local Lorentz symmetry.

We verified that one can choose the gauge parameters to  obtain anomaly representatives whose  ``triangular'' Q-supersymmetry anomalies vanish: this requires fixing 9 out of the 19 gauge parameters.  These are the anomaly representatives for which all the coefficients of the monomials of perturbative degree 3 involving the supersymmetry ghost $\zeta$ vanish: the corresponding (anomalous) Ward identities ensure that the diagrams involving  the divergence of the Q-supercurrent with two additional currents vanish.  

We checked that the remaining 10 parameters cannot be chosen to make all the coefficients of perturbative degree 4  monomials involving $\zeta$ vanish. Hence it is not possible to choose the anomaly representative in our class in such a way that the correlators of the divergence of the Q-supercurrent with three other  currents all vanish.\footnote{As we explained above, our class of equivalent anomalies is not the most general possible. We considered  all BRST trivial cocycles which can be written as the $\delta$ variation of {\it generalized}  connections and curvatures. It is a priori possible that by considering trivial terms which are the BRST variation of polynomials of {\it ordinary} connections and curvatures one could find other equivalent presentations of the same anomaly. In particular  our results do not  rule out  that by using these more general counterterms one could also make the  quartic and quintic Q anomaly vanish. It is also worth adding that our results  are also not in conflict with arguments based on the superspace formalism \cite{Kuzenko:2019vvi}, \cite{Bzowski:2020tue} affirming that there exists a choice of counterterms which makes the Q anomaly fully vanish. Indeed these works consider counterterms involving additional (auxiliary) fields beyond the ones  which we work with. It should be kept in mind however that the Q anomaly (non-)removability question is a perfectly well defined problem in our framework since the  BRST transformations close on our set of fields (vierbein, $U(1)_R$  gauge field and gravitino) without the need of auxiliary fields.  
}  
It is however  possible to choose 6 of the 10 gauge parameters to put to zero most of these monomials. We will present the corresponding form of the Q-anomaly of perturbative degree 4 in the next subsection.  With this choice of the representative one also puts to zero all triangular $U(1)_R$ anomalies involving two fermionic currents and all $U(1)_R$ anomalies of perturbative degree 4 (involving  3 extra currents). In this same gauge 
the triangular Weyl anomaly involving two supercurrents takes a particularly simple form.  
Furthermore, the remaining 4 gauge parameters do not affect the anomalies of perturbative degree 4: they could be fixed in principle by choosing renormalization conditions for the pentagon anomalous Feynman diagrams.

In the next subsection we present the generalized anomaly cocycle which satisfies all the renormalization conditions  we just stated, in which we fixed the last 4 parameters somewhat arbitrarily to maximize the vanishing monomials relevant for the pentagon anomalous Feynman diagrams.

The resulting generalized anomaly polynomial depends on the generalized connection $\bm{f}^{a} = \theta^a + f^a $ associated to special conformal transformations.   We will show in Subsection \ref{subsec:removingspecialconformal}   that one can further choose a representative in the same BRST class as  (\ref{superconfinvariantanomaly}) which 
is also independent of the ghost $\theta^a$ associated to special conformal transformations.  The reason is that the gauge connection $b$ and (a suitable completion of) the 1-form $ e_a\, \theta^a$  make up a so-called  BRST trivial doublet. Therefore one can add  a BRST exact term to the anomaly to eliminate both $b$ and $\theta^a$ from the anomaly cocycle: this is an example of a $\delta$-trivial term which cannot be written as the $\delta$-variation of a polynomial of generalized connections and curvatures.
 
 In conclusion there exists a family of anomaly  cocycles equivalent to (\ref{superconfinvariantanomaly})  independent of $\Omega^{ab}$ and $\theta^a$ which describes an effective action which is invariant under diffeomorphisms, local Lorentz transformations and local special conformal transformations. The  anomalous Ward identities associated to this cocycle encode the non-conservation  of the R-symmetry current ${\mathcal{J}}^\mu $ and of the supersymmetry current  ${\mathcal{S}}^\mu$, together with the non-vanishing of both the trace of the conserved stress-energy tensor, ${\mathcal{T}}_{\mu}{}^{\mu}$, and  the trace of the supercurrent, $\Gamma_\mu\,{\mathcal{S}}^\mu$.  This is the form in which the anomalies of superconformal gravity are usually presented \cite{Papadimitriou:2019gel}. 

\subsection{Removing  the Lorentz anomaly}
We want to investigate if there exists a generalized form $X_4$ of degree 4 such that the  cocycle 
\begin{equation}
    \tilde{Q}_{5}(\bm{A}, \bm{F})={Q}_{5}(\bm{A}, \bm{F}) + \delta\,  X_4,
\end{equation}
equivalent to the Chern-Simons superconformal invariant anomaly cocycle (\ref{superconfinvariantanomaly}), does not depend on the generalized Lorentz connection $\bm{\omega}^{ab}$.  One expects such a representative to exist because it is generally understood that Lorentz anomalies are equivalent to diffeomorphism anomalies: since there are no diffeomorphism anomalies in 4-dimensions the Lorentz anomaly should be removable \cite{Bardeen:1984pm}.  However we are not aware of a constructive  proof of existence of such a cocycle in the general superconformal context we are considering. Hence in the following we describe how to explicitly construct a Lorentz equivariant anomaly cocycle.

It is useful to  introduce a set of {\it commuting}  and {\it constant} ghosts $\bm{\kappa}^{ab}$  of degree +2  and  the  ``topological'' nilpotent operator $\partial_{\bm{\omega}}$ which shifts $\bm{\omega}^{ab}$ 
\begin{equation}
\partial_{\bm{\omega}}\, \bm{\omega}^{ab} = \bm{\kappa}^{ab},\qquad \partial_{\bm{\omega}}\, \bm{\kappa}^{ab} =0, \qquad \partial_{\bm{\omega}}^2=0. 
\end{equation}
The action of $\partial_{\bm{\omega}}$\; on all other fields is taken to be trivial.   The anti-commutator of $\delta$ and $\partial_{\bm{\omega}}$ is (minus) a (rigid) Lorentz transformation $\mathcal{\delta}^{Lorentz}_{\bm\kappa}$ with commuting parameter  $\bm{\kappa}^{ab}$:
\begin{equation}\label{deltadeltaL}
-\mathcal{\delta}^{Lorentz}_{\bm\kappa} = \{ \delta,\partial_{\bm{\omega}}\}.
\end{equation}
A Lorentz equivariant representative $\tilde{Q}_{5}$ of the ${Q}_{5}$ class is therefore a Lorentz-invariant cocycle satisfying
\begin{equation}\label{Q5equivariant}
\delta\, \tilde{Q}_{5}=   \partial_{\bm{\omega}}\,\tilde{Q}_{5} =\mathcal{\delta}^{Lorentz}_{\bm\kappa}\,\tilde{Q}_{5}=0.
\end{equation}

To solve (\ref{Q5equivariant}) it is convenient to introduce a filtration for $\delta$ on the space of polynomials in $\bm{A}$ and $\bm{F}$.  Let 
\begin{equation}\label{secequiv:Ndegree}
N\equiv N_ {\bm{A}} +  N_{\bm{F}}
\end{equation}
be the total  degree of a monomial $ \bm{A}^ {N_ {\bm{A}}}\,  \bm{F}^ {N_ {\bm{F}}}$. We can then decompose $\delta$
\begin{equation}
\delta \equiv \delta_0 + \delta_1
\end{equation}
as the sum of $\delta_0$  which commutes with  $N$ 
\begin{equation}
\delta_0\, \bm{A} = \bm{F}, \qquad \delta_0\, \bm{F} = 0,
\end{equation}
while $\delta_1$ increases $N$ by 1
\begin{align}
&\delta_1 \, \bm{A}  = - \bm{A}^2, \qquad \delta_1 \, \bm{F}  = - [\bm{A}, \bm{F}].
\end{align}
Let us also define the  operator $i_0$, which commutes with  $N$ 
\begin{equation}\label{trivialdelta0}
i_0\, \bm{A} = 0,  \qquad i_0\, \bm{F} =\bm{A}.
\end{equation}
It is immediate to verify that 
\begin{equation}
  N \equiv N_ {\bm{A}} +  N_{\bm{F}} = \{ \delta_0, i_0\}.
   \end{equation}
   Both $\delta_0$ and $i_0$ are nilpotent
   \begin{equation}
   \delta_0^2 = i_0^2 =0.
   \end{equation}
   Moreover
   \begin{equation}
  \{\delta_0, \delta_1\}=0, \qquad  [\delta_0, N]=0. 
  \end{equation}
The operator $l_1$
  \begin{equation}
  l_1 \equiv \{\delta_1, i_0\}
  \end{equation}
 increases the number of fields $N$ by 1 and acts trivially on connections
  \begin{equation}\label{l1action}
  l_1\, \bm{F} = \bm{A}^2, \qquad l_1\, \bm{A} =0.
  \end{equation}
Any polynomial $Q_5$ of total degree 5  can therefore be decomposed in the sum of  polynomials $Q_{5; N}$ of fixed degree $N$:
\begin{equation}
Q_5 = Q_{5; 3} + Q_{5; 4} + Q_{5; 5}.
\end{equation}
Evidently $Q_{5; N}$ contains  $5- N$ curvatures:
\begin{equation}
Q_{5; 3}  \sim \bm{A}\, \bm{F}\, \bm{F}, \qquad Q_{5; 4}  \sim \bm{A}\, \bm{A}\,\bm{A} \,\bm{F}, \qquad  Q_{5; 5}  \sim \bm{A}\, \bm{A}\,\bm{A} \, \bm{A}\, \bm{A}.
\end{equation}
$Q_5$ is a $\delta$-cocycle if and only if
\begin{align}\label{delta0delta1eqs}
&\delta_0\, Q_{5; 3} =0,\nonumber\\
&\delta_0\, Q_{5; 4}+\delta_1\, Q_{5; 3}  =0,\nonumber\\
&\delta_0\, Q_{5; 5}+\delta_1\, Q_{5; 4}  =0, \nonumber\\
&\delta_1\, Q_{5; 5}  =0.
\end{align}
Moreover two $\delta$-cocycles $\tilde{Q}_5$ and $Q_5$ are equivalent if and only if
\begin{align}
& \tilde{Q}_{5; 3} =Q_{5; 3}+ \delta_0 X_{4,3}, \nonumber\\
& \tilde{Q}_{5; 4} =Q_{5; 4}+ \delta_1 X_{4,3}+ \delta_0 X_{4,4}, \nonumber\\
& \tilde{Q}_{5; 5} =Q_{5; 5}+\delta_1 X_{4,4}.
\end{align}
Relation (\ref{trivialdelta0}) ensures that any $\delta_0$-closed monomial $Q_N$ with $N\not=0$ is $\delta_0$-exact:
\begin{equation}
N\,  Q_{N} = \{\delta_0, i_0\}\, Q_{N}  = \delta_0\, (i_0\, Q_N) \Rightarrow Q_{N} =\delta_0\, \left(\tfrac{1}{N}\,  i_0\, Q_N\right).
\end{equation}
Hence, given any $\delta_0$-closed polynomial
\begin{equation}
\delta_0\, Q_{5; 3} =0,
\end{equation}
we can extend it to a  $\delta$-cocycle by means of the formulae \footnote{$\delta_1\, Q_{5,5}=0$ thanks to (\ref{l1action}) and the fact that $Q_{5,5}$ does not contain curvatures.}
\begin{align}
& Q_{5; 4} = -\tfrac{1}{4}\, i_0\, \delta_1\, (Q_{5; 3}  ),\\
& Q_{5; 5} = -\tfrac{1}{5}\, i_0\, \delta_1\, (Q_{5; 4}  ).
\end{align}
Let us therefore start from the cubic polynomial $ Q_{5; 3}$ in (\ref{sec4: CS1})  associated to the superconformal invariant $Q_5\,(\bm{A}, \bm{F})$ (\ref{superconfinvariantanomaly}).  This polynomial does not include any pure ``Lorentz'' anomaly (in agreement with the fact that there is no 3-index totally  symmetric $SO(4)$ invariant tensor), which would have  the form
\begin{equation}
\bm{\omega}\, \bm{\tilde{R}}\, \bm{\tilde{R}}.
\end{equation}
Hence the terms in $ Q_{5; 3}$ proportional to  $\bm{\omega}$ contain at least one curvature  other than the Lorentz curvature  $\bm{\tilde{R}}$:
\begin{equation}\label{mixedLorentz}
 Q_{5; 3}  \sim \bm{\omega}\, \bm{\tilde{R}}\, \bm{F}^\prime, \quad \bm{\omega}\, \bm{F}^\prime\, \bm{F}^{\prime\prime},
\end{equation}
where $\bm{F}^\prime$ and $\bm{F}^{\prime\prime}$ denote generic curvatures associated to generators different from Lorentz.  Since
\begin{equation}
\bm{F}^\prime = \delta_0 \, \bm{A}^\prime
\end{equation}
and
\begin{equation}
\delta_0 \, \bm{\omega}^{ab} = \bm{\tilde{R}}^{ab},
\end{equation}
one can move, by adding $\delta_0$-exact terms,  the $\delta_0$ from  $\bm{F}^\prime=\delta_0 \,\bm{A}^\prime $ to hit the Lorentz connection and produce $\bm{\tilde{R}}^{ab}$.  Hence one can add to $ Q_{5; 3}$ a $\delta_0$-trivial term which  eliminates the $\bm{\omega}^{ab}$ dependence. Explicitly,  by choosing
\begin{align}
X_{4,3} =&+  \bm{\omega}{}_{ab}  \,\bar{\bm{\rho}}\,\Gamma^{ab}{}\,\Gamma_{5}{}\, \bm{\tilde\psi}{} -  \bm{\omega}{}_{ab}\,\overline{\tilde{\bm\rho}}\, \Gamma_{5}{} \Gamma^{ab}  \bm{\psi} \;+ \nonumber\\
&- \bm{\omega}{}_{ab} \,\bm{a}{}\,\bm{R}^{ab} - \tfrac{1}{2} \epsilon_{ab}{}^{cd}\, \bm{\omega}{}_{cd} \,\bm{b}{}\,  \bm{R}^{ab} - \epsilon_{a}{}^{bcd}\,  \bm{\omega}{}_{cd}\,\bm{f}{}_{b}\, \bm{T}^{a} -  \epsilon_{a}{}^{bcd}\, e_{b}\, \bm{\omega}{}_{cd}\bm{\tilde{T}}^{a},
\end{align}
one produces a trilinear $\delta_0$-cocycle $\tilde{Q}_{5,3}$ equivalent to the superconformal invariant $ Q_{5; 3}$ in (\ref{sec4: CS1}):
\begin{align} \label{Qtilde53equivariant}
\tilde{Q}_{5,3} \,=&\,  Q_{5; 3}+ \delta_0\, X_{4,3} 
\end{align}
which does not depend on the Lorentz generalized connection $\bm{\omega}$:
\begin{equation}
\partial_{\bm{\omega}} \, \tilde{Q}_{5,3}=0.
\end{equation}
 The task is now to show that there exists a $\delta$-closed  extension of $\tilde{Q}_{5,3}$ which is also Lorentz-equivariant. 
One starts by considering the quartic  extension of $\tilde{Q}_{5,3}$
\begin{equation}\label{Q54i0}
 \hat{Q}_{5; 4} = -\tfrac{1}{4}\, i_0\, (\delta_1\, \tilde{Q}_{5; 3}),
\end{equation}
  which satisfies the second of the equations (\ref{delta0delta1eqs}):
  \begin{equation}\label{Q54Qtilde53}
  \delta_0 \,  \hat{Q}_{5; 4} + \delta_1\, \tilde{Q}_{5; 3}  =0.
  \end{equation}
From (\ref{deltadeltaL}) we deduce
\begin{equation}
\{\delta_0, \partial_{\bm{\omega}}\}=0,\quad -\delta^{Lorentz}_{\bm{\kappa}} = \{\delta_1, \partial_{\bm{\omega}}\}.
\end{equation}
We can also introduce the bosonic operator $\partial_{\bm{\tilde{R}}}$ which shifts the Lorentz curvature by $\bm{\kappa}^{ab}$
\begin{equation}
\partial_{\bm{\tilde{R}}} \bm{\tilde{R}}^{ab}\equiv\bm{\kappa}^{ab},\qquad\partial_{\bm{\tilde{R}}} = [\partial_{\bm{\omega}}  , i_0], \qquad  [\partial_{\bm{\tilde{R}}}, \delta_0] = \partial_{\bm{\omega}}. 
\end{equation}
Since  $\tilde{Q}_{5; 3}$ is Lorentz-invariant,  $\delta_1\, \tilde{Q}_{5; 3} $  does not depend on $\bm{\omega}$:
\begin{equation}
 \partial_{\bm{\omega}}\, \delta_1\, (\tilde{Q}_{5; 3}  )= - \delta^{Lorentz}_{\bm{\kappa}} (\tilde{Q}_{5; 3}  )=0.
 \end{equation}
 The action of $i_0$ on $\delta_1 (\tilde{Q}_{5; 3} )$  in (\ref{Q54i0})  may introduce  an $\bm{\omega}$-dependence which is at most {\it linear}:
\begin{align}
 & \hat{Q}_{5; 4}  = \bm{\omega}^{ab}\, V_{ab}(\bm{A}^\prime, \bm{F}) +  \mathring{Q}_{5; 4} (\bm{A}^\prime, \bm{F})=\nonumber\\
&\qquad = \bm{\omega}^{ab}\, N_{ab;cd}(\bm{A}^\prime) \bm{\tilde{R}}^{cd}+\bm{\omega}^{ab}\, V^\prime_{ab}(\bm{A}^\prime, \bm{F}^\prime)   + Z_{ab}(\bm{A}^\prime)\,\bm{\tilde{R}}^{ab} +  Q^\prime_{5; 4} (\bm{A}^\prime, \bm{F}^\prime),
\end{align}
where $\bm{A}^\prime$ and $\bm{F}^\prime$ denote connections and curvatures  different from  Lorentz and where we used the fact that $\hat{Q}_{5; 4} $ is linear in  the generalized  curvatures.

From (\ref{Q54Qtilde53}) we obtain that 
 \begin{equation}
 \delta_0 \,  (\partial_{\bm{\omega}}\, \hat{Q}_{5; 4}) = \delta_0 \,  (\bm{\kappa}^{ab}\, V_{ab}(\bm{A}^\prime, \bm{F})) =0.
 \end{equation}
Therefore, since the  polynomial $V_{ab}(\bm{A}^\prime, \bm{F}) $ has degree $N=3$, we have
 \begin{align}
&V_{ab}(\bm{A}^\prime, \bm{F}) = N_{ab;cd}(\bm{A}^\prime) \bm{\tilde{R}}^{cd}+ V^\prime_{ab}(\bm{A}^\prime, \bm{F}^\prime) =\tfrac{1}{3}\, \delta_0 ( i_0\, (  V_{ab}(\bm{A}^\prime, \bm{F}) ) =\nonumber\\
&\qquad= \tfrac{1}{3}\, \delta_0 (\{ i_0,  \partial_{\bm{\omega}^{ab}}\}\,  \hat{Q}_{5; 4} ) =\tfrac{1}{3}\, \delta_0 (\partial_{\bm{\tilde{R}}^{ab}}\,  \hat{Q}_{5; 4} )=\nonumber\\
&\qquad = \tfrac{1}{3}\, \delta_0 (\bm{\omega}^{cd}\, N_{cd;ab}(\bm{A}^\prime) + Z_{ab}(\bm{A}^\prime) )=\nonumber\\
&\qquad =\tfrac{1}{3}\, \bm{\tilde{R}}^{cd}\, N_{cd;ab}(\bm{A}^\prime) - \tfrac{1}{3}\,\bm{\omega}^{cd}\, \, \delta_0\,  N_{cd;ab}(\bm{A}^\prime)+ \tfrac 1 3\, \delta_0\, Z_{ab}(\bm{A}^\prime).
 \end{align}
We conclude that
\begin{equation}
N_{cd;ab}(\bm{A}^\prime)=0,\qquad V^\prime_{ab}(\bm{A}^\prime, \bm{F}^\prime)=  \tfrac 1 3\, \delta_0\, Z_{ab}(\bm{A}^\prime),
\end{equation}
and hence
\begin{align}
&  \hat{Q}_{5; 4}  =  \tfrac{1}{3}\, \bm{\omega}^{ab}\, \delta_0\, Z_{ab}(\bm{A}^\prime)  +  Z_{ab}(\bm{A}^\prime)\,\bm{\tilde{R}}^{ab} +  Q^\prime_{5; 4} (\bm{A}^\prime, \bm{F}^\prime) = \nonumber\\
  & \qquad = - \tfrac{1}{3}\, \delta_0\,\bigl(\bm{\omega}^{ab}\, Z_{ab}(\bm{A}^\prime)) + \tfrac{4}{3}\,   Z_{ab}(\bm{A}^\prime)\,\bm{\tilde{R}}^{ab} +  Q^\prime_{5; 4} (\bm{A}^\prime, \bm{F}^\prime).
\end{align}
We can therefore pick
\begin{align}
\tilde{Q}_{5; 4} =&\, \tfrac{4}{3}\,   Z_{ab}(\bm{A}^\prime)\,\bm{\tilde{R}}^{ab} +  Q^\prime_{5; 4} (\bm{A}^\prime, \bm{F}^\prime)
\end{align}
as the quartic extension of $\tilde{Q}_{5; 4} $ which is both equivalent to  $\hat{Q}_{5; 4}$ and independent of $\bm{\omega}$.\\
The quintic extension
\begin{align}
\tilde{Q}_{5; 5} =&\, -\tfrac{1}{5}\, i_0\, \delta_1\, \tilde{Q}_{5; 4} 
\end{align}
is now also independent of $\bm{\omega}$. Indeed, $\partial_{\bm{\omega}}\, \tilde{Q}_{5; 5}$ is both $\delta_0$-closed and $i_0$-closed, and has $N=5$. Hence,
\begin{equation}
\partial_{\bm{\omega}}\, \tilde{Q}_{5; 5}= \tfrac{1}{5}\, \delta_0\, i_0\, \partial_{\bm{\omega}}\, \tilde{Q}_{5; 5}=  \tfrac{1}{5}\, \delta_0\, \{i_0 , \partial_{\bm{\omega}}\}\, \tilde{Q}_{5; 5} = \tfrac{1}{5}\, \delta_0\,  \partial_{\bm{\tilde{R}}}\, \tilde{Q}_{5; 5}=0,
\end{equation}
since $\tilde{Q}_{5; 5}$ contains no curvatures. 

Summarizing, the $\delta$-cocycle 
\begin{equation}
\tilde{Q}_5 = \tilde{Q}_{5,3} + \tilde{Q}_{5,4} + \tilde{Q}_{5,5}
\end{equation}
is both equivalent to the superconformal invariant $Q_5(\bm{A}, \bm{F})$ and Lorentz-equivariant. 
  
\subsection{Removing the special conformal anomaly}
\label{subsec:removingspecialconformal}
The BRST rules for the Weyl connection $b_\mu$ are
\begin{equation}
 \hat{s}\,b_\mu =\partial_\mu\,\sigma+2\,e_\mu{}^a\,\theta_a+2\,i\,(\overline{\psi}_\mu\,\eta +\overline{\zeta}\,\Tilde{\psi}_\mu).
\end{equation}
If we define a new ghost $\Tilde\theta_\mu$ associated to  special conformal transformations:
\begin{equation}\label{thetatildedef}
 \Tilde\theta_\mu\equiv \partial_\mu\,\sigma+2\,e_\mu{}^a\,\theta_a+2\,i\,(\overline{\psi}_\mu\,\eta+\overline{\zeta}\,\Tilde{\psi}_\mu), 
 \end{equation}
 the $(b_\mu,  \Tilde\theta_\mu)$ are, by construction, a trivial BRST doublet 
 \begin{equation}
 \hat{s}\, b_\mu =  \Tilde\theta_\mu,\qquad \hat{s}\, \Tilde\theta_\mu = \mathcal{L}_\gamma\, b_\mu.
 \end{equation}
 Relation   (\ref{thetatildedef}) can be inverted to express the original ghost  $\theta_a$  in terms of the new $\Tilde\theta_\mu$
\begin{equation}
\theta_a = \tfrac{1}{2}\, e^\mu{}_a\, \bigl[\Tilde\theta_\mu- \partial_\mu\,\sigma-2\,i\,(\overline{\psi}_\mu\,\eta+\overline{\zeta}\,\Tilde{\psi}_\mu\bigr)\bigr].
\end{equation}
The only ghost whose transformation rules contain $\Tilde\theta$ is the special supersymmetry ghost  $\eta$:
\begin{align}
 &\hat{s}\,\eta= i_\gamma(\Tilde{\psi})-\left(\tfrac{1}{4}\,\Omega^{ab}\,\Gamma_{ab}-\tfrac{1}{2}\,\sigma+\tfrac{3}{2}\,i\,\alpha\,\Gamma_5\right)\,\eta
 -   \bigl[\tfrac{i}{2}\, \partial_\mu\,\sigma+(\overline{\psi}_\mu\,\eta+\overline{\zeta}\,\Tilde{\psi}_\mu)\bigr]\,\Gamma^\mu\,\zeta+
  \tfrac{i}{2}\, \Tilde\theta_\mu\,\Gamma^\mu\,\zeta.
 \end{align}
If we redefine the special supersymmetry ghost as
 \begin{equation}
\tilde{\eta} =  \eta - \tfrac{i}{2}\, b_\mu\,\Gamma^\mu \,\zeta,
\end{equation}
 then  both $b$ and $\Tilde\theta$ disappear from the BRST transformation rules of all the other fields. 
 
 The implication of this is that if we put to zero $\Tilde\theta$ and $b$ in the Lorentz-equivariant cocycle we obtain an anomaly cocycle
 \begin{equation}\label{sec4:Qequiv5final}
Q^{equiv}_5 \equiv \tilde{Q}_5 \Big|_{b\to0\;;\Tilde\theta\to0}= \tilde{Q}_5 \Big|_{b\to0\;;\;\theta^a\to  -\frac{1}{2}\, e^{a\mu}\, \bigl[\partial_\mu\sigma+ 2\,i\,(\overline{\psi}_\mu\,\eta+\overline{\zeta}\Tilde{\psi}_\mu)\bigr]},
\end{equation}
which is equivalent to the Chern-Simons cocycle  $Q_5(\bm{A}, \bm{F})$, is Lorentz-equivariant and  does not contain either  $b$ or $\theta$. Note that for $b=0$, $\tilde{\eta}= \eta$.
\subsection{A special Lorentz equivariant cocycle}
In this section we present the Lorentz equivariant cocycle which satisfies the renormalization prescriptions we described
at the beginning of this Section.  To summarize:  

a) This cocycle preserves local Lorentz symmetry.  This fixes 10 parameters out of the original 29. 

b) This cocycle leads to zero  Q-supercurrent anomaly at perturbative degree 3, i.e.  correlators involving the divergence of the supercurrent with  two extra currents vanish.  This fixes 9 more gauge parameters. 

c) This cocycle has  no fermionic cubic corrections and no quartic corrections to the R-anomaly. It has a single contribution to the cubic conformal supersymmetry S-anomaly, involving, beyond the trace of the supercurrent, another supercurrent and the R-current.

Properties a), b) and c) still leave 4 free gauge parameters, which  affect  only anomaly terms of  perturbative degree 5. 

We divide the cocycle in terms cubic, quartic and quintic in the filtration degree $N$ in equation (\ref{secequiv:Ndegree}) which counts the number of generalized forms:\footnote{Let us remind, to avoid confusions,  that the degree $N$ is not the same as the perturbative degree. The reason is  that perturbative grading assigns degree 0 to the vierbein;  degree 1 to the other connections,  to the ghosts and to the ghost number 0  component of the generalized curvatures; degree 2 to the ghost number 1 components of the non-horizontal generalized curvatures.}
\begin{align} \label{eq:specialLorentzequiv3}
\tilde{Q}_{5; 3}=&\,12i \,\bar{\bm{\rho}} \,\Gamma^{a}{} \,\Gamma_5{} \,\bm{f}_{a} \,\bm{\rho} - 60i \bar{\bm{\rho}} \,\bm{\tilde{\psi}} \,\bm{\tilde{F}^{R}}{} + 15 \,\bm{a}{} \,(\bm{\tilde{F}^{R}}{})^2 - 12 \,\bar{\bm{\rho}} \,\Gamma_5{} \,\bm{\tilde{\psi}} \,\bm{\tilde{F}^{W}}{} \,+\nn\\
&+ 3 \,\bm{a}{} \,(\bm{\tilde{F}^{W}}{})^2 + 6 \bar{\bm{\rho}} \,\Gamma^{ab}{} \,\Gamma_5{} \,\bm{\tilde{\psi}} \,\bm{\tilde{R}}_{ab} -  \tfrac{3}{2} \,\bm{a}{} \,\bm{\tilde{R}}_{ab} \,\bm{\tilde{R}}^{ab} -  \tfrac{3}{4} \,\varepsilon_{abcd} \,\bm{b} \,\bm{\tilde{R}}^{ab} \,\bm{\tilde{R}}^{cd} \,+\nn\\
&+ 12i \,\overline{\bm{\tilde{\bm{\rho}}}} \,\Gamma^{a}{} \,\Gamma_5{} \,\bm{\tilde{\psi}} \,\bm{T}_{a} + 6 \,\varepsilon_{abcd} \,\bm{f}^{a} \,\bm{\tilde{R}}^{cd} \,\bm{T}^{b} + 12 \,\bm{a}{} \,\bm{T}^{a} \tilde{\bm{T}}_{a} \,,
\end{align}
\begin{align}
\tilde{Q}_{5; 4}=&-24 \,\overline{\bm{\tilde{\psi}}} \,\Gamma^{ab}{} \,\Gamma_5{}\,\bm{\rho} \,\bm{f}_{a} e_{b}  + 24 \,\overline{\bm{\tilde{\psi}}} \,\Gamma_5{} \,\bm{\rho}\,\bm{f}^{a} e_{a}  + 24 \bar{\bm{\rho}}  \,\Gamma^{a}{} \,\bm{\psi}\,\bm{f}_{a} \,\bm{a}{} + 12i \,\overline{\bm{\tilde{\psi}}} \,\Gamma^{a}{} \,\bm{\tilde{\psi}} \bar{\bm{\rho}} \,\Gamma_{a} \,\Gamma_5{} \,\bm{\psi} \,+\nn\\
&- 60 \,\overline{\bm{\tilde{\psi}}} \,\Gamma_5{} \,\bm{\psi} \,\bm{a}{} \,\bm{\tilde{F}^{R}}{} \,+ 30 \,\overline{\bm{\tilde{\psi}}} \,\Gamma^{a}{} \,\bm{\tilde{\psi}} \,\bm{\tilde{F}^{R}}{} e_{a} - 12i \,\overline{\bm{\tilde{\psi}}} \,\bm{\psi} \,\bm{a}{} \,\bm{\tilde{F}^{W}}{} - 6i \,\overline{\bm{\tilde{\psi}}} \,\Gamma^{ab}{} \,\bm{\psi} \,\bm{a}{} \,\bm{\tilde{R}}_{ab} +\nn\\
&- 6 \,\overline{\bm{\tilde{\psi}}} \,\Gamma^{ab}{} \,\Gamma_5{} \,\bm{\psi} \,\bm{b} \,\bm{\tilde{R}}_{ab} \,+ 6 \,\bm{f}_{c} e_{d} \,\bm{b} \,\varepsilon_{ab}{}^{cd} \,\bm{\tilde{R}}^{ab} + 3 \,\overline{\bm{\tilde{\psi}}} \,\varepsilon_{abcd} \,\Gamma^{a}{} \,\bm{\tilde{\psi}}\, e^{b} \,\bm{\tilde{R}}^{cd} \,+\nn\\
&+ 12i \,\overline{\bm{\tilde{\psi}}} \,\Gamma^{ab}{} \,\Gamma_5{} \,\bm{\tilde{\bm{\rho}}} \,\overline{\bm{\psi}} \,\Gamma_{ab} \,\bm{\psi} + 12 \,\overline{\bm{\tilde{\psi}}} \,\Gamma_5{}  \,\bm{\psi} \,\bm{f}^{a} \,\bm{T}_{a} + 12 \,\overline{\bm{\tilde{\psi}}} \,\Gamma^{a}{} \,\bm{\tilde{\psi}} \,\bm{a}{} \,\bm{T}_{a} \,+\nn\\
&+ 12 \,\bm{f}_{b} \,\bm{f}_{c} \,e_{d} \,\varepsilon_{a}{}^{bcd} \,\bm{T}^{a} + 12 \,\overline{\bm{\tilde{\psi}}} \,\Gamma_{ab}{} \,\Gamma_5{}  \,\bm{\psi}\,\bm{f}^{a} \,\bm{T}^{b} \,,
\end{align}
\begin{align} \label{eq:specialLorentzequiv5}
\tilde{Q}_{5; 5}=&-12 \,\bm{f}_{a} \,\bm{f}_{b} \,\bm{b} \, e_{c} \,e_{d}  \,\varepsilon^{abcd} - 12 \,\overline{\bm{\tilde{\psi}}}\,\Gamma^{a}{} \,\bm{\tilde{\psi}} \,\bm{f}_{b}\, e_{c}\, e_{d} \,\varepsilon_{a}{}^{bcd}  - 24i \,\overline{\bm{\psi}} \,\Gamma^{ab}{} \,\bm{\tilde{\psi}}\,\bm{f}_{a} \,\bm{a}{} \,e_{b}  \,+\nn\\
&+ 12i \,\overline{\bm{\psi}} \,\Gamma^{ab}{} \,\bm{\tilde{\psi}}\,\bm{b}\,\bm{f}_{c} \,e_{d}  \,\varepsilon_{ab}{}^{cd}  - 24i \,\overline{\bm{\psi}} \,\bm{\tilde{\psi}}\,\bm{f}^{a} \,\bm{a}{}\, e_{a}  \,+\nn\\
&-  \tfrac{48}{5} \,\overline{\bm{\psi}} \,\Gamma_5{}  \,\bm{\tilde{\psi}} \,\overline{\bm{\psi}} \,\Gamma^{a}{} \,\bm{\psi}\,\bm{f}_{a} + \tfrac{24}{5}i \,\overline{\bm{\psi}} \,\Gamma^{cd}{}  \,\bm{\tilde{\psi}} \,\overline{\bm{\psi}} \,\varepsilon_{abcd} \,\Gamma^{a}{} \,\bm{\psi} \,\bm{f}^{b}\,+\nn\\
&-  \tfrac{36}{5}i \,\overline{\bm{\psi}} \,\Gamma^{cd}{} \,\bm{\psi} \,\overline{\bm{\psi}} \,\varepsilon_{abcd} \,\Gamma^{a}{}  \,\bm{\tilde{\psi}}\,\bm{f}^{b} -  \tfrac{72}{5} \,\overline{\bm{\psi}} \,\Gamma_{ab}{} \,\bm{\psi} \,\overline{\bm{\psi}} \,\Gamma^{a}{} \,\Gamma_5{}  \,\bm{\tilde{\psi}} \,\bm{f}^{b}\,+\nn\\
&+ 18 \,\overline{\bm{\tilde{\psi}}} \,\Gamma^{a}{} \,\bm{\tilde{\psi}} \,\overline{\bm{\psi}} \,\Gamma_{a}{} \,\bm{\psi} \,\bm{a}{} + 3 \,\overline{\bm{\tilde{\psi}}} \,\Gamma^{ab}{} \,\bm{\tilde{\psi}} \,\overline{\bm{\psi}} \,\Gamma_{ab}{} \,\bm{\psi} \,\bm{a}{} \,+\nn\\
&+ \tfrac{96}{5} \,\overline{\bm{\tilde{\psi}}} \,\Gamma^{a}{} \,\bm{\tilde{\psi}} \,\overline{\bm{\psi}} \,\Gamma_5{} \,\bm{\tilde{\psi}} \,e_{a} + \tfrac{24}{5}i \,\overline{\bm{\tilde{\psi}}} \,\Gamma^{cd}{} \,\bm{\tilde{\psi}} \,\overline{\bm{\psi}} \,\varepsilon_{abcd} \,\Gamma^{a}{} \,\bm{\tilde{\psi}} \,e^{b} \,+\nn\\
&-  \tfrac{48}{5} \,\overline{\bm{\tilde{\psi}}} \,\Gamma_{ab}{} \,\bm{\tilde{\psi}} \,\overline{\bm{\psi}} \,\Gamma^{a}{} \,\Gamma_5{} \,\bm{\tilde{\psi}}\, e^{b} -  \tfrac{3}{2} \,\overline{\bm{\tilde{\psi}}} \,\varepsilon_{abcd} \,\Gamma^{ab}{} \,\bm{\tilde{\psi}} \,\overline{\bm{\psi}} \,\Gamma^{cd}{} \,\bm{\psi} \,\bm{b} \,.
\end{align}

In appendix \ref{appendix:lorentzcocycle} we present the ghost number 1 component of this cocycle ---  the anomaly  proper ---  written in terms of ordinary forms. 


\subsection{Bosonic anomalies}
Let us consider the quantum effective action,  a (non-local) functional of the independent fields of superconformal gravity:
\begin{equation}
\mathscr{W}[e,b,a,\psi] = -i\,\log{Z[e,b,a,\psi]}.
\end{equation}
We can define the corresponding currents
\begin{equation}
{\mathcal{T}}_a{}^\mu = e^{-1}\,\tfrac{\delta \mathscr{W}}{\delta e_\mu{}^a},\quad
{\mathcal{B}}^\mu = e^{-1}\,\tfrac{\delta \mathscr{W}}{\delta b_\mu},\quad
{\mathcal{J}}^\mu = e^{-1}\,\tfrac{\delta \mathscr{W}}{\delta a_\mu},\quad
{\mathcal{S}}^\mu = e^{-1}\,\tfrac{\delta \mathscr{W}}{\delta \bar{\psi}_\mu}.
\end{equation}
In the renormalization scheme in which the anomaly is described by the Lorentz-equivariant, $\theta$-independent cocycle (\ref{sec4:Qequiv5final}) we have 
\begin{align} \label{eq:mathcal_A_defs}
& s\, \mathscr{W}[e,b,a,\psi] = \int_{M_4}\,Q^{equiv}_5 =\nonumber\\
& \qquad \equiv \int\diff^4x\,e\,\bigl[\sigma\,\mathcal{A}_W + \alpha\,\mathcal{A}_R  + \bar{\zeta}\,\mathcal{A}_Q + \bar{\tilde\eta}\,\mathcal{A}_S\bigr],
\end{align}
which is equivalent to the Ward identities 
\begin{equations}\label{sec5:wardidentities}
& 0= \tfrac{1}{2}\,{\mathcal{T}}_{[ab]} - \tfrac{1}{4}\,\bar{\psi}_\mu\,\Gamma_{ab}\,{\mathcal{S}}^\mu, \\
 &0= \mathcal{B}_\mu, \\
&\mathcal{A}_W = - {\mathcal{T}}_\mu{}^\mu - \tfrac{1}{2}\,\bar{\psi}_\mu\,{\mathcal{S}}^\mu,  \\
& \mathcal{A}_R= -D_\mu\,{\mathcal{J}}^\mu + \tfrac{3}{2}\,i\,\bar{\psi}_\mu\,\Gamma_5\,{\mathcal{S}}^\mu,  \\
& \mathcal{A}_Q= D_\mu\,{\mathcal{S}}^\mu - 2\,\Gamma^a\,\psi_\mu\,{\mathcal{T}}_a{}^\mu + 2\,\Gamma_5\,\tilde{\psi}_\mu\,{\mathcal{J}}^\mu,   \\
&\mathcal{A}_S = - 2\,\Gamma_5\,\psi_\mu\,{\mathcal{J}}^\mu + i\,\Gamma_\mu\,{\mathcal{S}}^\mu,
\end{equations}
where the non-vanishing densities $\mathcal{A}_W $, $\mathcal{A}_R$, $\mathcal{A}_Q$ and $\mathcal{A}_S$ can be read-off  eq.  (\ref{eq:mathcal_A_defs}). 



In order to compare our Chern-Simons anomaly cocycle with the $a$ and $c$ anomalies of \cite{Deser:1993yx},\footnote{A recent work on the interpretation of the anomaly coefficients $a,\,c$ in  $d=4,\,\mathcal{N}=1$ SCFTs as central extensions of a higher Virasoro symmetry algebra is \cite{Bomans:2023mkd}.} we look at the  R-symmetry and Weyl anomalies, keeping the terms of perturbative degree 3.  These terms capture  anomalous Feynman diagrams with three  external currents.

The R-anomaly of perturbative degree 3  of our chosen Lorentz-equivariant cocycle is 
\begin{equation}\label{AWfirst}
\!\alpha\,\mathcal{A}^{(3)}_R = -  \tfrac{3}{2}\, \alpha \,{\tilde{R}}_{ab} \,{\tilde{R}}^{ab}  + 3\,\alpha \,(\tilde{F}^{W}{})^2 + 15\, \alpha \,(\tilde{F}^{R}{})^2.
\end{equation}
Note that the first term in this expression depends only on the Weyl tensor: the Ricci components of the Riemann tensors are encoded in the $f^a$ terms.  By replacing  the superconformal curvatures with the standard curvatures  
\begin{subequations} \label{eq:tildetogeo}
\begin{align}
 \Tilde{R}^{ab}&={R}^{ab}(\omega)+2\,e^{[a}f^{b]}-2\,i\,\overline{\psi}\,\Gamma^{ab}\,\Tilde{\psi},\\
 \Tilde{F}^W &=2\,e^a\,f_a-2\,i\,\overline{\psi}\,\Tilde{\psi}, \\
  \Tilde{F}^R &=\dd \,a-2\,\overline{\psi}\,\Gamma_5\,\Tilde{\psi}, \\
  \Tilde{T}^a &=D\,f^a-\overline{\Tilde{\psi}}\,\Gamma^a\,\Tilde{\psi}, \\
 \rho&=D\,\psi+i\,e^a\,\Gamma_a\,\Tilde{\psi}, \\
 \Tilde{\rho}&=D\,\Tilde{\psi}-i\,f^a\,\Gamma_a\,{\psi},  
\end{align}
\end{subequations}
all the $f^a$ dependence cancels out in (\ref{AWfirst}) to give the result
\begin{equation}\label{sec:comparisonAR}
\mathcal{A}^{(3)}_R=-  \tfrac{3}{2}\, {R}_{ab} \,{R}^{ab}  + 15\,(F^{R}{})^2,
\end{equation}
proportional to the so-called  $a$-anomaly.  Let us now turn to  the bosonic part of the  Weyl anomaly of perturbative degree 3:
\begin{equation}
\sigma\,\mathcal{A}^{(3), bos}_W = 
-  \tfrac{3}{4} \varepsilon_{abcd} \,{\tilde{R}}^{ab} \,{\tilde{R}}^{cd} \sigma   + 6\, f_{c}\, e_{d}\, \sigma \varepsilon_{ab}{}^{cd} \,{\tilde{R}}^{ab} - 12 \,\sigma\, \epsilon^{abcd}\, f_a\, f_b\, e_c\, e_d.
\end{equation}
Once again, the first term
depends only on the Weyl-tensor: by itself  it would be $c$-type anomaly. However,  after replacing superconformal curvatures with standard curvatures, eq.  (\ref{eq:tildetogeo}), 
the $f^a$ dependence cancels out and one obtains
\begin{equation}\label{sec:comparisonAW}
\sigma\,\mathcal{A}^{(3),bos}_W = -  \tfrac{3}{4} \,\varepsilon_{abcd} \,{R}^{ab} \,{R}^{cd} \,\sigma,  
\end{equation}
thereby confirming that our superconformal cocycle is equivalent to the $a$-anomaly.\footnote{The relative coefficients of the  $R^2$, $F^2$ and $R\, {\star R}$ tensor structures  which appear in the  $U(1)_R$ and Weyl anomalies of superconformal gravity are implicitly contained in the formulas derived in  \cite{Bonora:1984pn}, in which superconformal anomalies are expressed in terms of superfields.   In \cite{Anselmi:1997am} the same anomalies were written out in components and the relation between the coefficients of these tensor structures was explicitly exhibited, with a numerical error which was corrected in both  \cite{Schwimmer:2010za} and \cite{Cassani:2013dba}. \cite{Papadimitriou:2019gel}  rederives the correct result  solving the  Wess-Zumino consistency condition. Our results for the coefficients of the chiral anomaly and the Weyl anomaly --- eqs. (\ref{sec:comparisonAR}) and  (\ref{sec:comparisonAW}) ---  match the ones of   \cite{Schwimmer:2010za,Cassani:2013dba} and  \cite{Papadimitriou:2019gel}, after taking into account that our  $U(1)_R$ gauge field $a_\mu$  and the gauge parameter $\alpha$ are normalized differently with respect  to  \cite{Papadimitriou:2019gel}: $a_\mu^{\text{Ref.\cite{Papadimitriou:2019gel}}}= \frac{3}{2}\, a_\mu$, $\alpha^{\text{Ref.\cite{Papadimitriou:2019gel}}}= \frac{3}{2}\, \alpha$.  The numerical $a$-coefficient, as defined in  \cite{Papadimitriou:2019gel}, which gives the overall normalization of our cocycle  is $a= 12\, \pi^2$.}

Summarizing,  the dependence on the geometric Riemann tensor of the superconformal cocycle is  contained both in the superconformal curvature ${\tilde{R}}^{ab}$ and the special conformal connection $f^a$. Thanks to the constraints (the bosonic part of) ${\tilde{R}}^{ab}$ is essentially the Weyl tensor built with the geometric Riemann tensor $R^{ab}$, while $f^a$ encodes the Ricci components of $R^{ab}$. The superconformal Chern-Simons cocycle has the precise combinations of ${\tilde{R}}^{ab}$ and $f^a$ to produce the $a$-anomaly, thanks to the cancellation of the $f^a$ dependence in the cocycle.

Let us also consider the terms of perturbative degree 4 in both the $R$ and the Weyl anomaly and let us also include the fermionic terms.  It turns out that the correction at perturbative degree 4 in the R-anomaly vanishes, when going from the conformal curvatures to the Riemannian ones and taking into account the fermionic corrections:
\begin{equation}\label{sec:comparisonAR4}
\alpha\,(\mathcal{A}^{(3)}_R +\mathcal{A}^{(4)}_R)=\alpha\,\big[-  \tfrac{3}{2}\, {R}_{ab} \,{R}^{ab}  + 15\,(F^{R}{})^2\bigr],
\end{equation}
The Weyl anomaly up to perturbative degree 4 (thus neglecting the $O(\psi^4)$ terms which are of perturbative degree at least 5) turns out to be
\begin{equation}\label{sec:comparisonAW4}
\sigma\,(\mathcal{A}^{(3)}_W  +\mathcal{A}^{(4)}_W)= -  \tfrac{3}{4} \,\varepsilon_{abcd} \,{R}^{ab} \,{R}^{cd} \,\sigma  + 6\, i\, {D\,\bar\psi}\, \Gamma^\mu\,\Gamma_5\,D\psi\, \nabla_\mu\,\sigma  -12 \, a\, {D\,\bar\psi}\, \Gamma^\mu\, \psi \,\nabla_\mu\, \sigma.
\end{equation}
\subsection{Fermionic anomalies}

The full fermionic anomalies in appendix \ref{appendix:lorentzcocycle} include terms linear, cubic and quintic in the gravitino field.  We present below  the terms of these fermionic anomalies linear in the gravitino and up to perturbative degree 4. These expressions describe anomalous contributions to correlators involving the divergence or the trace of the supercurrent, one additional supercurrent and either 1 or 2 bosonic currents. 
\begin{align}
\overline{\zeta}\mathcal{A}^{(4)}_Q=&-5\, a^{\mu } \,\bar{\zeta} \,\Gamma_{\mu}{} D^{[\nu}\psi^{\rho]}{}\,(F^{R})_{\nu \rho } + 10\, a_{\mu } \,\bar{\zeta}\,  \,\Gamma^{\nu } D^{[\mu}\psi^{\rho]}{} \,(F^{R})_{\nu \rho }\,+\nn\\
&+ 10\,i\, a^{\mu } \,\bar{\zeta} \,\Gamma_{\nu } \Gamma_5{} D^{[\nu}\psi^{\rho]}{}\, \varepsilon_{\mu \rho \sigma \alpha } \,(F^{R})^{\sigma \alpha } - 6\, a^{\mu } \, \varepsilon_{\mu \rho \sigma \alpha } \,\bar{\zeta}\,\Gamma^{\nu } D^{[\rho}\psi^{\sigma]}{} S_{\nu }{}^{\alpha } \,+\nn\\
&+ 2\, a^{\mu } \, \varepsilon_{\nu \sigma \alpha \beta } \,\bar{\zeta}\,\Gamma^{\nu } D^{[\rho}\psi^{\sigma]}{} W_{\mu }{}^{\alpha }{}_{\rho }{}^{\beta } - 4\, a^{\mu } \, \varepsilon_{\mu \sigma \alpha \beta } \,\bar{\zeta}\,\Gamma^{\nu } D^{[\rho}\psi^{\sigma]}{} W_{\nu \rho }{}^{\alpha \beta } \,+\nn\\
&+ \tfrac{1}{2}\, a^{\mu } \, \varepsilon_{\mu \nu \alpha \beta }\, \bar{\zeta}\,\Gamma^{\nu } D^{[\rho}\psi^{\sigma]}{} W_{\rho \sigma }{}^{\alpha \beta },
\end{align}
where $W_{\mu\nu}{}^{\rho\sigma}$ is the Weyl tensor and  $S_{\mu\nu}$ is the Schouten tensor:
\begin{align}
W_{\mu \nu ; \rho \sigma}&=  R_{\mu \nu \rho \sigma }- \tfrac{1}{2} g_{\sigma \nu }\, R_{\mu \rho } + \tfrac{1}{2} g_{\rho \nu } \,R_{\mu \sigma } + \tfrac{1}{2} g_{\sigma \mu }\, R_{\nu \rho } -  \tfrac{1}{2} g_{\rho \mu }\, R_{\nu \sigma } -  \tfrac{1}{6} g_{\rho \nu } g_{\sigma \mu } \,R + \tfrac{1}{6} g_{\rho \mu } g_{\sigma \nu }\,R, \nn\\
S_{\mu\nu }&=  \frac{1}{2} \, R_{\mu\nu} - \frac{1}{12}\, g_{\mu\nu}\, R,
\end{align}

 $\mathcal{A}_S$ receives a contribution already at perturbative degree 3 \footnote{Our result for the $\mathcal{A}_S$ anomaly agrees
 with the one in \cite{Papadimitriou:2019gel},  with the same anomaly coefficient $a=12 \,\pi^2$  as the R and W anomalies. Our gravitino field $\psi$ and supersymmetry ghosts $\zeta$, $\eta$ are normalized differently with respect to  \cite{Papadimitriou:2019gel}: $\psi= \frac{1}{2}\,\psi^{\text{Ref.\cite{Papadimitriou:2019gel}}}$,  $\zeta= \frac{1}{2}\,\zeta^{\text{Ref.\cite{Papadimitriou:2019gel}}}$, $\eta= \frac{1}{2}\,\eta^{\text{Ref.\cite{Papadimitriou:2019gel}}}$.  The terms in  $\mathcal{A}^{(4)}_Q$ involving  two $U(1)_R$ fields also agree with the corresponding ones in  \cite{Papadimitriou:2019gel}, including the overall normalization.  This fixes completely the dependence of the anomaly on the remaining gauge parameters. However  the terms  in  $\mathcal{A}^{(4)}_Q$ involving the gravitational curvatures do not agree with \cite{Papadimitriou:2019gel}.}
 \begin{align}
\bar\eta\,(\mathcal{A}^{(3)}_S+\mathcal{A}^{(4)}_S)=&-15\,i \,\bar{\eta}\, D^{[\alpha}\psi^{\beta]}{} \varepsilon_{\alpha \beta \gamma \delta } \,(F^{R})^{\gamma \delta } + 30\, a^{\alpha } \,\bar{\eta} \,\Gamma_5{}\, \psi^{\beta }{} \, \varepsilon_{\alpha \beta \gamma \delta } \,(F^{R})^{\gamma \delta }\,+\nn\\
&- 12\,i \,\bar{\eta} \,\Gamma_{\alpha }{}^{\gamma } D^{[\alpha}\psi^{\beta]}{} R_{\beta \gamma } + 3\,i \,\bar{\eta}\, \,\Gamma_{\alpha \beta}{} D^{[\alpha}\psi^{\beta]}{} R + 3\,i \,\bar{\eta}\, \Gamma^{\gamma \delta } D^{[\alpha}\psi^{\beta]}{} R_{\alpha \beta \gamma \delta } \,+\nn\\
&+ 3\,i\, a^{\nu } \,\bar{\eta}\, \varepsilon_{\nu \beta \lambda \mu } \,\Gamma^{\gamma \delta } \psi^{\beta }{} R_{\gamma \delta }{}^{\lambda \mu }.
\end{align}

\section{Conclusions and open problems}\label{sec:conclusions}

We have remarked that the $d=4,\,\mathcal{N}=1\;$ Lie superconformal algebra admits a single invariant completely symmetric (in the graded sense) tensor with 3 indices in the super-adjoint representation. We have also shown that the corresponding invariant polynomial, cubic in the generalized curvatures of superconformal gravity,  vanishes ---    despite those generalized curvatures not being horizontal. Therefore the corresponding superconformal  secondary Chern-Simons class  is  an anomaly cocycle.   We computed this cocycle explicitly, in components and  to all orders in the independent propagating fields of superconformal gravity.   We showed  that it is equivalent to the so-called $a$-anomaly of superconformal gravity, a superconformal extension of the Euler Weyl anomaly of bosonic gravity. Our result is best viewed as an extension of the Stora-Zumino paradigm for producing anomaly cocycles out of secondary Chern-Simons classes --- generalizing it  to the case,  characteristic of supersymmetry and conformal invariance, in which generalized curvatures are not horizontal.  

Superconformal gravity is believed to possess a second independent anomaly known as the $c$-anomaly, a superconformal extension of the Weyl anomaly of bosonic gravity constructed from the Weyl tensor. Hence, it is natural to inquire whether the $c$-anomaly also lends itself to a Chern-Simons formulation.  The fact that the $d=4,\,\mathcal{N}=1\;$ Lie superconformal algebra admits a single 3-index completely symmetric (in the graded sense) invariant tensor --- which we proved to correspond to the $a$-anomaly --- would seem at first to rule this out.  
However, as we stressed throughout the paper, the superconformal curvatures must satisfy certain constraints. Therefore, although invariant polynomials of the curvatures  are necessarily BRST invariant, it is possible in principle that  non-invariant  but  BRST closed polynomials of generalized connections and curvatures exist, thanks to the constraints.  This could give rise to the emergence of extra BRST cohomology classes: it is worth noting, in this respect,  that  the superconformal formalism that we developed naturally gives rise to the (supersymmetrization of the) Weyl tensor out of which the $c$-anomaly is built. 

Another interesting open problem  is to provide a holographic 5-dimensional interpretation of our main result, the Chern-Simons formula for the superconformal anomaly eq. (\ref{superconfinvariantanomaly}).  If one extends all fields to 5 dimensions, the generalized Chern-Simons form in this equation does develop a non-vanishing 5-form $Q^{(5)}_0$ component. This  would be a candidate for a  5-dimensional Chern-Simons presentation of the 4-dimensional superconformal anomaly, thus unifying the  holographic descriptions of both Yang-Mills  \cite{Witten:1998qj} and Weyl anomalies  \cite{Imbimbo:1999bj}.  However, the constraints  which curvatures must satisfy are formulated in 4 dimensions: to substantiate the 5-dimensional reading of eq. (\ref{superconfinvariantanomaly}) one needs to  understand if and how these constraints can be extended to 5 dimensions.

\section*{Acknowledgments}
C.I. is deeply grateful to M. Fr\"ob for  many patient  and very helpful lessons about \textsc{FieldsX}. 

This work is  supported in part by the Italian Istituto Nazionale di Fisica Nucleare and by Research Projects, F.R.A. 2022 of the Universit\`a di Genova. The work of A.W. is supported by the UKRI Frontier Research Grant, underwriting the ERC Advanced Grant ``Gene\-ra\-liz\-ed Symmetries in Quantum Field Theory and Quantum Gravity''.

\appendix
\section{Relation between the Stora-Zumino formulation of anomalies and the so-called ``two-step  descent''}\label{appendix:SZ}

In this appendix we review in detail  the relationship between the   Stora-Zumino (SZ) formulation of anomalies  and the so-called ``two-step descent'' procedure \cite{Zumino:1983rz}. In essence,  the two ``descents'' are equivalent because they are algebraic in nature, not geometrical.  
The technical difference between the  formalisms is the following:  in the SZ formalism anomalies are described by a single {\it generalized} form, while in the ``two-step  descent'' procedure anomalies are captured by a collection of {\it ordinary}  forms.  In both cases one starts by considering connection and curvature with values in the {\it same} Lie (super)algebra. In the  ``two-step descent'' method,  the connection is an {\it ordinary} 1-form whose curvature is an {\it ordinary} 2-form:
\begin{equation}
F= \dd A + A^2,
\label{app:Fcurvature}
\end{equation}
where $\dd$ is the de Rham differential acting on ordinary forms.  In the  SZ-BRST framework the connection
\begin{equation}
\bm{A} = c + A,
\end{equation}
 is a {\it generalized form} of total degree (defined as the sum of form degree and ghost number)  equal to 1, whose curvature is a {\it generalized} form of total degree 2
\begin{equation}
\bm{F} = \delta \bm{A} + \bm{A}^2.
\label{app:Fcalcurvature}
\end{equation}
Here $\delta$ is the generalized nilpotent BRST coboundary operator, which  for  Yang-Mills  theories is written as follows
\begin{equation}\delta = s+ \dd,
\end{equation}
where $s$ is the nilpotent BRST operator acting on fields.   In the case of supergravity, in order to preserve nilpotency, one needs to  define  instead
\begin{equation}\label{app:deltasugra}
\delta= \hat{s}+ \dd - i_\gamma,
\end{equation}
where $\gamma^\mu = \bar\zeta\, \Gamma^\mu\, \zeta$ is the supersymmetry ghost bilinear and $\hat{s}$ is the BRST operator equivariant with respect to diffeomorphisms.\footnote{Anomalies of bosonic gravity  in the metric description can be obtained, in the dimensions in which they exist,  by starting with the same definition (\ref{app:deltasugra}) for $\delta$ involving  the equivariant $\hat{s}$ with the  superghost dependent term  $\gamma^\mu$  put to zero.}   

Both the ordinary and generalized curvature  satisfy the Bianchi identities  which are purely algebraic statements encoding the nilpotency of the relevant differentials:  
\begin{align} 
& \dd^2=0 \quad\Rightarrow\quad \dd F= - [A, F], \label{app:Bianchid}\\
& \delta^2=0 \quad\Rightarrow\quad \delta \bm{F} = - [\bm{A}, \bm{F}]. \label{app:Bianchidelta}
\end{align}
To construct the descent one picks  a completely (super)sym\-metric 3-index {\it invariant} tensor $d_{abc}$  of the (super)Lie algebra.   Correspondingly, one can define either an ordinary Chern polynomial
\begin{equation} P_3(F) = \mathrm{Tr}\, F^3 \equiv  d_{abc} \, F^a\, F^b\, F^c,
\end{equation}
or a generalized one
\begin{equation}P_3(\bm{F}) = \mathrm{Tr}\, \bm{F}^3 \equiv  d_{abc} \, \bm{F}^a\, \bm{F}^b\, \bm{F}^c.
\end{equation}
Both ordinary and generalized Chern polynomials are closed with respect to the relative differentials thanks to the Bianchi identities (\ref{app:Bianchid})-(\ref{app:Bianchidelta}):
\begin{equation}
\dd \, P_3(F) =0, \qquad \delta P_3(\bm{F}) =0.
\end{equation}
The second fact that the descent depends on is the triviality of the  cohomology of $\dd$ ($\delta$) on the space of non-zero degree polynomials  of  ordinary  (generalized) connections and ordinary (generalized) curvatures.  Again,  this fact is a purely algebraic property which descends merely from the  definitions of curvatures, which are identical for both ordinary (eq. (\ref{app:Fcurvature})) and generalized ones (eq. (\ref{app:Fcalcurvature})). Therefore triviality holds for either $\dd$ or $\delta$ for the same identical reason.\footnote{The triviality of the local cohomology of either $\dd$ or $\delta$  can be proven by means of standard filtration arguments from the fact that  curvatures are, by definition,  exact up to non-linear terms. } 
One concludes that  both  $P_3(F)$ and $P_3(\bm{F})$ are exact 
\begin{align}
& P_3(F) = \dd \, Q_5(A, F), \label{app:CSAF}\\
&P_3(\bm{F}) = \delta\,  Q_5(\bm{A}, \bm{F}),\label{app:CSAcalFcal}
\end{align}
and  the Chern-Simons polynomial of connections and curvatures $Q_5$ is a {\it universal} algebraic object: it only depends on the completely (super)symmetric invariant  3-index tensor $d_{abc}$.  It is the {\it same} polynomial for both the ordinary and generalized Chern polynomial. 

One difference between the two formalisms is the following. In the ``two-step'' descent one needs to make ordinary forms  depend on two ``unphysical''  extra-coordinates:   eq. (\ref{app:CSAF}) is empty in 4-dimensions since  both the ordinary 6-form  $P_3(F) $ and the 5-form  $Q_5(A, F) $ trivially vanish in   4-dimensional space-time. On the other hand, extending fields to higher dimensions is not  necessary in the SZ formalism  since generalized forms of degree greater than 4 do not identically vanish in 4 dimensions. 

The derivation of the  anomaly for Yang-Mills theories in the ordinary form formalism  relies on the fact that  the Yang-Mills curvature $F$ transforms in the adjoint representation of the Lie algebra under BRST (gauge) transformations:
\begin{equation}
s\, F = - [c , F]. 
\label{sFappendix}
\end{equation}
This is of course  a consequence of   the BRST (gauge) transformation rules for Yang-Mills  connections
\begin{equation}
s\, A = - D\,c.
\label{sAappendix}
\end{equation}
Hence  the Yang-Mills  ordinary Chern polynomial $P_3(F)$   (extended to 6-dimensions) is  $s$-invariant
$$ s\, P_3(F) =0. \label{app:sP3}$$ 
Since $\dd$ and $s$ anti-commute,  one has
$$ 0 = \dd\,  (s\,  Q_5(A, F)). $$
From the (local) triviality of $\dd$ one deduces
\begin{equations}
s\, Q_5(A, F) &= - \dd\, Q_{4,1} (c, A, F), \label{app:descentordinaryfirst} \\
s\, Q_{4,1}(c, A,F) &= - \dd\, Q_{3,2}(c, A, F).\label{app:descentordinarysecond} 
 \end{equations}
$Q_{4,1}(c, A, F)$ is a  4-form of ghost number 1   which  satisfies  the anomaly  consistency condition. By pulling this form back to  4-dimensional space-time, considered as a submanifold of higher-dimensional unphysical space, one recovers the 4-dimensional anomaly.

In the SZ formalism  both descent equations (\ref{app:descentordinaryfirst}) and (\ref{app:descentordinarysecond})  are contained in a single equation,   eq. (\ref{app:CSAcalFcal}), which captures the triviality of the generalized Chern polynomial. 
This is seen by first expanding the generalized Chern-Simons polynomial in powers of $c$
\begin{align}
Q_5(\bm{A}, \bm{F}) = Q_5(c +A, F) =& \;    Q_{5,0}(A, F)+Q_{4,1} (c, A, F)+Q_{3,2}(c, A, F)+\nn\\
& + Q_{2,3} (c, A, F)+Q_{1,4}(c, A, F)+Q_{0,5}(c). 
\end{align}
The ``descendants'' $Q_{n, 5-n}(c, A, F)$,  are $n$-forms of ghost num\-ber $5-n$.\footnote{$Q_{5,0}(A,F) = Q_5(A, F)$ is the original Chern-Simons polynomial.}  
One also observes that, in the case of Yang-Mills, the  BRST rules for both  connection (\ref{sFappendix}) and ghost $ s\, c = - c^2$ are summarized by the horizontality  equation
\begin{equation} 
\bm{F} = F,
\end{equation}
which implies 
\begin{equation}
P_3(\bm{F}) = P_3(F).
\end{equation}
 Hence eq. (\ref{app:CSAcalFcal})   becomes  for Yang-Mills theories: 
\begin{equations}
 P_3(F) &= \dd\, Q_{5,0}(A, F),\label{app:SZdes1}  \\
 0 &= \dd\, Q_{4,1}(c, A, F) + s\,Q_{5,0} (A, F), \label{app:SZdes2}\\
 0& = \dd\, Q_{3,2} (c, A, F) + s\,Q_{4,1} (c, A, F),\label{app:SZdes3}  \\
0& = \dd\, Q_{2,3}(c, A, F) + s\,Q_{3,2} (c, A, F), \label{app:SZdes4} \\
 0&= \dd\, Q_{1,4}(c, A, F) + s\,Q_{2,3} (c, A, F), \label{app:SZdes5} \\
0& = \dd\, Q_{0,5}(c) + s\,Q_{1,4} (c, A, F), \label{app:SZdes6} \\
 0&= s\, Q_{0,5}(c).
\end{equations}
which are  completely equivalent to the ``two-step  descent''  equations (\ref{app:descentordinaryfirst})-(\ref{app:descentordinarysecond}). 

As we remarked above,  there is no need  in the SZ framework to extend fields to higher-dimensions.  In 4-dimensions the first
two equations (\ref{app:SZdes1})-(\ref{app:SZdes2}) above are trivial and the  SZ descent actually starts from eq.  (\ref{app:SZdes3}) which is the anomaly consistency condition for $Q_{4,1} (c, A, F)$.  From this perspective  the SZ formalism {\it explains}  the connection between 4-dimensional anomalies and the 5-dimensional Chern-Simons polynomial and 6-dimensional Chern invariant. In the two-step approach this relation emerges, somewhat mysteriously,  by extending fields to unphysical higher dimensions.   The SZ formalism also makes trivial writing down the ``descendants''  $Q_{n, 5-n}(c, A, F)$ by simply expanding the universal Chern-Simons polynomial $Q_5(c+A, F)$ in powers of the ghost.

 Of course one has the option to extend fields to higher dimensions in the SZ framework too.  Notably, in the holographic context, one gives  ``physical'' meaning to extra-dimensions, by  thinking of (closed) 4-dimensional space-time  $M_4$ as the boundary of a 5-dimensional ``ball'' $B_5$.  In this ca\-se,  eq.(\ref{app:SZdes1})
is still trivial but eq.(\ref{app:SZdes2}) is not.  By integrating it on $B_5$ one obtains
\begin{equation}
\int_{M_4}  Q_{4,1}(c, A, F)  = - s\,\int_{B_5} Q_{5} (A, F).\label{app:hologranomaly}
\end{equation}
which states that the in\-te\-grated 4-dimensional anomaly is the BRST va\-ria\-tion of a 5-dimen\-sional local functional, the integral in the ``bulk''  of the ordinary Chern-Simons  polynomial. 

In supergravity (and conformal) theories horizontality of the generalized curvature does not hold.  The generalized curvature writes
\begin{equation}
\bm{F} = F + \lambda,
\end{equation}
where $\lambda$ is a 1-form of ghost number 1.  This is so because in the super-Lie algebra case the BRST variation of the
ordinary connection is not a mere gauge transformation since it also includes $\lambda$:
\begin{equation}
\hat{s}\, A = - D\,c  +\lambda
\end{equation}
and consequently  the ordinary curvature does not transform in the adjoint
\begin{equation}
\hat{s}\, F = - [c, F] - D\, \lambda.
\end{equation}
It follows that  the {\it ordinary} Chern polynomial is not $\hat{s}$-invariant
\begin{equation}
 \hat{s}\, P_3(F) = - \dd\, \mathrm{Tr} \, \lambda\, F^2. \label{app:sP3sugra}
 \end{equation}
Hence the ordinary ``two-step'' descent equations  (\ref{app:descentordinaryfirst})-(\ref{app:descentordinarysecond}) break down.  

The SZ formalism makes transparent the necessary and sufficient condition under which the generalized  Chern polynomial still encodes an anomaly. From eq. (\ref{app:CSAcalFcal}), one reads that  $\delta\, Q_5(\bm{A}, \bm{F})=0$  iff  
\begin{equation}
 P_3(\bm{F}) =0\label{app:P3horizontal}
\end{equation}
 in 4-dimensions.  We have seen that this is precisely what happens in  4-dimensional superconformal gravity,  notwithstanding the fact that  $\bm{F}$ is not horizontal. 
When (\ref{app:P3horizontal}) holds,   exactness of the generalized Chern polynomial (\ref{app:CSAcalFcal}) directly leads to the 4-dimensional descent equations
\begin{equations}
 0& = \dd\, Q_{3,2} (c, A, F) + \hat{s}\,Q_{4,1} (c, A, F),\label{app:SZdesgrav3}  \\
0& = \dd\, Q_{2,3}(c, A, F) + \hat{s}\,Q_{3,2} (c, A, F)- i_\gamma(Q_{4,1} (A, F)), \label{app:SZdesgrav4} \\
 0&= \dd\, Q_{1,4}(c, A, F) + \hat{s}\,Q_{2,3} (c, A, F))- i_\gamma(Q_{3,2} (A, F)), \label{app:SZdesgrav5} \\
0& = \dd\, Q_{0,5}(c) + \hat{s}\,Q_{1,4} (c, A, F)- i_\gamma(Q_{2,3} (A, F)), \label{app:SZdesgrav6} \\
 0&= \hat{s}\, Q_{0,5}(c)- i_\gamma(Q_{1,4} (A, F)).
\end{equations}
The Chern-Simons  descendant $Q_{4,1}(c, A, F)$ is therefore an anomaly of superconformal gravity just as it is  for Yang-Mills theories.

It should be emphasized that we proved the vanishing of the  generalized Chern polynomial, eq. (\ref{app:P3horizontal}),  by using the constraints of  conformal supergravity, which hold in 4-dimensional space-time.  Conformal supergravity  constraints do not have  obvious extensions to 5 dimensions.  If this extension were possible, while preserving  at the same time the vanishing of the generalized Chern polynomial,  then one could write the superconformal anomaly holographically as  in  eq. (\ref{app:hologranomaly}).  We leave to the future the investigation  of the validity of the holographic equation for the superconformal Chern-Simons anomaly. One attractive feature of the SZ formalism is that it connects anomalies to Chern-Simons polynomials without  making any reference to higher dimensions.

\section{\texorpdfstring{$d=4$, $\mathcal{N}=1$}{d4N1} Lie superconformal algebra}\label{appendixB}

In this appendix  we  review our conventions for the $d=4$, $\mathcal{N}=1$ superconformal algebra. The bosonic and fermionic generators, 
the corresponding gauge fields and BRST ghosts are listed in Table \ref{Table1}.
\begin{table}[h]
\begin{center}
\begin{tabular}{ | c | c | c c | c |} 
\hline
 Bosonic Symmetry & Generator & Gauge field & & Ghost  \\
 \hline
 Local Lorentz & $J_{ab}$ & spin connection & $\omega_\mu{}^{ab}$ & $\Omega^{ab}$ \\
 Weyl & $W$ & dilaton & $b_\mu$ & $\sigma$ \\
 $U(1)_R$ chiral R-symmetry & $R$ & $U(1)_R$-gauge field & $a_\mu$ & $\alpha$  \\
 Diffeomorphisms & $P_a$ & vierbein & $e_\mu{}^a$ & $\xi^\mu$  \\
 Special conformal & $K_a$ & conformal vierbein & $f_\mu{}^a$ & $\theta^a$  \\
 \hline
  Fermionic Symmetry & Generator & Gauge field & & Ghost  \\
 \hline
 Supersymmetry & $Q_\alpha$ & gravitino & $\psi_\mu{}^\alpha$ & $\zeta^\alpha$  \\
 Conformal supersymmetry & $S_\alpha$ & conformal gravitino & $\Tilde{\psi}_\mu{}^\alpha$ & $\eta^\alpha$ \\
 \hline
\end{tabular}
\caption{\label{Table1} $\mathfrak{su}(2,2|1)$ symmetries and generators, with their associated gauge fields and BRST ghosts.}
\end{center}
\end{table}

The (anti)-commutation relations defining the $d=4,\;\mathcal{N}=1$ superconformal algebra are:\footnote{We take spinor contractions in the $\searrow$ direction. Hence $\lambda^\alpha\chi_\alpha=-\lambda_\alpha\chi^\alpha$. E.g. $\zeta^\alpha(\Gamma^a)_{\alpha\beta}\zeta^\beta=-\zeta^\alpha(\Gamma^a)_\alpha{}^\beta\zeta_\beta=-\overline{\zeta}\,\Gamma^a\zeta$.}
\begin{equation*}
[J_{ab},J_{cd}]=\eta_{ac}\,J_{db}-\eta_{bc}\,J_{da} +\eta_{bd}\,J_{ca}-\eta_{ad}\,J_{cb}, \\[-3mm]
\end{equation*}
\begin{alignat*}{4}
& [J_{bc},P_{a}] &&= \eta_{ac}\,P_b-\eta_{ab}\,P_c, \qquad 
&& [J_{bc},K_{a}] &&= \eta_{ac}\,K_b-\eta_{ab}\,K_c, \\
 & [P_a,P_b] &&= 0, 
&& [K_a,K_b] &&= 0, \\
 & [W,P_a] &&= P_a, 
&& [W, K_a] &&= -K_a,  \\[-9mm]
\end{alignat*}
\begin{equation*}
[P_a,K_b] = 2\,(\eta_{ab}\,W+J_{ab}),\\
\end{equation*}
\begin{alignat*}{4}
 & [J_{ab},Q_\alpha] &&= \tfrac{1}{2}\,(\Gamma_{ab})_\alpha{}^\beta Q_\beta,
&& [J_{ab},S_\alpha] &&= \tfrac{1}{2}\,(\Gamma_{ab})_\alpha{}^\beta \,S_\beta, \\
 & [W,Q] &&= \tfrac{1}{2}\,Q,  
&& [W,S] &&= -\tfrac{1}{2}\,S, \\
 & [R ,Q_\alpha] &&= -\tfrac{3}{2}\,i\,(\Gamma_5)_\alpha{}^\beta\,Q_\beta,\qquad
&& [R,S_\alpha] &&= \tfrac{3}{2}\,i\,(\Gamma_5)_\alpha{}^\beta S_\beta, \\
 & [P_a,Q]&&=0,
&& [K_a,S]&&=0,\\
 & [P_a,S_\alpha]&&= i\,(\Gamma_a)_\alpha{}^\beta\,Q_\beta,
&& [K_a,Q_\alpha]&&=-i\,(\Gamma_a)_\alpha{}^\beta\,S_\beta,
\\
 & \{Q_\alpha,Q_\beta\}&&=-2\,(\Gamma^a)_{\alpha\beta}\,P_a,
&& \{S_\alpha,S_\beta\}&&=2\,(\Gamma^a)_{\alpha\beta}\,K_a, \\[-9mm]
\end{alignat*}
\begin{equation}
\{Q_\alpha,S_\beta\}=2\,i\,W\delta_{\alpha\beta}+2\,i\,(\Gamma^{ab})_{\alpha\beta}\,\tfrac{1}{2}\,J_{ab}+2\,(\Gamma_5)_{\alpha\beta}\,R.
\end{equation}
If we collectively denote such generators by $\{T_i\}$ with $1\leq i\leq 24$, $T_i$ is bosonic for $1\leq i\leq 16$ and fermionic for $17\leq i\leq 24$.  The \emph{grading} $|i|$ of $T_i$ is defined to be:
\begin{equation}
|i|=
    \begin{cases}
        0 \;\;(\text{mod}\;2), \hspace{5mm}\text{if $T_i$ is bosonic}\hspace{10mm} (1\leq i\leq 16), \\
        1 \;\;(\text{mod}\;2), \hspace{5mm}\text{if $T_i$ is fermionic}\hspace{5mm} (17\leq i\leq 24).
    \end{cases}
\end{equation}
Ghosts are fields which have \emph{opposite} statistics, (i.e. $\mathbb{Z}_2$ gradings) with respect to the generator $T_i$  to which they correspond:
\begin{equation} \label{eq:ghost_gradings}
    |c^i|=|i|+1.
\end{equation}
The superLie bracket  is written as 
\begin{equation}
 [T_j,T_k]= f_j{}^i{}_k\,T_i,
\end{equation}
where $[\cdot,\cdot]$ denotes the commutator or anti-commutator
\begin{equation} \label{eq:super-comm}
    [T_i,T_j]\equiv T_i\,T_j-(-)^{|i||j|}T_j\,T_i,
\end{equation}
  Hence
  \begin{equation} \label{eq:algebra_f_simm}
    f_j{}^i{}_k=-(-)^{|j||k|}f_k{}^i{}_j
\end{equation}
The super-Jacobi equation is 
equivalent to the statement that  $f_j{}^i{}_k$ is a super-invariant tensor:
\begin{equation} \label{eq:super_Jac_fijk}
    (-)^{|i||k|}f_i{}^l{}_j\,f_l{}^m{}_k+(-)^{|j||i|}f_j{}^l{}_k\,f_l{}^m{}_i  +(-)^{|k||j|}f_k{}^l{}_i\,f_l{}^m{}_j=0.
\end{equation}
Both ghosts $c^i$ and generators $T_i$ are graded, so that $ g\equiv c^i\, T_i$ is {\it odd}.  Hence
\begin{equation}
    [g ,g]=\Tilde{f}_j{}^i{}_k\,c^j\,c^k\,T_i,
\end{equation}
where the  ``Grassmann envelope structure constants" $\Tilde{f}_j{}^i{}_k$ are related to the structure constants $f_j{}^i{}_k$ as follows
\begin{equation} \label{appendixC:eq:f_tilde_def}
    \Tilde{f}_j{}^i{}_k=(-)^{|j|(|k|+1)}f_j{}^i{}_k.
\end{equation}
Hence
\begin{equation}
     \Tilde{f}_k{}^i{}_j= 
     (-)^{(1+|j|)(1+|k|)} \, \Tilde{f}_j{}^i{}_k.
\end{equation}
The structure constants of $\mathfrak{su}(2,2|1)$ are:
\begin{equation*}
f^{[ef]}_{[ab][cd]}=\eta_{[c[a}\,\delta^{[e}_{d]}\,\delta^{f]}_{b]}, \\[-6mm]
\end{equation*}
\begin{alignat*}{4}
 & f_{[bc],a}^d&&=\eta_{a[c}\,\delta^d_{b]},
&& f_{[bc],\Tilde{a}}^{\Tilde{d}}&&=\eta_{\Tilde{a}[c}\,\delta^{\Tilde{d}}_{b]},\\
 & f^b_{W,a} &&= \delta^a_{b},  
&& f^{\Tilde{b}}_{W,\Tilde{a}} &&= -\delta^{\Tilde{b}}_{\Tilde{a}},\\
 & f^{[cd]}_{a,\Tilde{b}} &&= \delta^{[c}_{a}\delta^{d]}_{\Tilde{b}},
&& f^{W}_{a,\Tilde{b}} &&= 2\,\eta_{a\Tilde{b}},\\
 & f^\beta_{[ab],\alpha} &&= \tfrac{1}{2}\,(\Gamma_{ab})_\alpha{}^\beta,  
&& f^{\Tilde{\beta}}_{[ab],\Tilde{\alpha}} &&= \tfrac{1}{2}\,(\Gamma_{ab})_{\Tilde{\alpha}}{}^{\Tilde{\beta}},\\
 & f^\alpha_{W,\beta} &&= \tfrac{1}{2}\,\delta^\alpha_\beta, 
&& f^{\Tilde{\alpha}}_{W,\Tilde{\beta}} &&= -\tfrac{1}{2}\,\delta^{\Tilde{\alpha}}_{\Tilde{\beta}},\\ 
 & f_{R,\alpha}^{\beta} &&= -\tfrac{3}{2}\,i\,(\Gamma_5)_{\alpha}{}^{\beta},\quad
&& f_{R,\Tilde{\alpha}}^{\Tilde{\beta}} &&= \tfrac{3}{2}\,i\,(\Gamma_5)_{\Tilde{\alpha}}{}^{\Tilde{\beta}},\\
 & f^\beta_{a,\Tilde{\alpha}} &&= i\,(\Gamma_a)_{\Tilde{\alpha}}{}^\beta,
&& f^{\Tilde{\beta}}_{\Tilde{a},{\alpha}} &&= -i\,(\Gamma_a)_{\alpha}{}^{\Tilde{\beta}},\\
 & f^a_{\alpha\beta}&&=-2\,(\Gamma^a)_{\alpha\beta},
&& f^{\Tilde{a}}_{\Tilde{\alpha}\Tilde{\beta}}&&=2\,(\Gamma^{\Tilde{a}})_{\Tilde{\alpha}\Tilde{\beta}},\\[-9mm]
\end{alignat*}
\begin{equation}
\begin{aligned}
&  f^W_{\alpha\Tilde{\beta}}=2\, i\,\delta_{\alpha\Tilde{\beta}},
&&  f^{ab}_{\alpha\Tilde{\beta}}=2\,i\,(\Gamma^{ab})_{\alpha\Tilde{\beta}},
&&& f^R_{\alpha\Tilde{\beta}}=2\,(\Gamma_5)_{\alpha\Tilde{\beta}}.
\end{aligned}
\end{equation}


The mass dimensions of ghosts and gauge fields in $d=4,\,\mathcal{N}=1$ conformal supergravity are fixed by their BRST transformations taking into account that $s$ is dimensionless. For the standard bosonic YM symmetries for which  $s \, c=-c^2$, one has
    \begin{equation}
        [\Omega^{ab}]=[\sigma]=[\alpha]=0.
     \end{equation}
From the BRST transformation rules for the diffeomorphism ghost $\xi^\mu$ we obtain
\begin{equation}
	[\gamma^\mu]=[\xi^\mu]=-1.
\end{equation}
From the BRST transformations for the supersymmetry ghosts we deduce
\begin{equation}
 [\zeta]=-\tfrac{1}{2},\qquad  [\eta] =\tfrac{1}{2},\qquad [\theta]= 1.
\end{equation}
For tensorial connections and curvatures we have
\begin{equation} 
        [A_\mu]=[c]+1, \qquad         [F_{\mu\nu}]=[A_\mu]+1.
\end{equation}
Note that the physical fields $e_\mu{}^a$, $a_\mu$, $b_\mu$ and $\psi_\mu{}^\alpha$ have canonical mass dimensions, $0, 1, 1, \tfrac{1}{2}$ respectively.\footnote{Restoring the gravitational constant, which we put to 1, graviton and gravitino would get the familiar mass dimensions, $1$ and $\tfrac{3}{2}$.}  Instead the composite fields $f_\mu{}^a, \omega_\mu{}^{ab}, \tilde{\psi}_\mu{}^\alpha$ have non-canonical higher mass dimensions $2, 1,\tfrac{3}{2}$.  Table \ref{Table2} lists the $W$ and $R$ charges of the fields of the theory, that we took into account in Section \ref{sec:sol_for_constraints} to construct the possible forms for the $\lambda_0^i$'s.
\begin{table}[h]
\begin{center}
\begin{tabular}{|c|c|c|}
\hline
$\bm{A}$& $W\text{-weight}$ & $R\text{-charge}$ \\
\hline
$\quad e^a$ & $\quad\;\;\;1$ & $\quad\;\;\;0$ \\
$\quad \bm{\omega}^{ab}$ & $\quad\;\;\;0$ & $\quad\;\;\;0$ \\
$\quad \bm{b}$ & $\quad\;\;\;0$ & $\quad\;\;\;0$ \\
$\quad \bm{a}$ & $\quad\;\;\;0$ & $\quad\;\;\;0$ \\
$\quad \bm{f}^a$ & $\quad-1$ & $\quad\;\;\;0$ \\
$\quad \bm{\psi}^\alpha$ & $\quad\;\;\;\tfrac{1}{2}$ & $\quad-\tfrac{3}{2}$ \\
$\quad\bm{\tilde{\psi}}^\alpha$ & $\quad-\tfrac{1}{2}$ & $\quad\;\;\;\tfrac{3}{2}$ \\
\hline
\end{tabular}
\caption{\label{Table2} Weyl weights and R-charges of the connections.}
\end{center}
\end{table}

\section{Supertrace for Lie superalgebras} \label{appendix:super-Lie-algebras}

In this appendix  we  review a few general properties of supertraces for  Lie superalgebras, relevant for superconformal anomalies. For more details see \cite{KAC19778,Berezin1987a}. 

If $V=V_{\Bar{0}}\oplus V_{\Bar{1}}$ is a $\mathbb{Z}_2$ graded vector space,
a \emph{homogeneous basis} of $V=V_{\Bar{0}}\oplus V_{\Bar{1}}$, where $m=\dim{V_{\Bar{0}}}$ and $n=\dim{V_{\Bar{1}}}$ is of the form
\begin{equation}
    \{e^{(b)}_1,...e^{(b)}_m,\;e^{(f)}_{m+1},...e^{(f)}_{m+n}\},
\end{equation}
where the superscripts $^b$ and $^f$ stand for ``bosonic" (even) and ``fermionic" (odd) respectively.

One can define linear \emph{representations} of Lie superalgebras, by associating to each algebra generator an element of $End(V)=l(V)_{\Bar{0}}\oplus l(V)_{\Bar{1}}$, which are matrices in a given basis, and by defining the super-Lie bracket in terms of the graded-commutator of matrices.

The matrices representing bosonic $B \in l(V)_{\Bar{0}}$ and fermionic $F \in l(V)_{\Bar{1}}$ operators respectively have the form
\begin{equation}
    B=
        \begin{pmatrix}
            B_{bb} &\rvline & 0 \\
            \hline
            0 &\rvline & B_{ff}
        \end{pmatrix},
    \qquad
    F=
        \begin{pmatrix}
            0 &\rvline & F_{bf} \\
            \hline
            F_{fb} &\rvline & 0
        \end{pmatrix}.
\end{equation}
Let $A\in l(V)$ be a generic operator. In block-diagonal form it is written as
\begin{equation}
    A=\begin{pmatrix}
            \alpha &\rvline & \gamma \\
            \hline
            \delta &\rvline & \beta
        \end{pmatrix}.
\end{equation}
The \emph{supertrace} of $A$ is defined as
\begin{equation}
    \str(A)=\tr(\alpha)-\tr(\beta).
\end{equation}
The supertrace is  independent of the choice of (homogeneous) basis. It has the following properties
\begin{itemize}
    \item Consistency: $\str(BF)=0=\str(FB)$ where $B\in l(V)_{\Bar{0}}$ and $F\in l(V)_{\Bar{1}}$.
    \item Supersymmetry: $\str(TA)=(-)^{|T||A|}\str(AT)$ \;\;$\forall\, T,A\in l(V)$.
    \item Invariance: $\str([T,A])=0$ \;\;$\forall\; T,A\in l(V)$ where $[\cdot,\cdot]$ denotes the super-Lie bracket.
\end{itemize}

If $\{T_i\}_{i\in I}$ are a linear representation $R$ of a super-Lie-algebra one can define a superinvariant tensor as follows
\begin{equation} \label{eq:Kijk}
    K^R_{ijk}=2\,\str_R(T_iT_jT_k)=\str_R([T_i,T_j]T_k)+\str_R(\{T_i,T_j\}T_k)=C(R)f_{ijk}+d_{ijk}(R).
\end{equation}
    $K^R_{ijk}$ satisfies the following equation, consequence of the properties of the supertrace:\footnote{
    $0=\str([T_l,T_iT_jT_k])=\str([T_l,T_i]T_jT_k)+(-)^{|l||i|}\str(T_i[T_l,T_j]T_k)+(-)^{|l|(|i|+|j|)}\str(T_iT_j[T_l,T_k]).$
}
\begin{equation} \label{eq:invariance_equation}
f_m{}^l{}_i\,K^R_{ljk}+(-)^{|l||i|}\,f_m{}^l{}_j\,K^R_{ilk}+(-)^{|l|(|i|+|j|)}\,f_m{}^l{}_k\,K^R_{ijl}=0.
\end{equation}
$C(R)$ is the \emph{index} of the representation, such that
\begin{equation}
    \str_R(T_iT_j)=C(R)g_{ij}
\end{equation}
and $g_{ij}$ is the Lie superalgebra Cartan-Killing metric, defined as \footnote{For $\mathfrak{su}(2,2|1)$ this  metric is non-degenerate.}
\begin{equation}
    g_{ij}=\str_{\mathrm{super\text{-}adj}}(T_iT_j).
\end{equation}
$f_{ijk}$ are therefore related to the structure constants $f_i{}^l{}_j$ as follows:
\begin{equation}
    f_{ijk}=f_i{}^l{}_j\,g_{lk}.
\end{equation}
 $f_{ijk}$  does not depend on the representation $R$ because the Cartan-Killing metric is the \emph{unique} rank-two (super)-symmetric invariant tensor for a given Lie-superalgebra.  $f_{ijk}$  is completely  ``anti-symmetric'' in the graded sense, that is
\begin{equation}
    f_{jik}=-(-)^{|i||j|}f_{ijk},  \qquad f_{ikj}=-(-)^{|k||j|}f_{ijk}.
\end{equation}
 $d_{ijk}(R)$ is instead  completely ``symmetric"  in the graded sense
\begin{equation} \label{eq:dijk_simm1}
    d_{jik}(R) =(-)^{|i||j|}d_{ijk}(R),  \qquad d_{ikj}(R)=(-)^{|k||j|} d_{ijk} (R).
\end{equation}
The tensor $d_{ijk}(R)$ could in principle, for a generic superalgebra,  depend on the representation.
However we computed  solutions to the invariance equation (\ref{eq:invariance_equation})  for the  $d=4,\,\mathcal{N}=1$ superconformal algebra and we obtained two linearly independent solutions: one which coincides (up to a multiplicative constant) with the lowered structure constants $f_{ijk}$ and another solution with precisely the symmetry properties of $d_{ijk}(R)$. Hence there is a \emph{unique} rank three invariant tensor of  $\mathfrak{su}(2,2|1)$ with its symmetry properties, up to a multiplicative constant. Therefore,
\begin{equation}\label{eq:dijk}
d_{ijk}(R) = 2\, A(R) \, d_{ijk},
\end{equation}
where $d_{ijk}$ is independent of the representation $R$.  

$d_{ijk}$ is related to the tensor $\Tilde{d}_{ijk}$ which defines the invariant polynomial $P_3(\bm{F})$,  (see eq. (\ref{sec4:P3dtilde})), as follows 
\begin{equation} \label{appendix:eq:str_P3}
 \!\!\str_R(\bm{F}^3)=  A(R)\, (-)^{|i||j|+|i||k|+|j||k|}\,d_{ijk}\, \bm{F}^i\,\bm{F}^j\,\bm{F}^k \equiv   A(R)\, \Tilde{d}_{ijk}\bm{F}^i\bm{F}^j\bm{F}^k= A(R)\, P_3(\bm{F}),
\end{equation}
that is 
\begin{equation}
 \Tilde{d}_{ijk} = (-)^{|i||j|+|i||k|+|j||k|}\,d_{ijk}.
 \end{equation}
The sign factor relating $d_{ijk}$ to $\Tilde{d}_{ijk}$, which is caused by the fact that both the curvatures and generators are graded, is invariant under exchange and cyclic permutation of its indices. Therefore it does not change the symmetry properties (\ref{eq:dijk_simm1}). 
It is important to keep in mind that  the ``invariance'' equation  satisfied by the tensor  $\Tilde{d}_{ijk}$ ---  which ensures BRST invariance of $P_3(\bm{F})$ ---   is different, although equivalent, to eq. (\ref{eq:invariance_equation}) valid for ${d}_{ijk}$:
\begin{equation} \label{appendix:eq:invariance_equation_Tilde}
f_m{}^l{}_i\,\tilde{d}_{ljk}+(-)^{|j||m|}f_m{}^l{}_j\,\tilde{d}_{ilk}+(-)^{|m||k|+|m||j|}\,f_m{}^l{}_k\,\tilde{d}_{ijl}=0.
\end{equation}
The coefficient $A(R)$  in equation (\ref{eq:dijk})  is  the \emph{anomaly coefficient}:  it describes the contribution to the superconformal anomaly of matter in a representation $R$ of the superconformal algebra.

\section{The special Lorentz equivariant anomaly cocycle}\label{appendix:lorentzcocycle}

In this section we write the  ghost number 1, 4-form components of the special Lorentz equivariant  anomaly cocycle, eqs. (\ref{eq:specialLorentzequiv3})-(\ref{eq:specialLorentzequiv5}). We separate the components associated to each  ghost.  

Let us introduce the combinations
\begin{equations}
&\theta^{(\sigma)}{}^a \equiv -\tfrac{1}{2}\, e^{a\mu}\, \partial_\mu\sigma, \label{thetasubsa}\\
&\theta^{(\zeta)}{}^a \equiv  - i\, e^{a\mu}\,\overline{\zeta}\,\Tilde{\psi}_\mu,\\
& \theta^{(\eta)}{}^a\equiv - i\, e^{a\mu}\,\overline{\psi}_\mu\,\eta, \label{thetasubsc}
\end{equations} 
corresponding to the replacement 
\begin{equation}\label{sec:fullanomalythetasubstitution}
\theta^a\to  -\tfrac{1}{2}\, e^{a\mu}\, \bigl[\partial_\mu\,\sigma+ 2\,i\,(\overline{\psi}_\mu\,\eta+\overline{\zeta}\,\Tilde{\psi}_\mu\bigr)\bigr],
\end{equation}
which eliminates the trivial BRST doublet $b$ and $\Tilde\theta$ from the Lorentz-equivariant cocycle (\ref{sec4:Qequiv5final}).  
Therefore, after performing the substitutions (\ref{thetasubsa})-(\ref{thetasubsc}), the K-anomaly contributes  to the Weyl, the Q and the S anomaly. To obtain explicit results from the formulae below one needs to replace  $\tilde{\psi}$ and $f^a$ with their expressions (\ref{sec3:psitildesol}) and  (\ref{fabsolved}) in terms of the fundamental fields $e^a$, $\psi$ and $a$ (after having put $b$ to zero). 


\subsection{Cubic anomalies}
\begin{align}
\alpha\,\mathcal{A}_R^{(3)}=&15 \,\alpha \,({\tilde{F}^{R}}{})^2 + 3 \,\alpha \,({\tilde{F}^{W}}{})^2 -  \tfrac{3}{2}\, \alpha \,{\tilde{R}}_{ab} \,{\tilde{R}}^{ab} \,,
\end{align}
\begin{align}
\sigma\,\mathcal{A}_W^{(3)}=&-12 \,\sigma\,f_{a}\, f_{b}\, e_{c}\, e_{d}\,  \,\varepsilon^{abcd} + 6 \,\sigma\,f_{c}\, e_{d}\,  \,\varepsilon_{ab}{}^{cd} \,{\tilde{R}}^{ab} -  \tfrac{3}{4} \,\varepsilon_{abcd} \,\sigma\,{\tilde{R}}^{ab} \,{\tilde{R}}^{cd} \,, 
\end{align}
\begin{align}
\theta^a\,(\mathcal{A}_K^{(3)})_a=&-24 \,\theta_{a}\,\overline{\tilde{\psi}} \,\Gamma_5{} \,{\rho}\,e^{a} - 12i \,\theta_{a}\,\bar{{\rho}} \,\Gamma^{a}{} \,\Gamma_5{}\,{\rho}
\,+\nn\\
& - 24 \,\theta_{b}\,\overline{\tilde{\psi}} \,\Gamma^{ab}{} \,\Gamma_5{} \,{\rho}\,e_{a} - 12 \,\theta_{d} \,\overline{\tilde{\psi}}\,\Gamma^{a}{} \tilde{\psi} \,e_{b}\, e_{c}\,\varepsilon_{a}{}^{bcd}  \,,
\end{align}
\begin{align}
\overline{\zeta}\,\mathcal{A}_Q^{(3)}=0 \,,
\end{align}
\begin{align}
\bar{\eta}\,\mathcal{A}_S^{(3)}=&24 \,\bar{\eta}\,\Gamma_5{}\,{\rho} \,f^{a}\, e_{a} - 24 \,\bar{\eta}\,\Gamma^{ab}{} \,\Gamma_5{} \,{\rho} \,f_{a}\, e_{b} - 24 \,\bar{\eta}\,\Gamma^{a}{} \tilde{\psi} \, f_{b}\, e_{c}\, e_{d} \,\varepsilon_{a}{}^{bcd}  \,+\nn\\
&+ 60 \,\bar{\eta}\,\Gamma^{a}{} \tilde{\psi} \,{\tilde{F}^{R}}{} e_{a} + 6 \,\bar{\eta}\,\varepsilon_{abcd} \,\Gamma^{a}{} e^{b} \tilde{\psi} \,{\tilde{R}}^{cd} + 6 \,\bar{\eta}\,\Gamma_5{} \,\Gamma^{ab}{}\,{\rho} \,{\tilde{R}}_{ab}  \,+\nn\\
&+ 60i \,\bar{\eta}\,{\rho}\,{\tilde{F}^{R}}{}  + 12 \,\bar{\eta}\,\Gamma_5{} \,{\rho}\,{\tilde{F}^{W}}{} \,.
\end{align}

\subsection{Quartic anomalies}
\begin{align}
\alpha\,\mathcal{A}_R^{(4)}=&-24i \,\alpha\,\overline{\tilde{\psi}}\,\psi \,  f^{a} e_{a}  - 12i \,\alpha \,\overline{\tilde{\psi}}\,\psi \,{\tilde{F}^{W}}{}  + 24 \,\alpha  \,\bar{{\rho}} \,\Gamma^{a}{} \,\psi \,f_{a} \,+\nn\\
&+ 24i \,\alpha \,\overline{\tilde{\psi}} \,\Gamma^{ab}{} \psi\, f_{a}\, e_{b}  - 60 \,\alpha \,\overline{\tilde{\psi}} \,\Gamma_5{} \psi\,{\tilde{F}^{R}}{}  - 6i \,\alpha \,\overline{\tilde{\psi}} \,\Gamma^{ab}{} \psi \,{\tilde{R}}_{ab} \,,
\end{align}
\begin{align}
\sigma\,\mathcal{A}_W^{(4)}=&12i\,\sigma \,\overline{\tilde{\psi}}\,\Gamma^{ab}{} \psi\, f_{c}\, e_{d}\,  \,\varepsilon_{ab}{}^{cd}  - 6\,\sigma \,\overline{\tilde{\psi}} \,\Gamma_5{} \,\Gamma^{ab}{} \psi \,{\tilde{R}}_{ab}  \,,
\end{align}
\begin{align}
\theta^a\,(\mathcal{A}_K^{(4)})_a=&24i\, \theta_{a}\,\overline{\tilde{\psi}}\,\psi \,a\, e^{a}\,  + 24 \,\theta_{a}\,a\,\bar{{\rho}} \,\Gamma^{a}{} \psi + 24i \,\theta_{b}\,\overline{\tilde{\psi}}\,\Gamma^{ab}{} \psi \,a\, e_{a} \,,
\end{align}
\begin{align}
\overline{\zeta}\,\mathcal{A}_Q^{(4)}=&-24 \,\bar{\zeta} \,\Gamma^{a}{}\,{\rho} \,a\, f_{a} + 24i \,\bar{\zeta} \,\tilde{\psi} \,a\, f^{a}\, e_{a} + 12i  \,\bar{\zeta}\,\tilde{\psi} \,a\,{\tilde{F}^{W}}{}  + \tfrac{96}{5} \,\overline{\tilde{\psi}} \,\Gamma^{a}{} e_{a} \tilde{\psi} \,\bar{\zeta} \,\Gamma_5{} \tilde{\psi} \,+\nn\\
&+ \tfrac{48}{5} \,\bar{\zeta} \,\Gamma^{a}{} e^{b} \tilde{\psi}\,\overline{\tilde{\psi}} \,\Gamma_{ab}{} \,\Gamma_5{} \tilde{\psi}  -  \tfrac{48}{5} \,\bar{\zeta} \,\Gamma^{a}{} \,\Gamma_5{} e^{b} \tilde{\psi}\,\overline{\tilde{\psi}} \,\Gamma_{ab}{} \tilde{\psi}  + 24i \,\bar{\zeta} \,\Gamma^{ab}{} \tilde{\psi}\,a\, f_{a}\, e_{b}  \,+\nn\\
&+ 60  \,\bar{\zeta} \,\Gamma_5{}\tilde{\psi} \,a\,{\tilde{F}^{R}}{}  - 6i  \,\bar{\zeta} \,\Gamma^{ab}{} \tilde{\psi} \,a\,{\tilde{R}}_{ab} - 12i \,\bar{\zeta} \,\Gamma_{a}{} \,\Gamma_5{} \,{\rho}\,\overline{\tilde{\psi}} \,\Gamma^{a}{} \tilde{\psi}  + 24i \,\bar{\zeta} \,\Gamma_{ab}{} \psi \,\overline{\tilde{\psi}} \,\Gamma^{ab}{} \,\Gamma_5{} \,{\tilde{{\rho}}}  \,+\nn\\
&- 3 \,a \,{\tilde{R}}^{ab} \,(\lambda_0^J)_{ab} + 6 \,\overline{\tilde{\psi}} \,\Gamma^{ab}{} \,\Gamma_5{} \,{\rho} (\lambda_0^J)_{ab} + 3 \,\overline{\tilde{\psi}} \,\varepsilon_{abcd} \,\Gamma^{a}{} e^{b} \tilde{\psi}\, (\lambda_0^J)^{cd}\,,
\end{align}
\begin{align}
\bar{\eta}\,\mathcal{A}_S^{(4)}=&\tfrac{24}{5}i\,\bar{\eta}\,\varepsilon_{abcd} \,\Gamma^{a}{} \psi \,\overline{\tilde{\psi}} \,\Gamma^{cd}{} e^{b} \tilde{\psi} \, + \tfrac{48}{5}i \,\bar{\eta}\,\Gamma^{cd}{} e^{b} \tilde{\psi}\,\overline{\tilde{\psi}} \,\varepsilon_{abcd} \,\Gamma^{a}{} \psi \, \,+\nn\\
&- 24i \,\bar{\eta}\,\psi\,a\, f^{a}\, e_{a}  - 12i  \,\bar{\eta}\,\psi\,{\tilde{F}^{W}}{}a -  \tfrac{192}{5} \,\bar{\eta}\,\Gamma^{a}{} e_{a} \tilde{\psi} \,\overline{\tilde{\psi}} \,\Gamma_5{} \psi -  \tfrac{96}{5}\,\bar{\eta}\,\Gamma_5{} \psi \,\overline{\tilde{\psi}} \,\Gamma^{a}{} e_{a} \tilde{\psi}  \,+\nn\\
&-  \tfrac{96}{5} \,\bar{\eta}\,\Gamma_{ab}{} e^{b} \tilde{\psi}\,\overline{\tilde{\psi}} \,\Gamma_5{} \,\Gamma^{a}{} \psi  -  \tfrac{48}{5}\,\bar{\eta}\,\Gamma_5{} \,\Gamma^{a}{} \psi \,\overline{\tilde{\psi}} \,\Gamma_{ab}{} e^{b} \tilde{\psi}  \,+\nn\\
&+ 24i \,\bar{\eta}\,\Gamma_{a}{} \tilde{\psi}\,\bar{{\rho}} \,\Gamma^{a}{} \,\Gamma_5{} \psi  + 24i \,\bar{\eta}\, \Gamma^{ab}{} \psi\,a\, f_{a}\, e_{b}  \,+\nn\\
&- 60  \,\bar{\eta}\,\Gamma_5{}\psi \,a\,{\tilde{F}^{R}}{} \psi - 6i  \,\bar{\eta}\,\Gamma^{ab}{} \psi\,a \,{\tilde{R}}_{ab} + 12i \,\bar{\eta}\,\Gamma_{ab}{} \,\Gamma_5{} \,{\tilde{{\rho}}}\,\bar{\psi} \,\Gamma^{ab}{} \psi \,.
\end{align}

\subsection{Quintic anomalies}
\begin{align}
\alpha\,\mathcal{A}_R^{(5)}=&18 \,\alpha \,\overline{\tilde{\psi}} \,\Gamma^{a}{} \tilde{\psi} \,\bar{\psi} \,\Gamma_{a}{} \psi + 3 \,\alpha \,\overline{\tilde{\psi}}\,\Gamma^{ab}{} \tilde{\psi} \,\bar{\psi} \,\Gamma_{ab}{}  \psi \,,
\end{align}
\begin{align}
\sigma\,\mathcal{A}_W^{(5)}=&- \tfrac{3}{2}\,\sigma  \,\overline{\tilde{\psi}} \,\varepsilon_{abcd} \,\Gamma^{ab}{} \tilde{\psi} \,\bar{\psi} \,\Gamma^{cd}{} \psi \,,
\end{align}
\begin{align}
\theta^a\,(\mathcal{A}_K^{(5)})_a=&\tfrac{48}{5} \,\theta_{a}\,\overline{\tilde{\psi}} \,\Gamma_5{} \psi \,\bar{\psi} \,\Gamma^{a}{} \psi  + \tfrac{24}{5}i \,\theta^{b} \,\overline{\tilde{\psi}} \,\Gamma^{cd}{} \psi \,\bar{\psi} \,\varepsilon_{abcd} \,\Gamma^{a}{} \psi  \,+\nn\\
&-  \tfrac{36}{5}i \,\theta^{b}\,\overline{\tilde{\psi}} \,\varepsilon_{abcd} \,\Gamma^{a}{} \psi \,\bar{\psi} \,\Gamma^{cd}{} \psi  + \tfrac{72}{5} \,\theta^{b}\,\overline{\tilde{\psi}} \,\Gamma^{a}{} \,\Gamma_5{} \psi \,\bar{\psi} \,\Gamma_{ab}{} \psi \,,
\end{align}
\begin{align}
\overline{\zeta}\,\mathcal{A}_Q^{(5)}=&- \tfrac{48}{5} \,\bar{\zeta} \,\Gamma_5{} \tilde{\psi}\,\bar{\psi} f^{a} \,\Gamma_{a}{} \psi  -  \tfrac{48}{5}\,\bar{\zeta} \,\Gamma_{ab}{} \,\Gamma_5{} \tilde{\psi} \,\bar{\psi} f^{a} \,\Gamma^{b}{} \psi  \,+\nn\\
&+ 36 \,\bar{\zeta} \,\Gamma_{a}{} \psi\,a \,\overline{\tilde{\psi}} \,\Gamma^{a}{} \tilde{\psi}  + 6\,\bar{\zeta} \,\Gamma_{ab}{} \psi \,a \,\overline{\tilde{\psi}} \,\Gamma^{ab}{} \tilde{\psi}  + 12i \,\bar{\psi} \,\Gamma^{ab}{} \,\Gamma_5{} \psi \,\overline{\lambda_0^S} \,\Gamma_{ab}{} \tilde{\psi} \,+\nn\\
&+ \tfrac{96}{5}\,\bar{\zeta} f^{a} \,\Gamma_{a}{} \psi \,\overline{\tilde{\psi}} \,\Gamma_5{} \psi  + \tfrac{72}{5}\,\bar{\zeta} f^{a} \,\Gamma^{b}{} \tilde{\psi} \,\bar{\psi} \,\Gamma_{ab}{} \,\Gamma_5{} \psi  + \tfrac{72}{5}\,\bar{\zeta} f^{a} \,\Gamma^{b}{} \,\Gamma_5{} \tilde{\psi} \,\bar{\psi} \,\Gamma_{ab}{} \psi  \,+\nn\\
&-  \tfrac{96}{5}\,\bar{\zeta} f^{a} \,\Gamma^{b}{} \psi \,\overline{\tilde{\psi}} \,\Gamma_{ab}{} \,\Gamma_5{} \psi  + \tfrac{144}{5} \,\bar{\zeta} \,\Gamma_{ab}{} \,\Gamma_5{} \psi\,\overline{\tilde{\psi}} f^{a} \,\Gamma^{b}{} \psi  \,+\nn\\
&-  \tfrac{144}{5} \,\bar{\zeta} \,\Gamma_{ab}{} \psi\,\overline{\tilde{\psi}} f^{a} \,\Gamma^{b}{} \,\Gamma_5{} \psi  - 6i \,a \,\overline{\tilde{\psi}} \,\Gamma^{ab}{} \psi \,(\lambda_0^J)_{ab} \,,
\end{align}
\begin{align}
\bar{\eta}\,\mathcal{A}_S^{(5)}=&\,36\,\bar{\eta}\,\Gamma_{a}{} \tilde{\psi} \,a \,\bar{\psi} \,\Gamma^{a}{} \psi  + 6 \,\bar{\eta}\,\Gamma_{ab}{} \tilde{\psi}\,a \,\bar{\psi} \,\Gamma^{ab}{} \psi  \,+\nn\\
&+ \tfrac{36}{5}i \,\bar{\eta}\,\varepsilon_{abcd} f^{a} \,\Gamma^{b}{} \psi \,\bar{\psi} \,\Gamma^{cd}{} \psi -  \tfrac{24}{5}i \,\bar{\eta}\,\Gamma^{cd}{} \psi  \,\bar{\psi} \,\varepsilon_{abcd} f^{a} \,\Gamma^{b}{} \psi \,+\nn\\
&+ \tfrac{48}{5} \,\bar{\eta}\,\Gamma_5{} \psi\,\bar{\psi} f^{a} \,\Gamma_{a}{} \psi  + \tfrac{72}{5} \,\bar{\eta}\,\Gamma_5{} f^{a} \,\Gamma^{b}{} \psi\,\bar{\psi} \,\Gamma_{ab}{} \psi \,.
\end{align}

\bibliographystyle{JHEP}
\bibliography{ir}

\end{document}